\documentclass[fleqn,usenatbib]{mnras}

\usepackage{newtxtext,newtxmath}

\usepackage[T1]{fontenc}

\DeclareRobustCommand{\VAN}[3]{#2}
\let\VANthebibliography\thebibliography
\def\thebibliography{\DeclareRobustCommand{\VAN}[3]{##3}\VANthebibliography}

\usepackage{graphicx}	
\usepackage{amsmath}	
\usepackage{multirow}
\usepackage{diagbox}
\usepackage{subcaption}
\usepackage{cleveref}
\usepackage{multirow}
\usepackage{makecell}
\usepackage{booktabs}
\usepackage{geometry}
\usepackage{adjustbox}
\usepackage[table]{xcolor}
\usepackage{url}
\usepackage{natbib}
\usepackage[bb=stixtwo]{mathalpha}
\usepackage[superscript]{cite}
\usepackage{hyperref}
\usepackage{scrextend} 
\usepackage{pdflscape}
\usepackage{rotating}
\usepackage{longtable}
\usepackage{tabularx}
\usepackage{tikz}
\usepackage{arydshln}

\deffootnote[1.5em]{1.5em}{1em}{\textsuperscript{\thefootnotemark}\,}

\title{Polycyclic Aromatic Hydrocarbons in Exoplanet Atmospheres: A Detectability Study}

\author[R. Arenales-Lope et al.]{
Rosa Arenales-Lope$^{1,2}$,
Karan Molaverdikhani$^{1,2}$,
Dwaipayan Dubey$^{1,2}$,
Barbara Ercolano$^{1,2}$,
\newauthor Fabian Grübel$^{1,2}$
and Christian Rab$^{1,3}$
\\
$^{1}$ Universitäts-Sternwarte, Fakultät für Physik, Ludwig-Maximilians-Universität München, Scheinerstr. 1, 81679 München, Germany\\
$^{2}$ Exzellenzcluster ‘Origins’, Boltzmannstr. 2, D-85748 Garching, Germany\\
$^{3}$ Max-Planck-Institut für Extraterrestrische Physik, Giessenbachstr. 1, 85748 Garching, Germany
}

\date{Accepted XXX. Received YYY; in original form ZZZ}

\pubyear{2015}

\begin{document}
\label{firstpage}
\pagerange{\pageref{firstpage}--\pageref{lastpage}}
\maketitle

\begin{abstract}
In this paper, we explore the detectability of polycyclic aromatic hydrocarbons (PAHs) under diverse planetary conditions, aiming to identify promising targets for future observations of planetary atmospheres. Our primary goal is to determine the minimum detectable mass fractions of PAHs on each studied planet. We integrate the one-dimensional self consistent model petitCODE with petitRADTRANS, a radiative transfer model, to simulate the transmission spectra of these planets. Subsequently, we employ the PandExo noise simulator using the NIRSpec PRISM instrument aboard the JWST to assess the observability. Then, we conduct a Bayesian analysis through the MULTINEST code. Our findings illustrate that variations in C/O ratios and planet temperatures significantly influence the transmission spectra and the detectability of PAHs. 
Our results show that planets with [Fe/H]=0 and 1, C/O=0.55, and temperatures around 1200~K are the most promising for detecting PAHs, with detectable mass fractions as low as 10$^{-7}$, or one thousandth of the ISM abundance level. For colder planets with lower metallicities and C/O ratios, as well as hotter planets with carbon-rich atmospheres, PAHs can be detected at abundances around 10$^{-6}$. These results aid our strategy for selecting targets to study PAHs in the atmospheres of exoplanets.
\end{abstract}



\section{Introduction}
Polycyclic Aromatic Hydrocarbons (PAHs) are complex organic molecules predominantly found in star-forming regions, interstellar clouds, and around aging stars. These molecules are primarily composed of interconnected benzene rings. Exposed to ultraviolet (UV) light from nearby stars, PAHs become excited and release energy by emitting infrared light at wavelengths that are characteristic of their molecular structure \citep{LEGER1993473, Allamandola1985}. Their distinctive infrared features have been identified across a broad spectrum of astronomical environments, including planetary nebulae, protoplanetary nebulae, reflection nebulae, HII regions, circumstellar envelopes, and external galaxies, marking them as significant constituents within these systems \citep{Tielens1999}.

Photoelectric emission from PAHs is very efficient process to convert infrared radiation into gas heating and may have a substantial impact on atmospheric loss rates \citep{gorti2009, mitani2020}. These processes are critical in the thermochemical dynamics of gas atmospheres within newborn exoplanets and protoplanetary discs \citep{gorti2009, Ercolano2022ObservationsExoplanets}. In addition, theoretical models postulate a crucial role for PAHs in the synthesis of amino acids and nucleotides, which are the fundamental building blocks of proteins and RNA, respectively \citep{Bernstein2002,Ehrenfreund2006,Ehrenfreund2007,rapacioli2006,EhrenfreundCharnley2000,wakelam2008,galliano2008}. This synthesis underscores the potential association between PAHs and prebiotic chemistry, suggesting their involvement in biogenesis.

\citet{giese2022} found that PAHs did not react with ammonia or ammonia salts at temperatures up to 150°C. Yet, they noted that PAHs could undergo pyrolysis at higher temperatures or be activated using Fenton's reagent, leading to the creation of functionalized aromatic compounds.

PAHs have also been detected in our solar system, for example, in the nitrogen-rich atmosphere of Saturn’s moon Titan \citep{Lopez-Puertas2013LargeAtmosphere, Dinelli2013}. This suggests that PAHs might be present in the atmospheres of exoplanets as well. However, detecting these molecules in exoplanetary atmospheres has posed a considerable challenge due to technological limitations.
The deployment of space-based observatories such as the James Webb Space Telescope\footnote{\url{https://jwst-docs.stsci.edu/jwst-near-infrared-spectrograph}}(JWST)\citep{Gasman2022InvestigatingJWST,Beichman2014} and upcoming missions such as Twinkle\footnote{\url{https://bssl.space/twinkle/}}\citep{Edwards2019, Stotesbury2022} and Ariel\footnote{\url{https://www.cosmos.esa.int/web/ariel}}\citep{Pascale2018,tinetti2021ariel}, along with ground-based telescopes like the Extremely Large Telescope\footnote{\url{https://elt.eso.org}}(ELT) \citep{Houlle2021}, represents a significant improvement in our observational capabilities.

\citealt{Dubey_2023} investigated ex-situ PAH formation in the thermalized atmospheres of irradiated hot-Jupiters around Sun-like stars. They concluded that planets with an effective temperature of around 1300 K, higher C/O ratios, and increased metallicity are crucial for facilitate PAH formation under thermalized conditions, with C/O exhibiting dominant influence.

Building upon these findings, our study explores the detectability of PAHs under various planetary conditions through a detailed parameter space analysis. This analysis aims to identify conditions that are optimal for their detection. Using simulated observations from the Time Exposure Calculator for JWST (PanDExo) \citep{batalha2017pasp} and petitRADTRANS (pRT) code \citep{molliere2019,molliere2020,Alei_2022}. Additionally, Bayesian Inference tools such as MultiNest \citep{feroz2008multinest,feroz2009multinest,feroz2019multinest} and its \texttt{Python} interface called PyMultinest \citep{Buchner2014b} aid in refining our understanding of atmospheric composition and its implications for the habitability of exoplanets.

The paper is organized into three sections: Section~\ref{sec:methodology} details our approach to simulating JWST observations and the analytical process for data interpretation. Section~\ref{sec:results} presents our findings on PAH detection and abundance in exoplanetary atmospheres,  discusses the significance of these findings for exoplanetary science, and suggests directions for future research. Finally, our conclusions are presented in Section~\ref{sec:conclusions}.

\section{Methodology}\label{sec:methodology}

In this section, we detail our approach to characterizing exoplanet atmospheres and simulating their transmission spectra for subsequent analysis. Our methodology proceeds in the following sequence:

\begin{enumerate}
    \item First, we use petitCODE \citep{Mollière_2015,Molliere2017} for generating fifteen self-consistent models to calculate atmospheric abundances and temperature structure.
    \item We then introduce three levels of PAH abundances in the models, resulting in $3\times$15=45 forward models, Section \ref{sec:self_consistent_simulations}.
    \item Next, we employ pRT to model the transmission spectra based on these forward models.
    \item Afterward, we use PandExo to simulate observations with JWST and estimating uncertainties in these transmission spectra, Section \ref{sec:simulating_jwst_observations}.
    \item Finally, we use pyMULTINEST through pRT and  to retrieve the atmospheric abundances from these simulated observations and evaluate the potential detectability of PAHs, Section \ref{sec:retrieval_framework}. In order to perform this, we run three retrieval models with PAH (PAH-included models) and three retrieval models without PAH (baseline models), bringing the total number of retrievals to 45$\times$(3+3)= 270.
\end{enumerate}

To assess the detectability of PAHs in each of the 45 forward models, we compare the results of the three baseline models with their corresponding PAH-included models. This involves calculating Bayes factors to perform model selection and determine if PAHs are favoured for each forward model.

\subsection{Self-consistent simulations and synthetic spectra}
\label{sec:self_consistent_simulations}

First, we use petitCODE, a self-consistent model, to compute forward model atmospheres. petitCODE calculates the temperature structure and molecular abundances assuming equilibrium chemistry for planets with various temperatures (800K, 1200K, and 1600K), C/O ratios (0.3, 0.55, and 1.0), and metallicities ([Fe/H]= -1 (0.1$\times$ Solar), 0 ($\times$ Solar), and 1 (10$\times$ Solar)). Table \ref{tab:spec_param}  provides an overview of the parameters used in the forward models. \citealp{Dubey_2023} found that PAH abundance does not vary significantly with increasing metallicity, concluding that the C/O ratio is the dominant factor influencing PAH abundance. However, to ensure a comprehensive analysis, we have included a case where the C/O ratio is held constant (0.55) while varying metallicity ([Fe/H] = -1, 0, and 1). This allows us to isolate and examine the effect of metallicity in detail.

Next, we use the petitRADTRANS (pRT) package, a \texttt{Python} tool designed for the spectral characterization of exoplanet atmospheres, to model the transmission and emission spectra. pRT provides a fast, accurate, and versatile platform for calculating both transmission and emission spectra. It handles opacities in two ways: correlated-k ($\lambda/ \Delta \lambda$ = 1000) and line-by-line ($\lambda/ \Delta \lambda$ = 10$^6$) spectra. 
For this study, we employ correlated-k spectra to accelerate the retrievals while maintaining accuracy \citep[e.g.][]{molliere2020}. Clouds are incorporated into the models as continuum opacities \citep{molliere2020}, with parameters including mean particle size, cloud mass fraction (X$\mathrm{_{cloud}}$), particle size distribution ($\sigma_g$), and the settling parameter f$\mathrm{_{sed}}$ \citep[see e.g.][]{Ackerman2001}. We select C/O ratios and [Fe/H] based on the findings of \citep{Dubey_2023}, which indicate that for a solar C/O ratio, the abundance of PAHs remains relatively constant despite variations in metallicity.

The model atmospheres include all major contributor opacities to the infrared spectra: H$_2$O, CO (HITEMP; see \citealt{rothman2010}), HCN (ExoMol; see \citealt{Barber2014}), Ti, (Kurucz database; see \citealt{Kurucz2017}), CO$_2$ \citep{Yurchenko2020}, CH$_4$ \citep{Yurchenko2017}, AlH \citep{Yurchenko2018}, MgH, NH$_3$ \citep{Coles2019}, H$_2$S \citep{Azzam2016}, two hydrocarbons (C$_2$H$_2$ and C$_2$H$_4$; see \citealt{Chubb2020} and \citealt{Mant2018}), Ca, Na, K \citep{Allard2019}, and two strong optical absorbers (TiO and VO; see \citealt{Allard1996}). We use H$_2$-H$_2$ \citep{Borysow2001,Borysow2002} and H$_2$-He H$_2$-He \citep{Borysow1988,Borysow1989a, Borysow1989b} as CIA and H$_2$ \citep{Dalgarno1962}, He \citep{Dalgarno1965}, and N$_2$ \citep{Thalman2014,Thalman2017} as Rayleigh opacity species.

For PAHs, we utilize absorption cross-sections of neutral circumcoronene, composed of 54 carbon and 18 hydrogen atoms, without including scattering effects \citep{Li2001InfraredMedium,2007ApJ...657..810D}. These molecules are treated as condensate species with vertically constant abundances, ranging from $10^{-5}$ to $10^{-7}$. This range is chosen based on findings by \citet{Tielens2008} and \citet{gredel_2011}, who estimated the abundance of neutral PAHs in the ISM at $\mathrm{\approx 3 \times 10^{-7}}$ relative to hydrogen nuclei \citep{Tielens2008, gredel_2011}. All together, 45 forward models are generated for further processing and assessment.

\begin{table}
  \centering
   \caption{Parameters employed for the generation of 45 synthetic observations across various atmospheric set-ups. X$\mathrm{_{PAH}}$ denotes the mass fraction of PAHs, as defined by \citealt{Li2001InfraredMedium}; [Fe/H] represents the metallicity; f$\mathrm{_{sed}}$ is the sedimentation factor; and $\mathrm{\sigma_g}$ specifies the log-normal distribution of particle sizes.}
  \begin{tabular}{cc|cc} 
  \hline \hline
     Parameter& Value& Parameter& Value\\ 
     \hline
     Temp & 800,1200, 1600 K& f$\mathrm{_{sed}}$& 2.0\\ 
     R$\mathrm{_{p}}$& 0.83R$\mathrm{_{Jup}}$& $\mathrm{\sigma_{g}}$& 1.05\\ 
     C/O& 0.3, 0.55, 1.0& log(g) (cm$s^{-2}$)& 3.02\\ 
     $\mathrm{\left[Fe/H\right]}$& -1, 0, 1 & log(P/bar)& -2\\ 
 log$\mathrm{\left(X_{PAH}\right)}$& -5, -6, -7& &\\
    \hline
  \end{tabular}

  \label{tab:spec_param}
\end{table}

\subsection{Simulating JWST observations}
\label{sec:simulating_jwst_observations}

Following the computation of the 45 forward model spectra, we emulate JWST observations by estimating uncertainties using PanDExo.

In our setup, we utilize the NIRSpec PRISM because of its broad wavelength coverage (0.6 to 5.3 microns), which allows us to capture the spectral slope in the optical and 3$\mathrm{\mu}$m features from PAHs. Since these molecules have broad and diffuse spectral features \citep{Tielens2008}, low-resolution instruments like the PRISM are more effective for detecting these signals, as they do not require the fine spectral detail that higher-resolution instruments provide. Additionally, as the PRISM is more suitable for faint targets \citep{price2024}, brighter stars would require alternative configurations, such as NIRISS or NIRSpec G395M/H, to avoid saturation.

Configured specifically for NIRSpec PRISM's subarray, PandExo requires instrument-dependent parameters, for which we set a saturation level of 80$\%$, a noise floor of zero and one transit of 3 hours \citep{Ercolano2022ObservationsExoplanets} and 4 hours of out-of-transit baseline. These settings are chosen to optimize the dataset for subsequent retrieval. Although we recognize the inherent noise floor in telescope observations, we assume an ideal scenario with a noise floor set to zero. This assumption is made to isolate and analyze the performance of the retrieval algorithms without additional noise interference, allowing us to better understand the theoretical limits and the impact of other observational parameters. The elements of this dataset include the wavelengths, the resultant spectrum with stochastic noise, and the error featuring the noise floor. We have set a star with an apparent magnitude of 9.5 (K-band). The star's brightness affects spectral uncertainties and the detectability of PAHs. This brightness is comparable to stars like WASP-39 and WASP-6 \citep[][accepted]{grubel2024detectability}, both commonly used in exoplanet studies and suitable for PAH detection under these observational conditions. These planets provide representative examples of the type of systems where PAH features could be identified.

Additionally, we use a binning tool in the PandExo package to accelerate the retrieval process by downsizing the dataset. The spectral resolution per bin is set at a resolution of 100. This processed data, with error bars provided by PandExo, is then used for molecular retrievals employing pRT.

\begin{figure*}
\centering
\includegraphics[width=\textwidth]{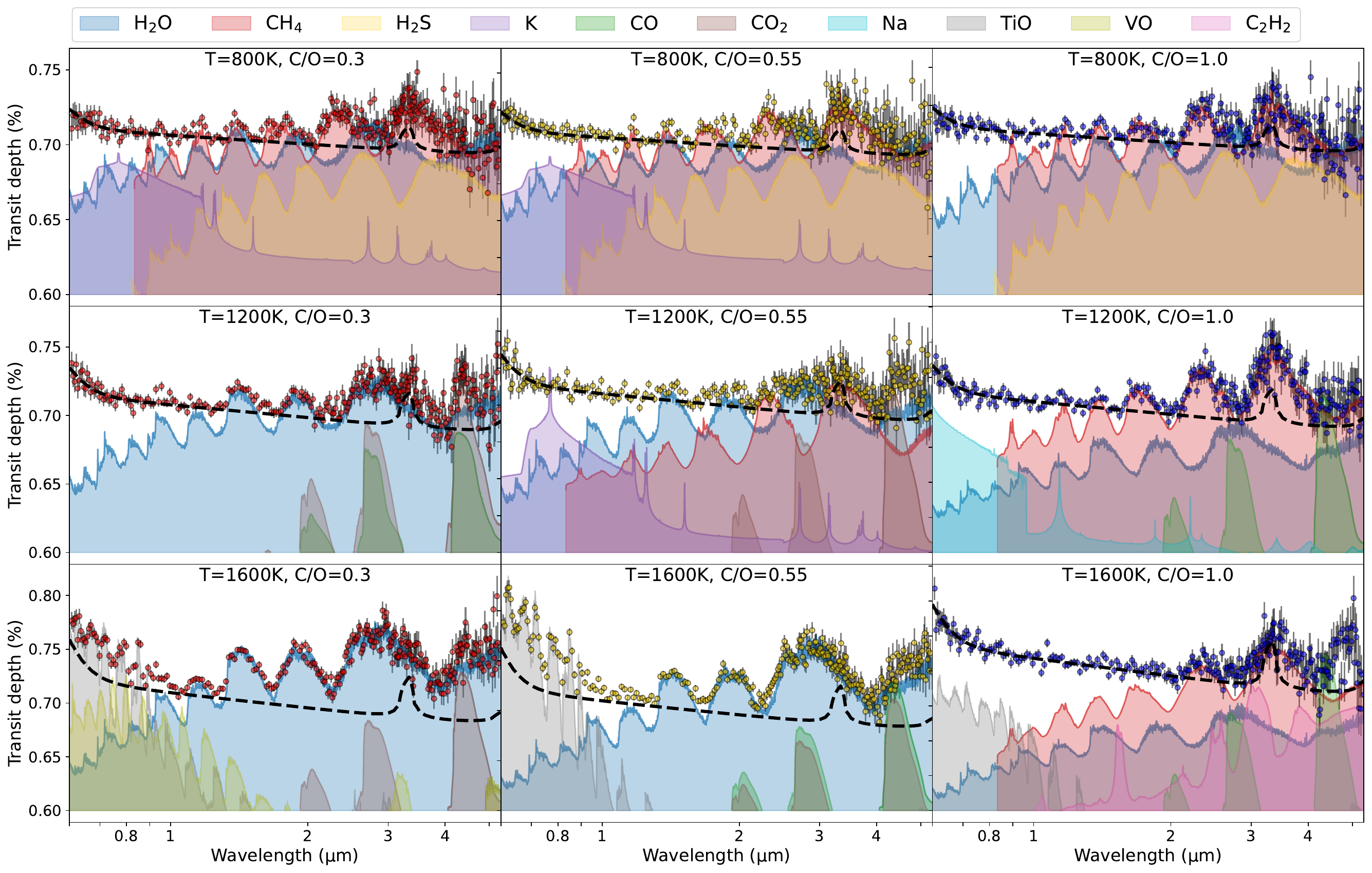}
\caption{Synthetic spectra for $\mathrm{X_{PAH}}$=10$^{-5}$ at temperatures of 800~K (top), 1200~K (middle), and 1600~K (bottom). The forward-modeled spectra (depicted as straight lines) and PandExo-simulated observations (represented by dotted points) for each planet are illustrated. The C/O ratios of 0.3 (left), 0.55 (center) and and 1.0 (right) are displayed in red, in red, yellow and blue, respectively. Various colors are assigned to indicate the contribution of each molecule to the model spectrum. While the primary features are attributed to H$_2$O, CH$_4$, CO, and CO$_2$, the spectrum also showcases a diverse range of other molecules.}

\label{fig:forward_model}
\end{figure*}

\begin{figure*}
\centering
\includegraphics[width=\textwidth]{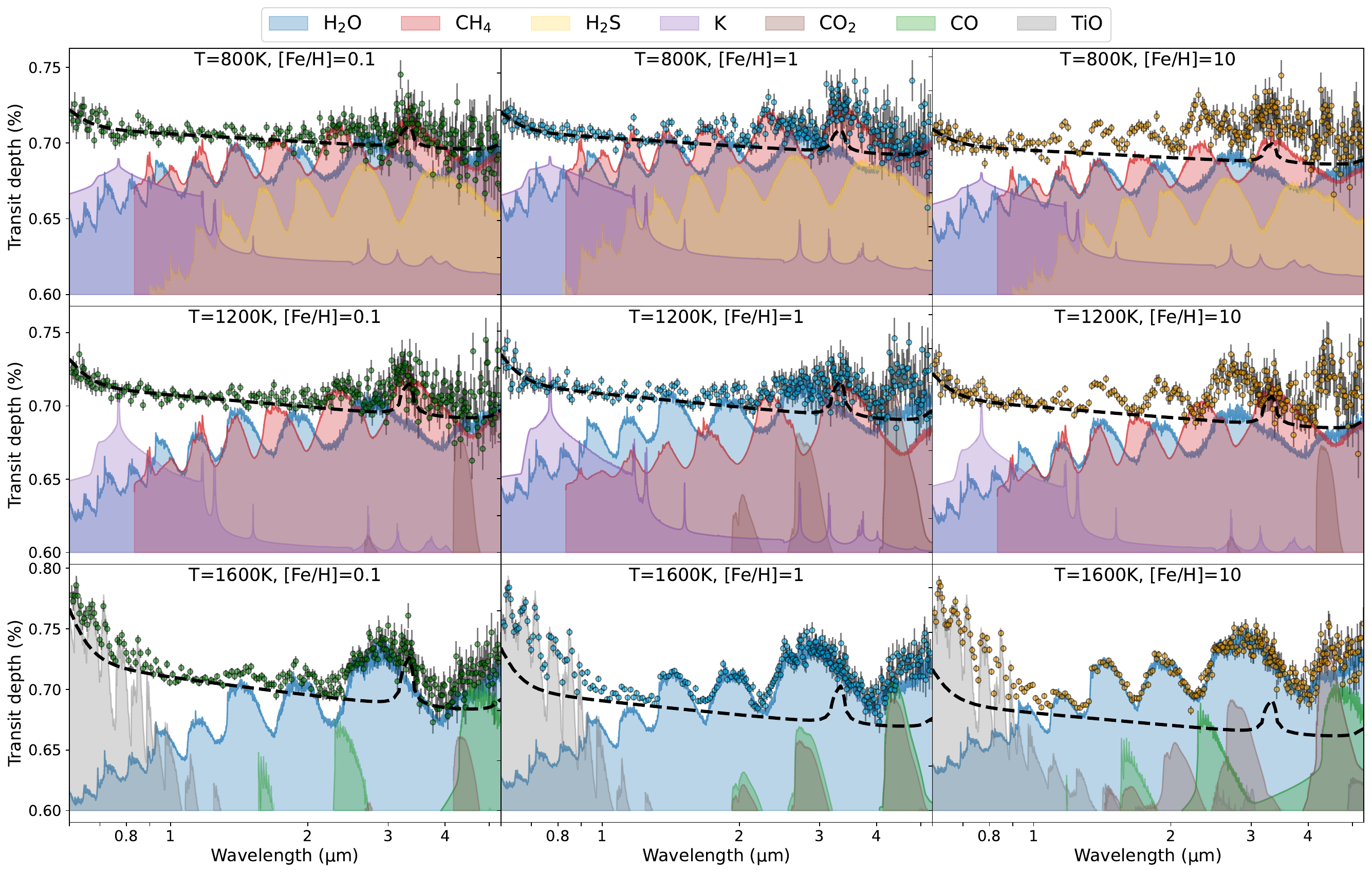}
\caption{Similar to Fig. \ref{fig:forward_model}, synthetic spectra for $\mathrm{X_{PAH}}$=10$^{-5}$ at temperatures of 800~K (top), 1200~K (middle), and 1600~K (bottom) with constant C/O (0.55) and varying metallicities ([Fe/H]) are displayed for 0.1 (left) in green, 1 (middle) in cyan and 10 (right) in yellow.}

\label{fig:3metal_5}
\end{figure*}

An illustrative representation of the simulated spectra for a PAH abundance of 10$^{-5}$, along with their major opacity contributors are shown in Figures \ref{fig:forward_model} and \ref{fig:3metal_5}. For PAH abundance of 10$^{-6}$ and 10$^{-7}$, see Figures \ref{fig:forward_model_6} and \ref{fig:3metal_6}, and Figures \ref{fig:forward_model_7} and \ref{fig:3metal_7}, respectively.

\subsection{Retrieval Framework}
\label{sec:retrieval_framework}

\begin{table}
    \centering
    \caption{Priors for retrievals. $\mathcal{U}$ stands for a uniform distribution, with the two parameters being the range boundaries and X$_i$ is the mass fraction of absorber species i. The unit of $\textit{Temp}$ is K, the unit of $\mathrm{R_p}$ is $\mathrm{R_{Jup}}$, and the unit of $\mathrm{P_{cloud}}$ is bar}
    \begin{tabular}{cc|cc}
    \hline \hline
         Parameter & Prior & Parameter & Prior \\  \hline
         Temp & $\mathcal{U}$(200,2200) & $\mathrm{\log P_{cloud}}$ & $\mathcal{U}$(-8,11) \\
         R$\mathrm{_{p}}$ & $\mathcal{U}$(0.2,1.2) &$\mathrm{\kappa_0}$&$\mathcal{U}$(-4,14) \\
         $\mathrm{log}$ g &$\mathcal{U}$(2.5,6.5) & $\mathrm{\gamma_{scat}}$&$\mathcal{U}$(-20,22) \\
         log(X$_{i}$) & $\mathcal{U}$(-20,0) & log(X$\mathrm{_{H_2O}}$) & $\mathcal{U}$(-5,0) \\
    \hline
    \end{tabular}

    \label{tab:param_retrieval}
\end{table}

In our retrieval process, we use the integrated pyMULTINEST algorithm, an implementation of the MultiNest method, integrated within the pRT code. Our goal is to identify the model that best fits the observed data and extract critical parameters, such as PAH abundance, chi-squared ($\chi^2$), and maximum marginal likelihood ($\ln Z$).

We start by running retrievals for each individual molecule to evaluate their impact, selecting the molecule with the best $\chi^2$ value. We then iteratively add the next most impactful molecule, as determined by the chi-squared improvement, and continue this process. This systematic approach ensures that we incorporate the most significant molecules into our retrieval studies, continuing until there is no substantial difference between consecutive chi-squared values and PAH mass fraction posteriors.

This method allows us to build a comprehensive model that accurately represents the observed data while focusing on the most influential molecular contributors to the exoplanet atmosphere's spectral characteristics.

Previous studies have highlighted the potential for systematic biases in signal-to-noise ratio (S/N) for observations made by the JWST when simplifying assumptions, such as using isothermal or one-dimensional models, are applied in the analysis of transmission spectra \citep{Rocchetto2016,MacDonald2020,Welbanks2024}. Despite these biases, our methodology employs an isothermal transmission model to simplify the retrieval process. The implications of this approach will be discussed in Section \ref{sec:results}.

We explore three model set-ups: 

\textbf{Model 1:} This model exclusively incorporates PAHs as the primary haze opacity source. Their spectral features are characterized by optical slopes which can often be attributed to phenomena such as super-Rayleigh scattering \citep{Ohno_2020} and a prominent 3.3 $\mu$m feature, which is likely produced by C-H stretching modes \citep{2007ApJ...657..810D}.
It aims to assess whether the presence of hazes could be inferred from the atmospheric data. However, a positive detection under this model would not exclusively indicate the presence of PAHs and the detected aerosols could be any other particles with similar optical properties.

\textbf{Model 2:} This setup includes PAHs along with power-law clouds, characterized by the initial opacity ($\kappa_0$), the scattering slope ($\gamma_{\text{scat}}$), and a baseline wavelength ($\lambda_0$). The opacity adds to the scattering cross-section is determined by the equation:
\begin{equation}
    \kappa = \kappa_0 \left(\frac{\lambda}{\lambda_0}\right)^{\gamma_{\text{scat}}}
\end{equation}
where $\kappa_0$ (in cm$^2$/g) represents the opacity at a reference wavelength of $\lambda_0 = 0.35\, \mu m$, and $\gamma_{\text{scat}}$ dictates the wavelength dependence. A positive PAH detection in this model would suggest PAHs are favored over a parameterized Rayleigh slope model .\\
\textbf{Model 3:} This model further extends Model 2 by incorporating a Grey cloud deck ($\mathrm{P_{cloud}}$) into the retrieval analysis. This mimics the presence of clouds, characterised by reducing or eliminating absorption features \citep{Chubb_2022}. A PAH detection under this setup would indicate the data is still favouring PAH over power-law clouds and Grey cloud decks. 

We configure our models with several fixed parameters: stellar radius (R$_{*}$), the settling parameter (f$\mathrm{_{sed}}$), the lognormal distribution width for particles ($\mathrm{\sigma_{g}}$), and the effective particle radius. Additionally, there are free parameters such as surface gravity (log$ (g)$) and planet radius (R$\mathrm{_{P}}$). The priors of these free parameters are summarized in Table \ref{tab:param_retrieval}.

\begin{table}
    \centering
    \caption{Baseline models utilized on the retrievals for the different planet's set-ups.}
    \begin{tabular}{ccc}
    \hline \hline
         Model & Baseline & Extra parameters  \\ 
         \hline
         1 & Cloud-free & - \\ 
         2 & Power-Law cloud & $\mathrm{\gamma_{scat}}$, $\mathrm{\kappa_0}$ \\
         3 & Power-Law cloud, Cloud deck & $\mathrm{\gamma_{scat}}$, $\mathrm{\kappa_0}$, P$\mathrm{_{cloud}}$\\
         \hline \hline
    \end{tabular}
    \label{tab:baselines}
\end{table}
Our objective is to assess whether the inclusion of PAHs offers a superior match to observed atmospheric data. This assessment is quantified using the Bayesian factor \citep{Trotta2008BayesCosmology}. We conducted a total of 270 retrieval analyses for each model, spanning 45 different planetary scenarios for both baseline and PAH-included setups.

\subsection{Detection Significance}
\label{sec:detection_significance}

To evaluate the detectability of PAHs at specific mass fractions, initially we examine the resulting reduced $\chi^2$ of each baseline models and models augmented by PAHs. This analysis serves to gauge the goodness-of-fit of each model to the observed data. Subsequently, we employ Bayesian evidence, denoted as \(\ln Z\). This numerical measure, derived via the Multinest algorithm \citep{feroz2008multinest,feroz2009multinest,feroz2019multinest}, reflects the likelihood of a model given the observed data.\\
The Bayes factor, (\(\ln B_{01}\)), provides a comparative metric of evidence strength between two models:
\begin{equation}
    \centering
    \ln B_{01} = \ln Z_1 - \ln Z_2  
    \label{eq:bayes}
\end{equation}

Here, \(\ln Z_1\) and \(\ln Z_1\) denote the Bayesian log-evidence for the baseline model and an alternative model, respectively. This factor is crucial for assessing the relative significance of models, with the detection significance ($\sigma$) being derivable from it \citep{Trotta2008BayesCosmology}.
The detection significance is a key indicator of our model's reliability and robustness in detecting PAHs. A higher \(\ln B_{01}\) suggests stronger evidence for the preferred model, enhancing the confidence in our detection outcomes and guiding the interpretation of results and decision-making regarding PAH presence.  We categorized the detection significance according to \citet{Trotta2008BayesCosmology,trotta2017bayesian} and \citet{refId0} as very strong for $\ln B_{01}$ greater than 11, strong for values between 5 and 11, moderate for values between 2.5 and 5, weak for values between 1 and 2.5, and inconclusive for valuer less than 1.

\section{Results and Discussion} \label{sec:results}
\begin{figure*}  
  \centering  
  \begin{minipage}{0.9\linewidth}
    \includegraphics[width=\textwidth]{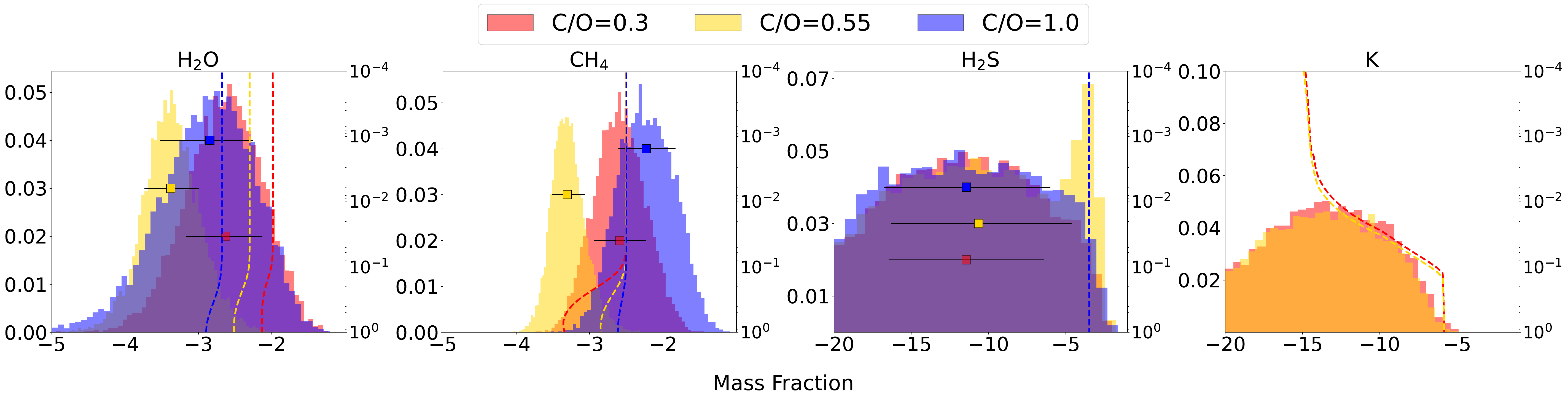}
    \subcaption{T=800K.}
    \label{fig:800-5-results_dist}
  \end{minipage}
  \begin{minipage}{0.9\linewidth}
    \includegraphics[width=\textwidth]{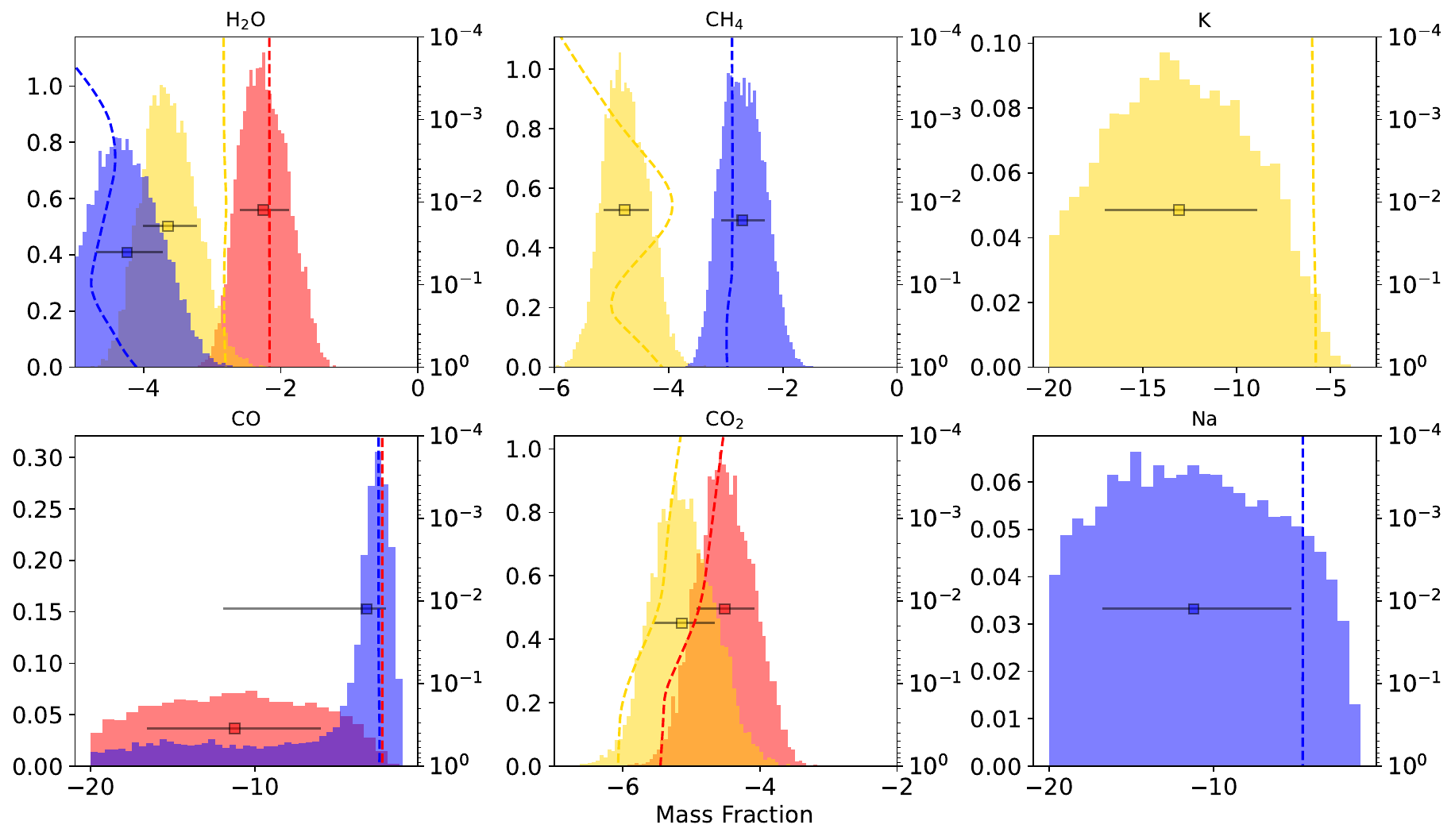}
    \subcaption{T=1200K}
    \label{fig:1200-5-results_dist}
  \end{minipage}
  \begin{minipage}{0.9\linewidth}
    \includegraphics[width=\textwidth]{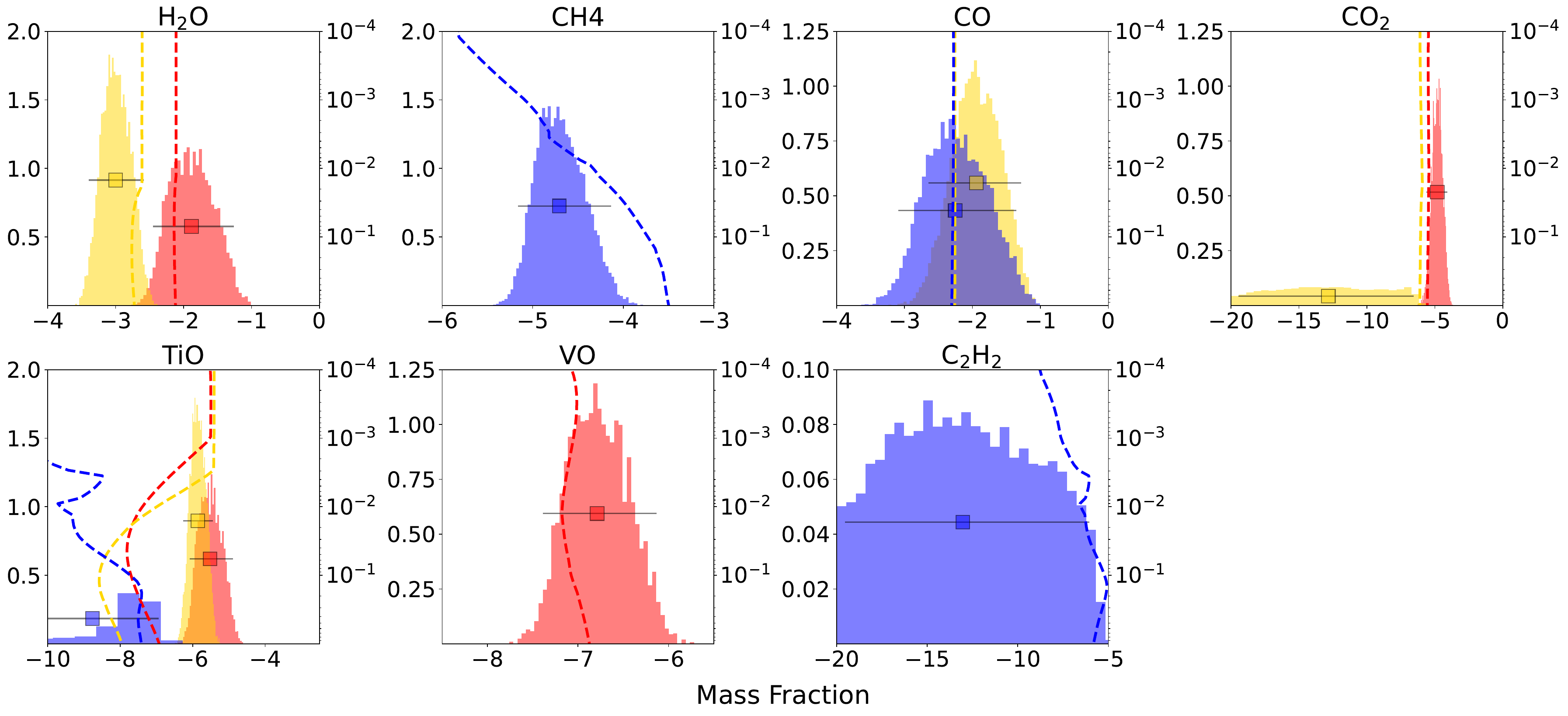}
    \subcaption{T=1600K}
    \label{fig:1600-5-results_dist}
  \end{minipage}
  \caption{Retrieved posterior probability distributions (left y-axis) and pressure profiles (right y-axis) for abundances of different molecule's planet set-up described in Section \ref{sec:methodology} for $\mathrm{X_{PAH}=10^{-5}}$. C/O ratios 0.3, 0.55 and 1.0 are depicted in red, yellow, and blue, respectively. Abundances inputs at different pressures are shown in dashed lines. The error bar denotes each distribution's median and corresponding 1$\sigma$ interval. The abundance estimates are shown in Table \ref{tab:all_results_jwst_5}}
  \label{fig:posteriors}
\end{figure*}

Our primary aim is to quantify if PAHs are observable in each 45 forward models. We present detailed results for PAH abundances of $10^{-5}$, $10^{-6}$, and $10^{-7}$. The tables and figures for the latter two abundances are provided in the appendix. 

Table \ref{tab:all_results_jwst_5} shows the retrieved posterior probabilities and their input values from the 27 forward models, taking their altitudinal mean values at the photosphere level (1-10$^{-3}$ bar) for a PAH concentration of $10^{-5}$ (Tables \ref{tab:all_results_jwst_6} and \ref{tab:all_results_jwst_7} are for $10^{-6}$ and $10^{-7}$, respectively).

The posterior distributions of all retrieved PAH abundances for the three mentioned abundances are summarized separately in Table \ref{tab:pahs_jwst} and illustrated in Figure \ref{fig:allfigures}.

We present the detection significance in terms of 'sigma' in Table \ref{tab:sigma-detection}, for each retrieval set-up, comparing it with the specific PAH-included abundance ($\ln \mathrm{X_{PAH}}$), as opposed to retrievals without PAH. Strong detections are attributed to cases where $\ln B_{01} \geq 5$ or 'sigma' $\geq 3.6\sigma$ \citep{Trotta2008BayesCosmology}.

\subsection{Retrieval Consistency Verification}

Retrieval consistency analysis involves comparing the retrieved parameters with their input values (truth) to determine if they are consistent. This targeted approach aims to test the consistency of the retrieval process, offering insight into its reliability. Here, we explore Model 1 that features PAHs as the sole haze contributor. Figure \ref{fig:posteriors}, \ref{fig:posteriors6}, and \ref{fig:posteriors7} display the molecular posterior distributions corresponding to X$\mathrm{_{PAH} = 10^{-5}}$, 10$^{-6}$, and 10$^{-7}$, respectively, for T=800~K (a), 1200~K (b), and 1600~K (c), and three different C/O ratios, along with the input parameters shown as dashed lines.\\

\textbf{Temperature at 800~K:}
\\
We evaluate the retrieval precision of key atmospheric constituents displayed in the Table \ref{tab:T800K} — water vapor (H$_2$O), methane (CH$_4$), hydrogen sulfide (H$_2$S), and potassium (K) — from the spectra for the three C/O ratios considered, as shown in the first row of Figures 
\ref{fig:forward_model}, \ref{fig:forward_model_6} and \ref{fig:forward_model_7} (X$\mathrm{_{PAH}}$=10$^{-5}$, 10$^{-6}$ and 10$^{-7}$, respectively). For H$_2$O, the detection significance generally falls within the 1-$\sigma$ confidence interval for C/O=0.3 and 1.0 across all PAH abundances. However, a small deviation is observed at a X$\mathrm{_{PAH}}$=10$^{-5}$ and C/O=0.55, where the posterior estimate falls within 3$\sigma$. CH$_4$ retrieval accuracy at a C/O=0.3, with PAH abundances of $10^{-5}$ and $10^{-6}$, remains within the 2-$\sigma$ confidence interval, indicating a modest variance. Yet, for a C/O=1.0 and a X$\mathrm{_{PAH}}$=$10^{-6}$, the deviation exceeds a 3-$\sigma$ threshold, highlighting a substantial discrepancy. H$_2$S estimations consistently align within the 2-$\sigma$ range for all tested C/O ratios and the higher X$\mathrm{_{PAH}}$, with precision improving to within 1-$\sigma$ for C/O ratios of 0.3 and 0.55 at the lowest PAH abundance. Notably, potassium retrievals showcase high precision, consistently falling within the 1-$\sigma$ interval across all examined conditions. Nevertheless, the abundances of CO and Na, at C/O ratios of 0.3 and 1.0 respectively, are underestimated. This occurs due to the spectral interference from other species. Specifically, the CO is obscured by the presence of CO$_2$, which shares similar spectral features, while Na is masked by PAHs as their optical slope overlaps. 
\\

\textbf{Temperature at 1200~K:}
\\
In this case, we assess the retrieval accuracy of water vapor, carbon monoxide (C/O=0.3, 1.0), carbon dioxide (C/O=0.3, 0.55), methane (C/O=0.55, 1.0), potassium (C=0.55) and sodium (C/O=1.0), shown in 
Table \ref{tab:T800K}. The corresponding spectra are shown in the second row of Figures \ref{fig:forward_model}, \ref{fig:forward_model_6}, and \ref{fig:forward_model_7} (X$\mathrm{_{PAH}}$=10$^{-5}$, 10$^{-6}$ and 10$^{-7}$, respectively).
At a X$\mathrm{_{PAH}}$= $10^{-5}$, H$_2$O abundances are accurately estimated within the 1$\sigma$ confidence level for 0.3 and 1.0, but not at 0.55 (C/O), where a significant underestimation within 3$\sigma$ was observed. CH$_4$ abundances were well-aligned within 1$\sigma$ for C/O ratios of 0.55 and 1.0. Potassium and carbon monoxide (CO) estimations for specific C/O ratios demonstrate high accuracy, falling within the 1$\sigma$ interval. 
\\

\textbf{Temperature at 1600~K:}
\\
Finally, we assess the retrieval accuracy of water vapor, carbon monoxide (C/O=0.3, 1.0), carbon dioxide (C/O=0.3, 0.55), methane (C/O=055, 1.0), potassium (C=0.55) and sodium (C/O=1.0), shown in 
Table \ref{tab:T800K}. The corresponding spectra are shown in the third row of Figures \ref{fig:forward_model}, \ref{fig:forward_model_6} and \ref{fig:forward_model_7} (X$\mathrm{_{PAH}}$=10$^{-5}$, 10$^{-6}$ and 10$^{-7}$, respectively).
The retrieved abundances of CO and CO$_2$ show slight variations with decreasing PAH input abundance. For CO, estimations fall within the 1$\sigma$ confidence interval for C/O ratios of 0.55 and 1.0 at a PAH abundance of $10^{-5}$, and within 2$\sigma$ for abundances of $10^{-6}$ and $10^{-7}$. In contrast, CO$_2$ retrieved abundances are significantly lower than the input values across all PAH abundances, likely due to the overlapping spectral features in the NIR, where CO's features dominate and obscure those of CO$_2$.

The retrieval accuracy for H$_2$O ranges from within 1$\sigma$ for a C/O ratio of 0.55 to 3$\sigma$ for a C/O=0.3 at a PAH abundance of $10^{-5}$. At lower PAH abundances ($10^{-6}$ and $10^{-7}$), the accuracy of H$_2$O estimations shifts significantly. The retrieved abundances of titanium oxide (TiO) and vanadium oxide (VO) remain relatively constant, with slight increases in their values as PAH abundance decreases, typically within a 1-2$\sigma$ range.

Conversely, acetylene (C$_2$H$_2$) is consistently underestimated at a C/O ratio of 1.0 for all PAH concentrations. This underestimation is likely due to the overlap of its spectral feature around 3.3-3.5 $\mu$m with that of the PAHs, causing its feature to be masked by the PAH features.

\subsection*{Temperature, Surface Gravity, and Planetary Radius}
\begin{figure*}

  \begin{minipage}[b]{\linewidth}
    \centering
    \includegraphics[width=\textwidth]{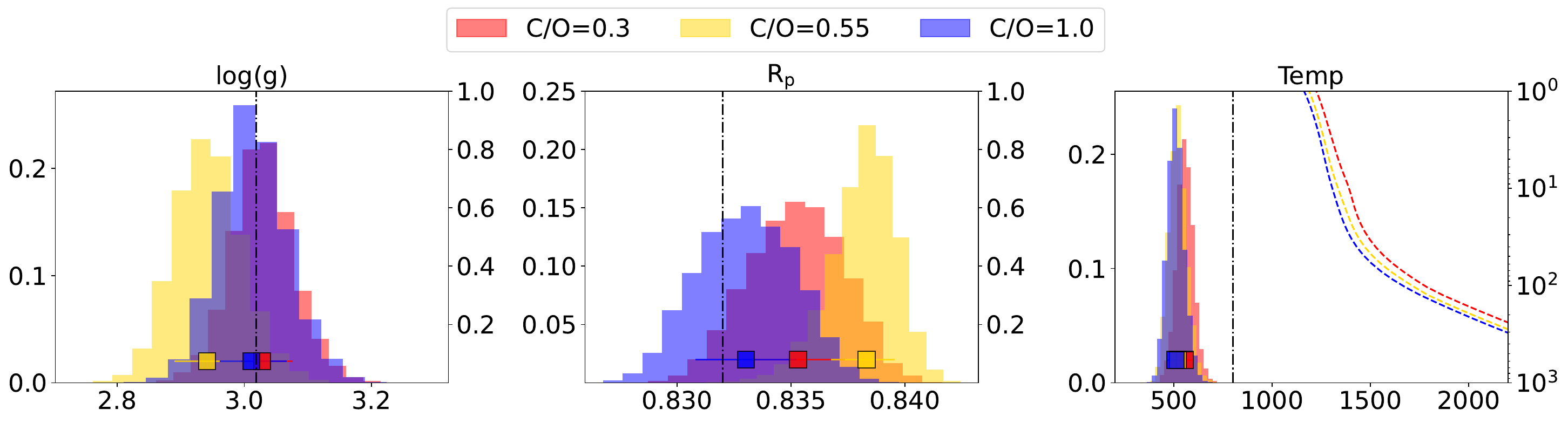}

  \end{minipage}
  \begin{minipage}[b]{\linewidth}
    \centering
    \includegraphics[width=\textwidth]{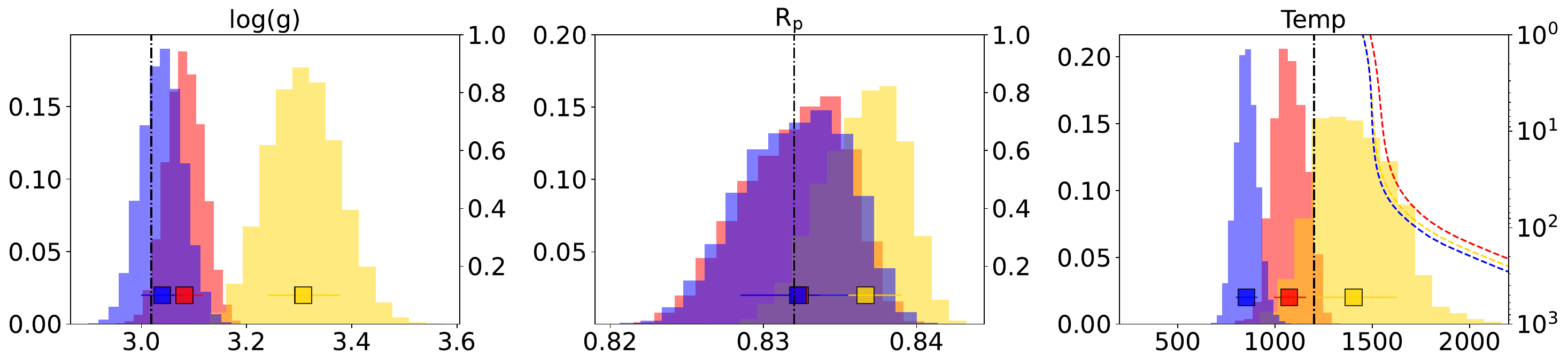}

  \end{minipage}
  \begin{minipage}[b]{\linewidth}
    \centering
    \includegraphics[width=\textwidth]{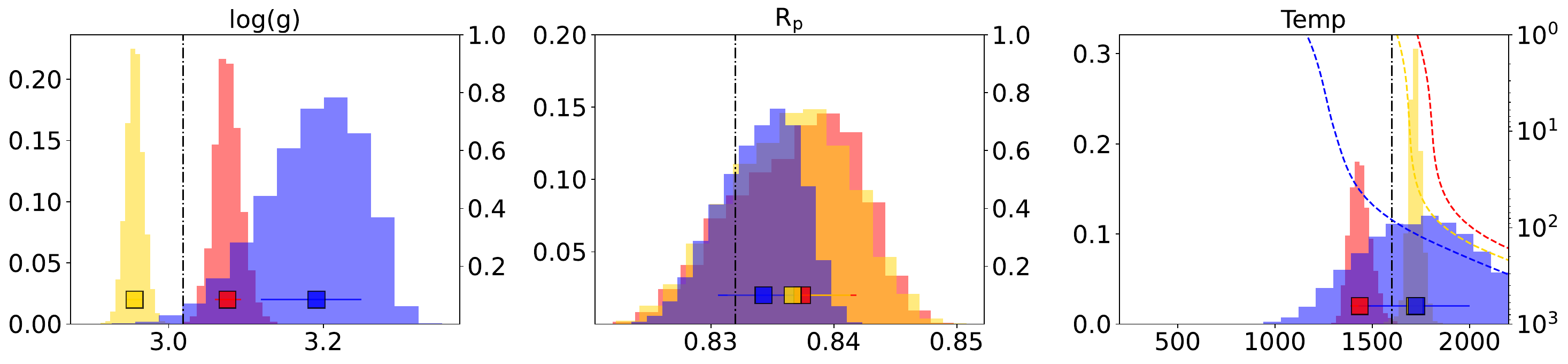}

  \end{minipage}
    \caption{Retrieved posterior probability distributions of log(g), radius of the planet (R$\mathrm{_{P}}$) and the temperature, for planets with input temperatures of 800 (top), 1200 (middle) and 1600K (bottom), C/O from 0.3 (red), 0.55 (yellow) and 1.0 (purple) and X$\mathrm{_{PAH}}$ of 10$^{-5}$. The dashed black lines correspond to the input parameters.}
    \label{fig:all-5-temps}
\end{figure*}

We explore the disparities in temperature estimates, planetary radius, and log gravity ($\log(g)$) across various carbon-to-oxygen (C/O) ratios, at temperatures of 800~K, 1200~K, and 1600~K, under PAH abundances of 10$^{-5}$ (Fig \ref{fig:all-5-temps}), 10$^{-6}$ (Fig. \ref{fig:800-6-temps}, \ref{fig:1200-6-temps}, \ref{fig:1600-6-temps}), and 10$^{-7}$(Fig. \ref{fig:800-7-temps},\ref{fig:1200-7-temps}, \ref{fig:1600-7-temps}), as described below.

For PAH abundances at the level of 10$^{-5}$, we observe a consistent underestimation of temperature estimates at 800~K for all C/O ratios, indicating a systematic bias in our retrieval process. Specifically, at 800~K, the planetary radius is overestimated for C/O ratios of 0.3 and 0.55, while for 1.0, the estimates align within the 1-$\sigma$ confidence interval. Furthermore, the posterior distributions of $\log(g)$ are accurately estimated within the 16-84$\%$ interval for C/O ratios of 0.3 and 1.0.

At a temperature of 1200~K, the variability in temperature estimations is minimal, with consistent results observed for a C/O ratio of 0.3. The planetary radius estimates for C/O ratios of 0.3 and 1.0 fell within one sigma, indicating a closer alignment with expected values, whereas for C/O=0.55, the estimates are notably overestimated. Additionally, $\log(g)$ estimations are accurate within one sigma exclusively for a C/O ratio of 1.0.

When examining conditions at 1600~K, the temperature for a C/O ratio of 1.0 is accurately estimated within the one sigma range, albeit with a wide distribution spanning 1000 to 2400~K, indicating a broader uncertainty in these estimates.

Upon extending our analysis to a temperature of 1200K in PAH abundances of 10$^{-5}$, 10$^{-6}$, and 10$^{-7}$, we find temperature estimates for C/O ratios of 0.3 and 0.55 to be within 1-2 sigma, suggesting a closer alignment with expected values. Conversely, for a C/O ratio of 1.0, the estimates are consistently more than 3 sigma below the expected value for all PAH abundances, highlighting a significant underestimation.

At 1600~K, for a PAH abundance of 10$^{-5}$, temperature estimates for all C/O ratios demonstrate better consistency, falling within 2 sigma. However, for 10$^{-6}$ PAHs, only the C/O ratio of 0.3 is within 1 sigma; other ratios show significantly larger discrepancies, exceeding 3 sigma. For 10$^{-7}$ PAHs, except for C/O=0.55, which is higher than expected, the C/O ratios of 0.3 and 1.0 accurately match within the 1 sigma confidence level.

Our analysis underscores the significant deviation of temperature posteriors from input values across various conditions, attributing this discrepancy primarily to the isothermal assumption in the transmission model employed \citep{Rocchetto2016, MacDonald2020, Welbanks2024} .

As detailed in Section \ref{sec:methodology}, we employ an isothermal transmission model for the sake of simplicity. This model is effective for all scenarios considered in this study with the exception of the underestimation of 800 and 1200~K. Future research could explore a non-isothermal profile. However, since the primary focus of this work is not on retrieving the abundances of all molecules but on PAHs, which input abundances are constant with altitude, we use an isothermal profile. Thus, further discussion on this point will not be pursued.

\subsection{PAH Detection}

\begin{figure*}
  \centering
  \begin{subfigure}{.258\textwidth}
    \includegraphics[width=0.96\linewidth]{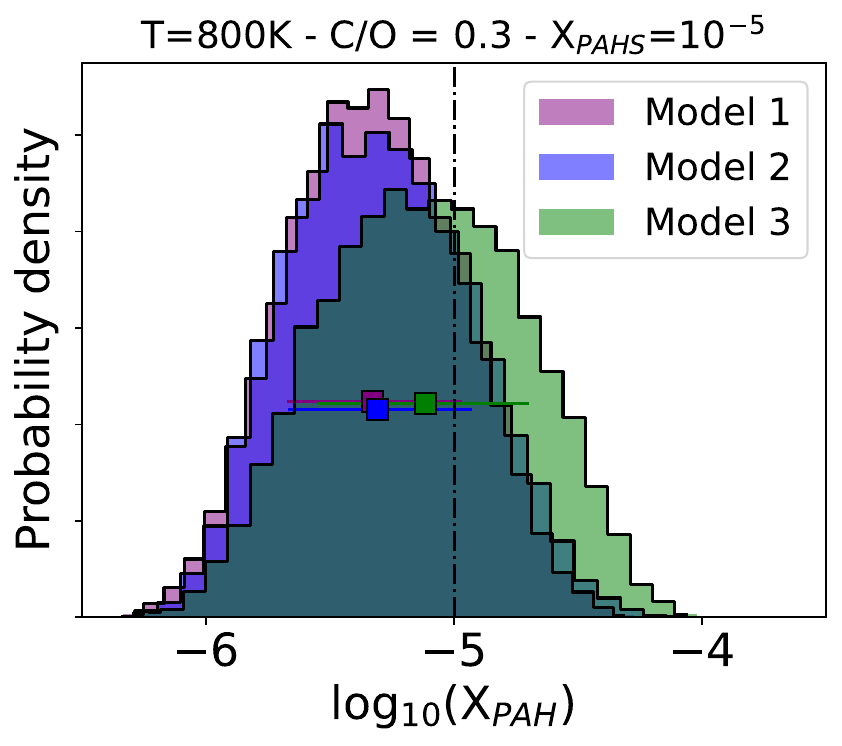}
    \caption{}
  \end{subfigure}
  \begin{subfigure}{.24\textwidth}
    \includegraphics[width=0.96\linewidth]{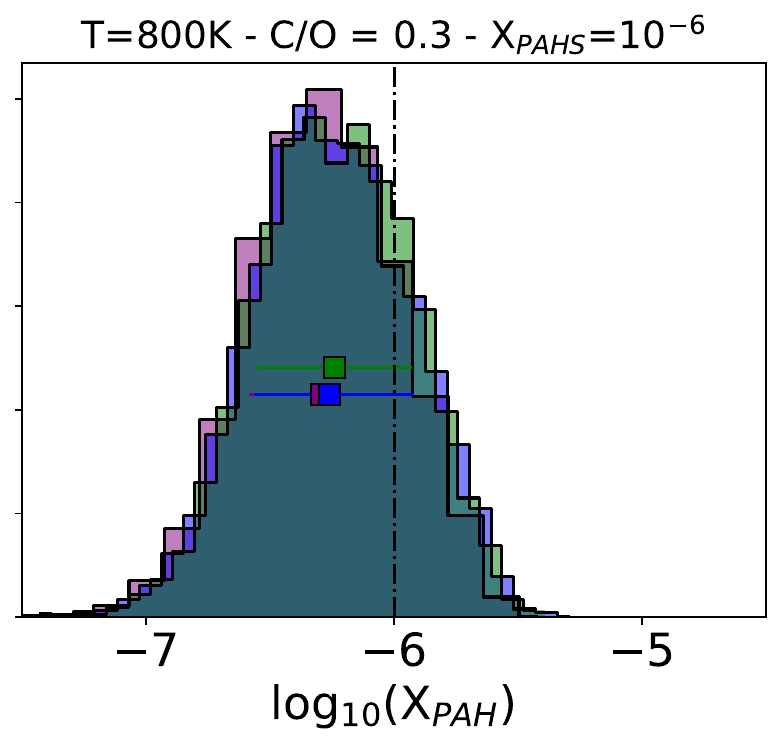}
    \caption{}
  \end{subfigure}
  \begin{subfigure}{.24\textwidth}
    \includegraphics[width=0.96\linewidth]{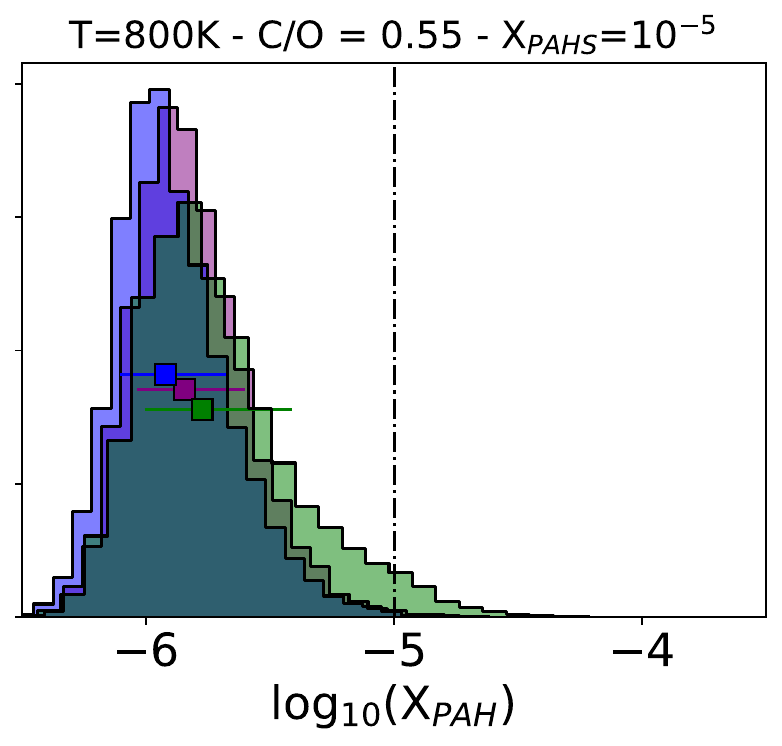}
    \caption{}
  \end{subfigure}
  \begin{subfigure}{.24\textwidth}
    \includegraphics[width=0.96\linewidth]{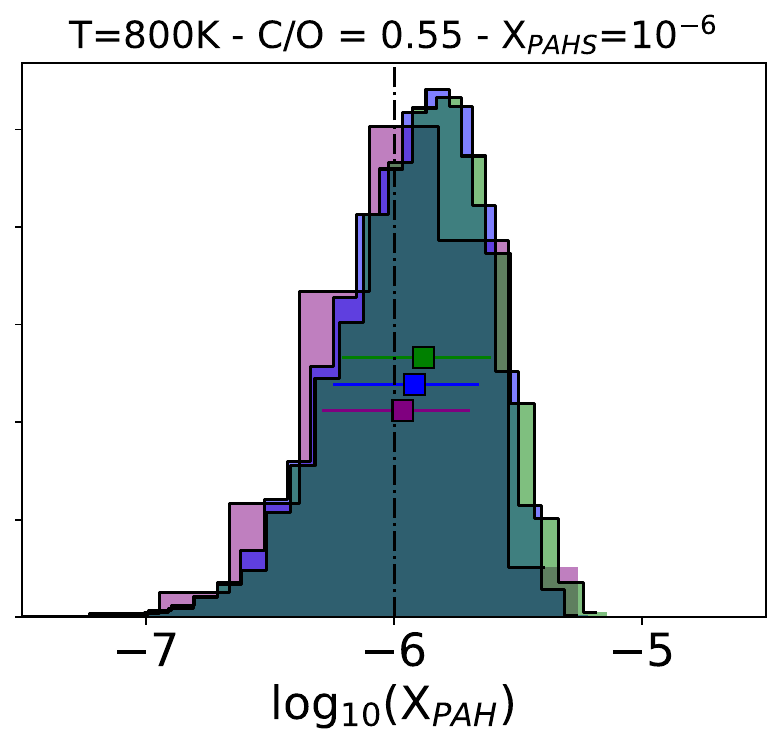}
    \caption{}
  \end{subfigure}

  \begin{subfigure}{.258\textwidth}
    \includegraphics[width=0.96\linewidth]{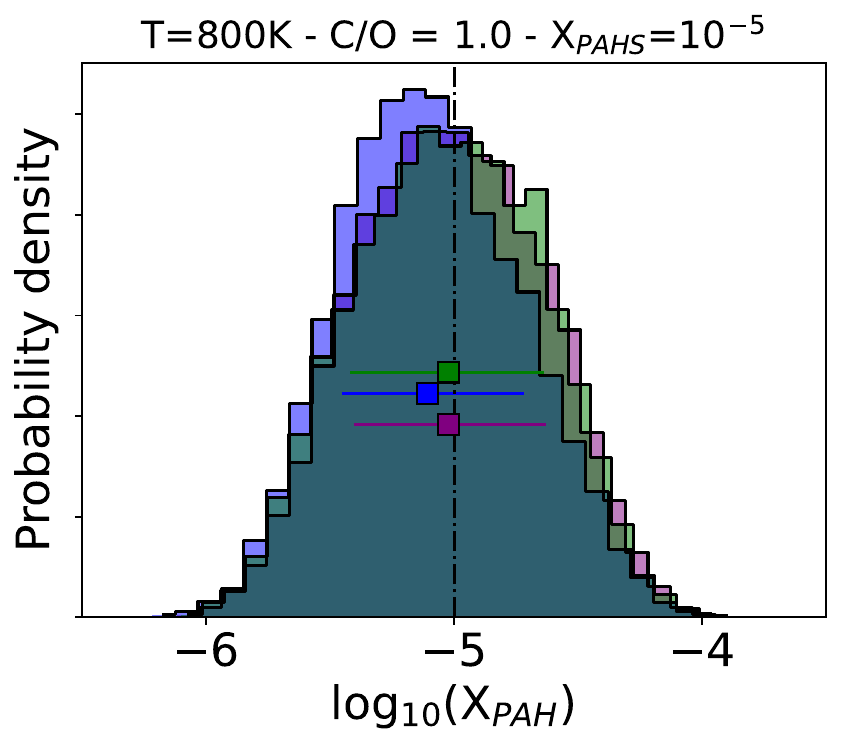}
    \caption{}
  \end{subfigure}
  \begin{subfigure}{.24\textwidth}
    \includegraphics[width=0.96\linewidth]{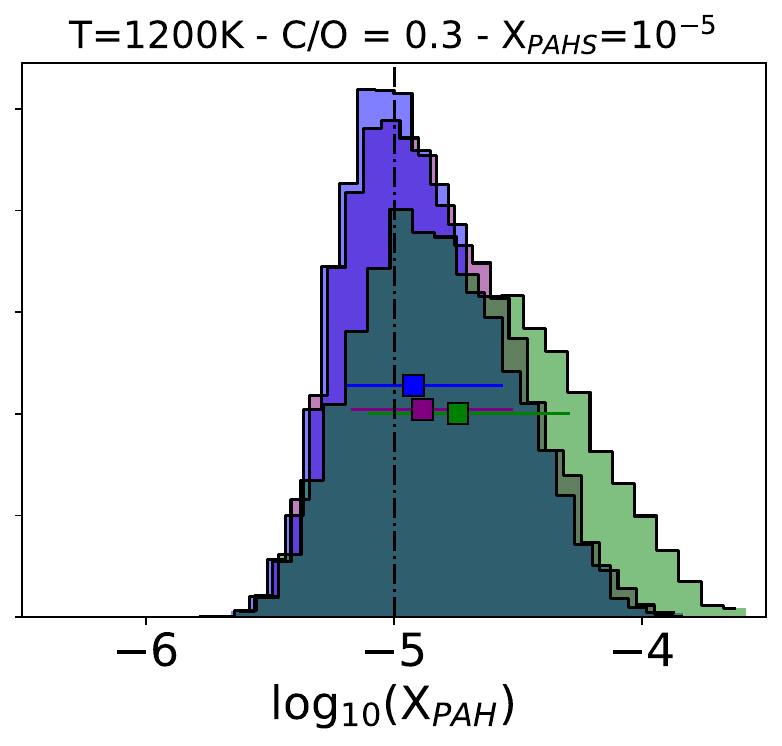}
    \caption{}
  \end{subfigure}
  \begin{subfigure}{.24\textwidth}
    \includegraphics[width=0.96\linewidth]{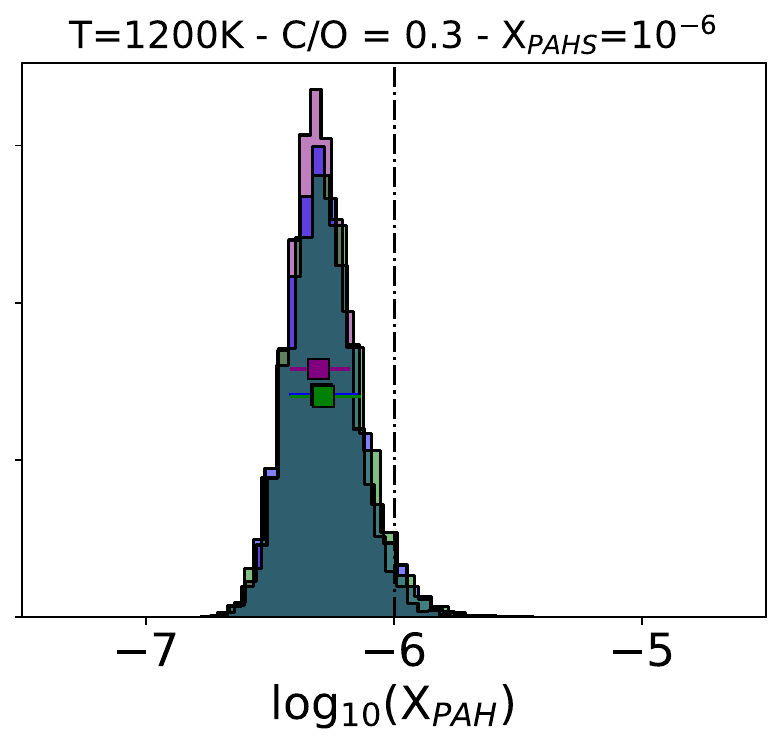}
    \caption{}
  \end{subfigure}
  \begin{subfigure}{.24\textwidth}
    \includegraphics[width=0.96\linewidth]{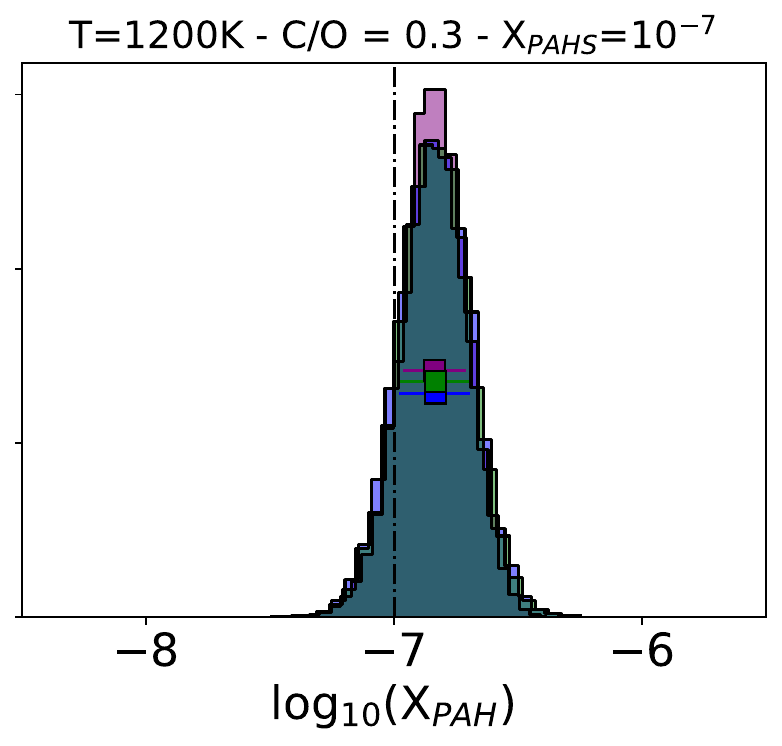}
    \caption{}
  \end{subfigure}

  \begin{subfigure}{.258\textwidth}
    \includegraphics[width=0.96\linewidth]{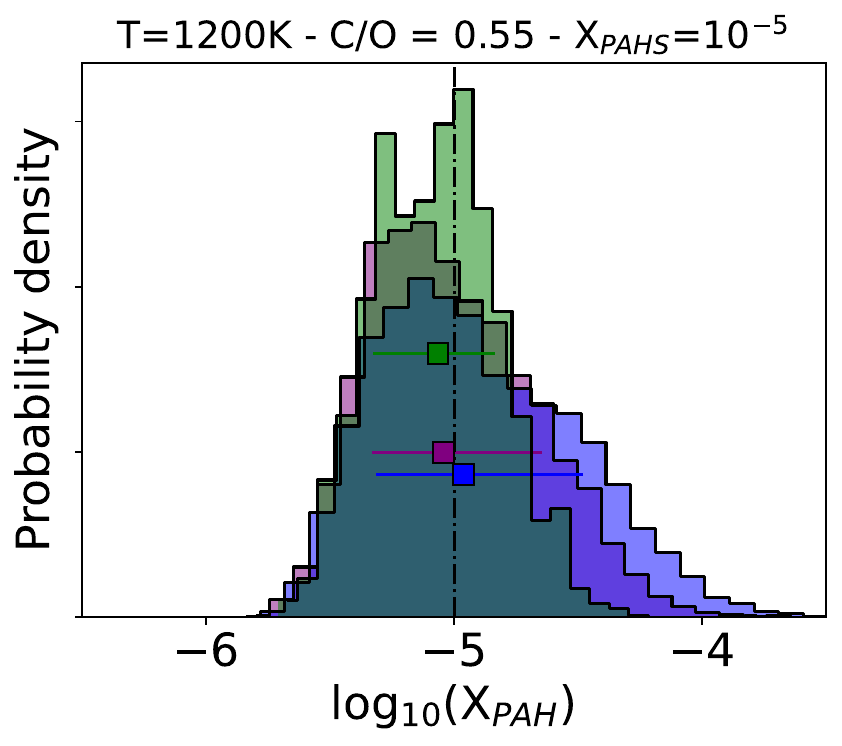}
    \caption{}
  \end{subfigure}
  \begin{subfigure}{.24\textwidth}
    \includegraphics[width=0.96\linewidth]{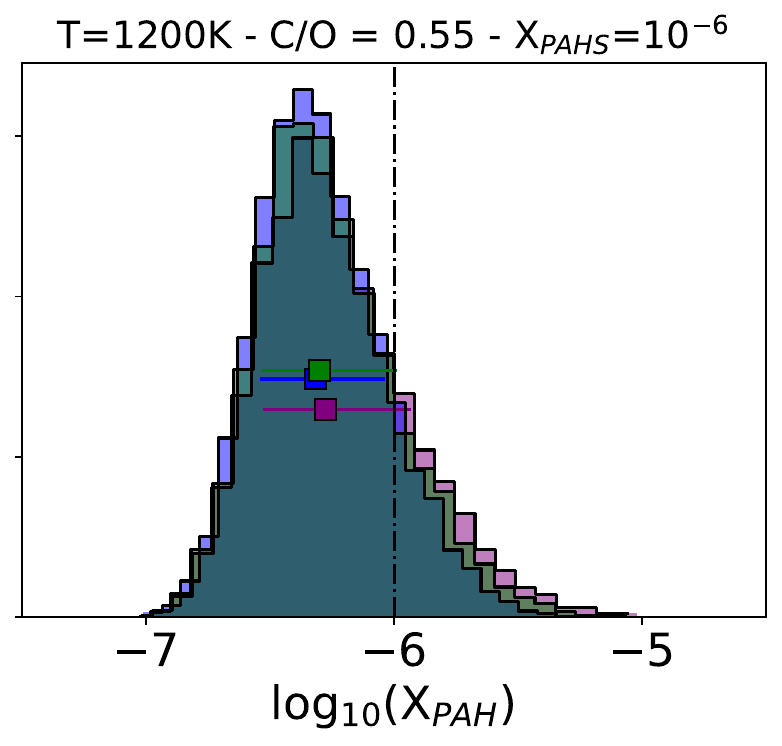}
    \caption{}
  \end{subfigure}
  \begin{subfigure}{.24\textwidth}
    \includegraphics[width=0.96\linewidth]{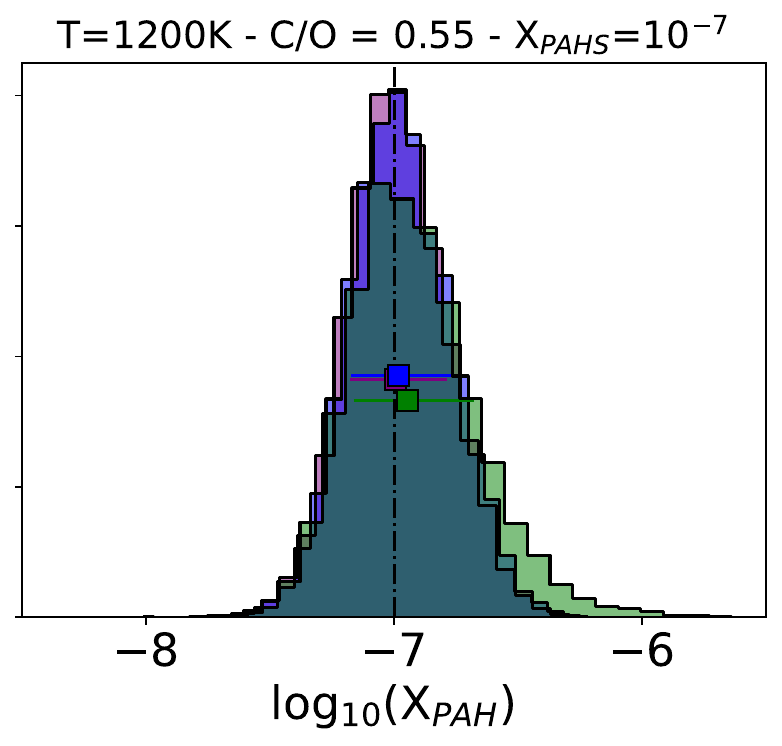}
    \caption{}
  \end{subfigure}
  \begin{subfigure}{.24\textwidth}
    \includegraphics[width=0.96\linewidth]{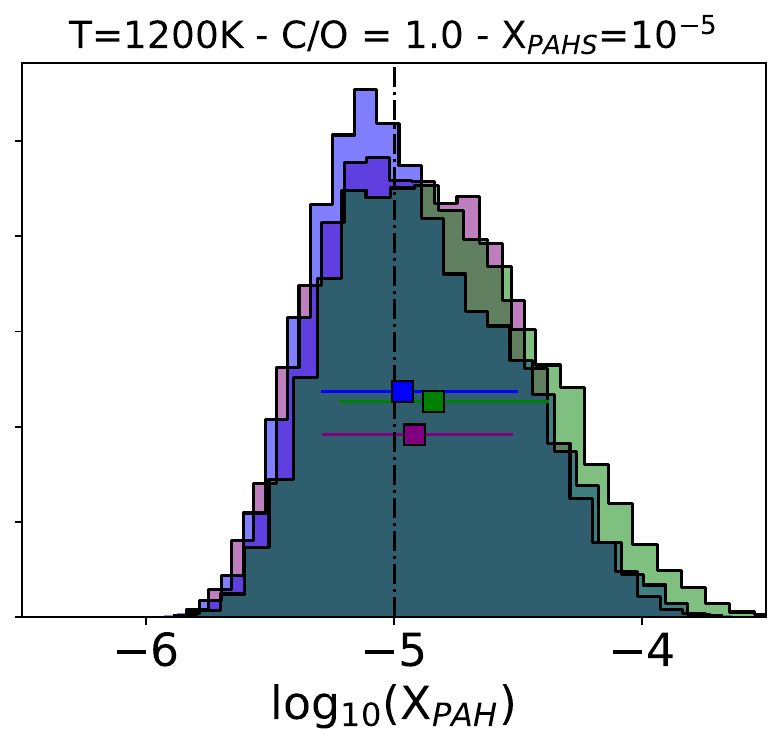}
    \caption{}
  \end{subfigure}

  \begin{subfigure}{.258\textwidth}
    \includegraphics[width=0.96\linewidth]{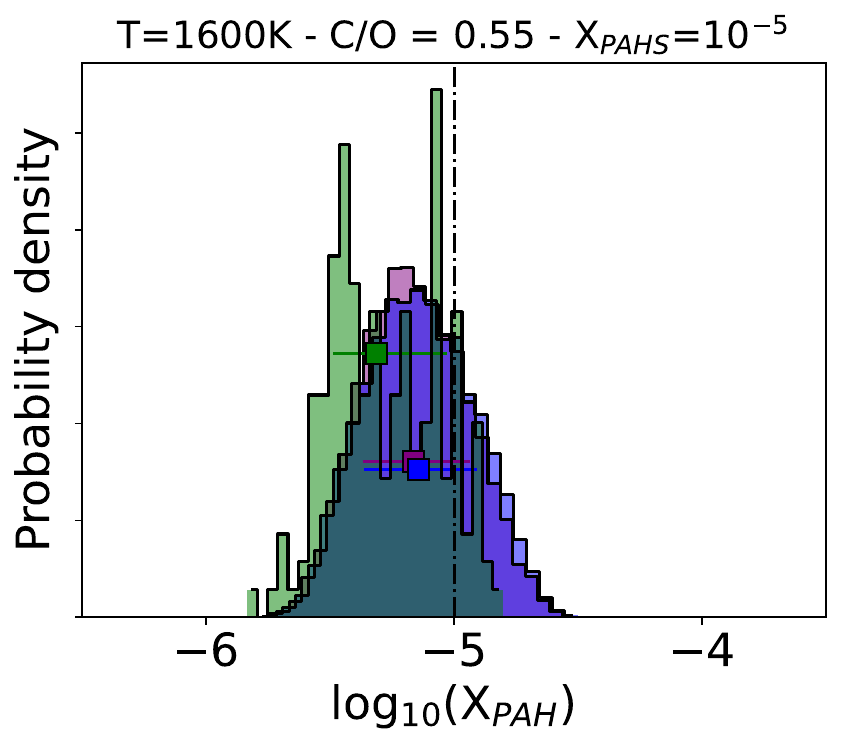}
    \caption{}
  \end{subfigure}
  \begin{subfigure}{.24\textwidth}
    \includegraphics[width=0.96\linewidth]{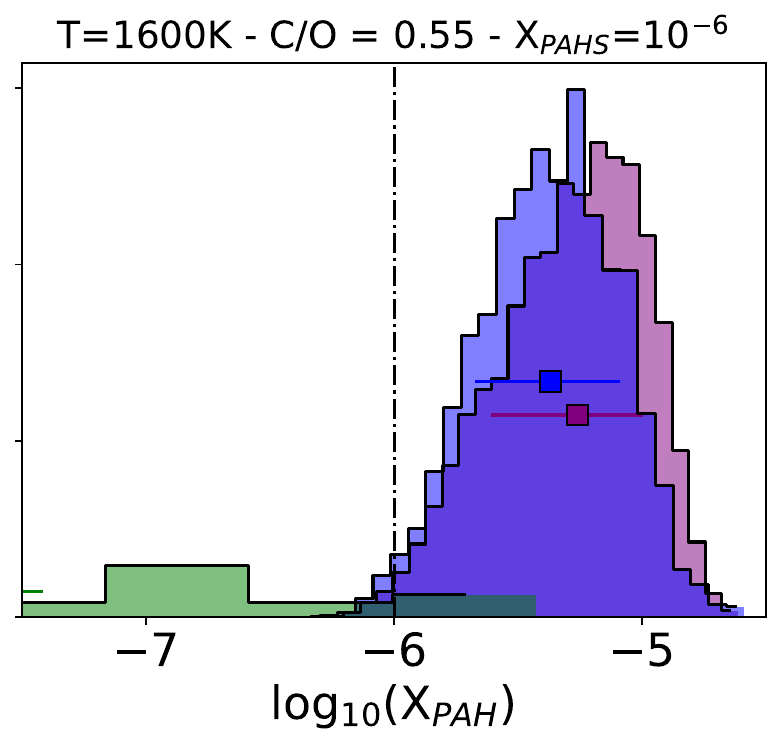}
    \caption{}
  \end{subfigure}
  \begin{subfigure}{.24\textwidth}
    \includegraphics[width=0.96\linewidth]{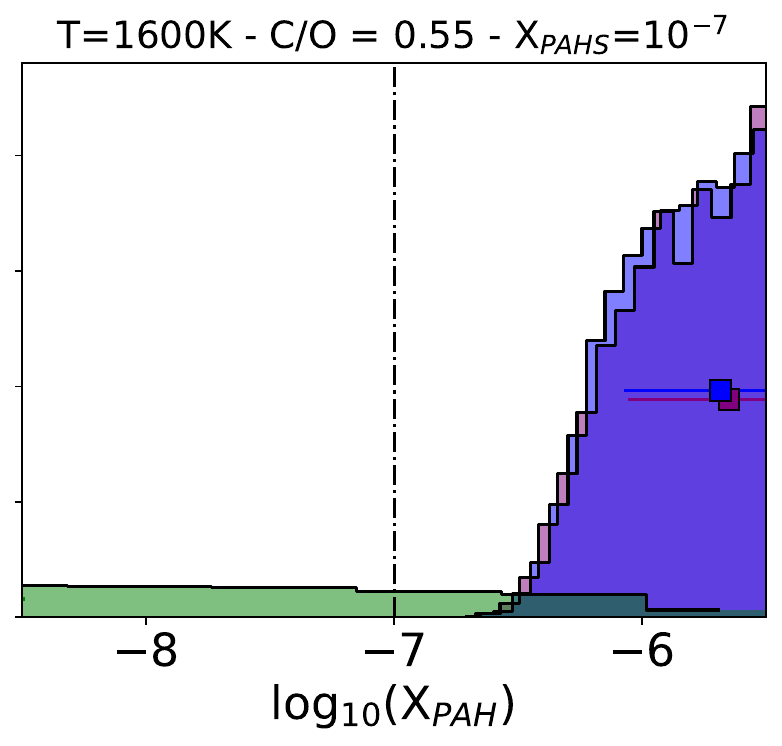}
    \caption{}
  \end{subfigure}
  \begin{subfigure}{.24\textwidth}
    \includegraphics[width=0.96\linewidth]{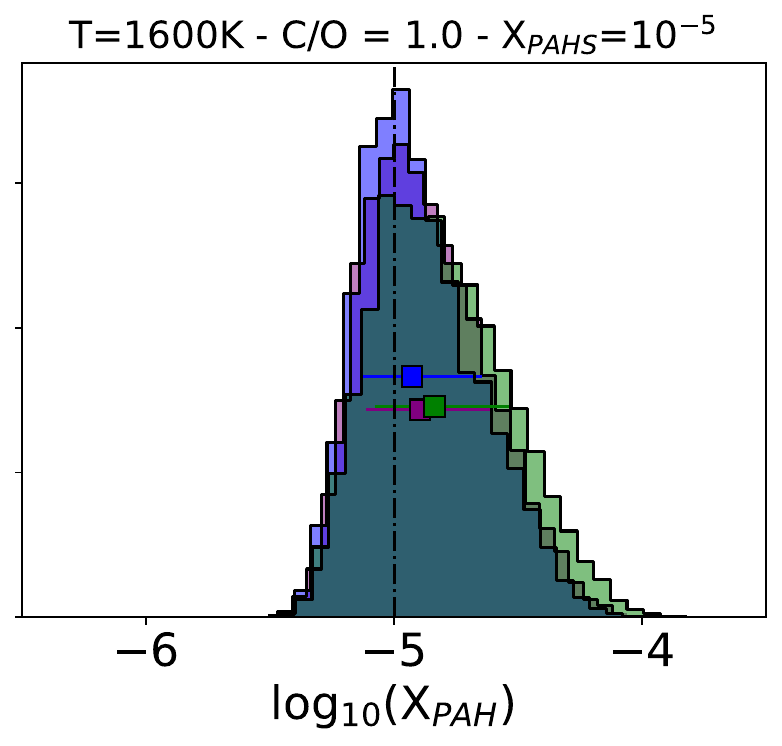}
    \caption{}
  \end{subfigure}
  \begin{subfigure}{.258\textwidth}
    \includegraphics[width=0.96\linewidth]{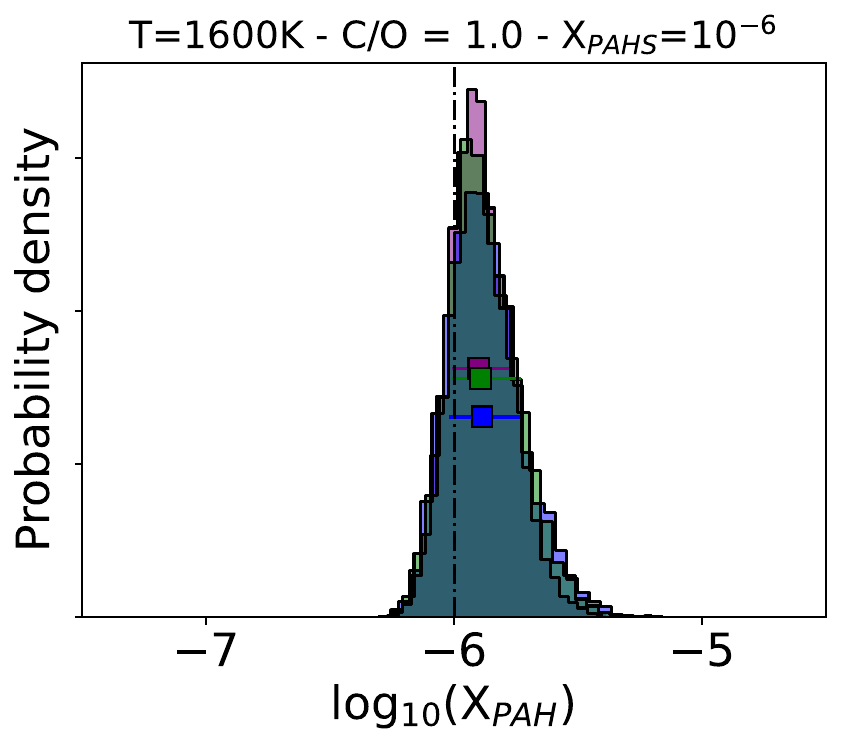}
    \caption{}
  \end{subfigure}

  \caption{PAH posterior distributions retrieved for three abundances ($10^{-5}$, $10^{-6}$ and $10^{-7}$) for the cases where they were constrained within 1-3$\sigma$. \textbf{Model 1:} PAHs as only haze contribution to the atmosphere, in pink. \textbf{Model 2:} PAHs and Power law cloud in blue. \textbf{Model 3:} PAHs, Power law cloud and Cloud deck in green. The squares of varying colors correspond to the retrieved PAHs values, each accompanied by its 1$\sigma$ confidence level. The PAH input values are desplayed in each case as a vertical dash-dotted line.}
  \label{fig:allfigures}
\end{figure*}

A total of 45 scenarios were considered, encompassing three abundance levels for each of the fifteen planets (as shown in Fig. \ref{fig:forward_model} and \ref{fig:3metal_5}, Fig. \ref{fig:forward_model_6} and \ref{fig:3metal_6}, and Fig. \ref{fig:forward_model_7} and \ref{fig:3metal_7}). The derived outcomes, reflecting the possible abundances of PAHs across this planetary diversity, are encapsulated in Table \ref{tab:all_results_jwst_5} and further detailed in Tables \ref{tab:all_results_jwst_6} and \ref{tab:all_results_jwst_7}. Subsequently, a Bayesian analysis is conducted (Table~\ref{tab:sigma-detection} and ~\ref{tab:sigma-detection-metal}), where the Bayes factor ($|ln B_{01}|$), and its translation to 'sigma'-significance (see Eq. 27 from \citealt{Trotta2008BayesCosmology}), is calculated for each model comparison, distinguishing between scenarios with and without the presence of C$_{54}$H$_{18}$. 

The primary objective is to demonstrate the detectability of PAHs for a given abundance using transmission spectroscopy. We propose three main criteria when assessing reliability of retrievals using synthetic data: (i) the retrieved PAH abundance must be within three standard deviations of the input abundance, (ii) The posterior distribution of the retrieved abundances are bounded and (iii) the Bayes factor must exceed a specified threshold (>5) or 'sigma' (>3.6$\sigma$), signaling a robust detection\citep{Trotta2008BayesCosmology}.

\begin{table*}
    \centering
    \caption{Comparative Analysis of "sigma" significance across atmospheric models for various planets. This figure presents a systematic comparison of Bayes factors for different planetary atmospheric set-ups, delineated into three models: (1) an atmosphere with exclusively PAHs contributing to haze; (2) an atmosphere featuring both PAHs and a power law cloud distribution; (3) an expanded version of Model 2 including a cloud deck. Calculations of B$_{01}$ for Models 2 and 3 were limited to planets with significant detection evidence in Model 1. The strength of detection is visually differentiated by shades of pink, categorizing the detection probability as very strong ("sigma" > 5$\mathrm{\sigma}$), strong ("sigma" > 3.6$\mathrm{\sigma}$), moderate ("sigma" > 2.7$\mathrm{\sigma}$), weak ("sigma" > 2.0$\mathrm{\sigma}$), or inconclusive ("sigma" < 2$\mathrm{\sigma}$).}
    \renewcommand{\arraystretch}{1.5}
    \begin{tabular}{c|cccc|ccc|ccc}
        
        & & \multicolumn{3}{c}{T=800K} & \multicolumn{3}{c}{T=1200K} & \multicolumn{3}{c}{T=1600K} \\
        \hline 
        \diagbox{C/O}{$\mathrm{\log X_{PAH}}$} & & -5 & -6 & -7 & -5 & -6 & -7 & -5 & -6 & -7 \\ \hline

        0.3 & Model 1 & \cellcolor{purple!20} 8.89$\mathrm{\sigma}$ & \cellcolor{purple!10} 4.59$\mathrm{\sigma}$ & \cellcolor{purple!5} 0.93$\mathrm{\sigma}$ & \cellcolor{purple!20} 15.95$\mathrm{\sigma}$ &\cellcolor{purple!20}10.14$\mathrm{\sigma}$ & \cellcolor{purple!20}6.14$\mathrm{\sigma}$ & \cellcolor{purple!5}1.58$\mathrm{\sigma}$ & \cellcolor{purple!5}0.90$\mathrm{\sigma}$  & \cellcolor{purple!5}0.90$\mathrm{\sigma}$\\
         & Model 2 & \cellcolor{purple!20} 8.96$\mathrm{\sigma}$ & \cellcolor{purple!10} 4.70$\mathrm{\sigma}$ & N/A &\cellcolor{purple!20} 15.94$\mathrm{\sigma}$ & \cellcolor{purple!20} 10.21$\mathrm{\sigma}$ & \cellcolor{purple!20} 6.24$\mathrm{\sigma}$ & N/A & N/A & N/A \\
         & Model 3 & \cellcolor{purple!20} 7.64$\mathrm{\sigma}$  & \cellcolor{purple!10} 4.59$\mathrm{\sigma}$ & N/A & \cellcolor{purple!20}15.34$\mathrm{\sigma}$ & \cellcolor{purple!20}9.79$\mathrm{\sigma}$ & \cellcolor{purple!20}6.16$\mathrm{\sigma}$ & N/A & N/A & N/A \\
        \hline
        0.55 & Model 1 & \cellcolor{purple!20} 8.92$\mathrm{\sigma}$ & \cellcolor{purple!10} 4.65$\mathrm{\sigma}$ & \cellcolor{purple!5}0.93$\mathrm{\sigma}$ & \cellcolor{purple!20}10.30$\mathrm{\sigma}$ & \cellcolor{purple!20} 6.85$\mathrm{\sigma}$ & \cellcolor{purple!20} 5.12$\mathrm{\sigma}$ & \cellcolor{purple!20} 12.61$\mathrm{\sigma}$ & \cellcolor{purple!20} 8.01$\mathrm{\sigma}$ & \cellcolor{purple!20} 6.47$\mathrm{\sigma}$ \\
          & Model 2 & \cellcolor{purple!20} 9.39$\mathrm{\sigma}$ & \cellcolor{purple!10} 4.70$\mathrm{\sigma}$ & N/A & \cellcolor{purple!20}10.34$\mathrm{\sigma}$ & \cellcolor{purple!20}6.72$\mathrm{\sigma}$ & \cellcolor{purple!20} 5.06$\mathrm{\sigma}$ & \cellcolor{purple!20} 15.27$\mathrm{\sigma}$ &  \cellcolor{purple!20} 7.98$\sigma$ & \cellcolor{purple!20} 6.66$\sigma$  \\
        & Model 3 & \cellcolor{purple!20} 8.83$\sigma$ & \cellcolor{purple!10} 4.49$\mathrm{\sigma}$ & N/A  &\cellcolor{purple!20}10.36$\mathrm{\sigma}$ & \cellcolor{purple!20}6.93$\mathrm{\sigma}$ & \cellcolor{purple!20} 5.02$\mathrm{\sigma}$ & \cellcolor{purple!20} 6.61$\mathrm{\sigma}$ & \cellcolor{purple!5} 1.67$\sigma$ & \cellcolor{purple!5} 0.96$\sigma$ \\ 
          \hline
        1.0 & Model 1 &\cellcolor{purple!20} 9.56$\mathrm{\sigma}$ & \cellcolor{purple!5}0.98$\mathrm{\sigma}$ & \cellcolor{purple!5}1.22$\mathrm{\sigma}$ & \cellcolor{purple!20} 11.22$\mathrm{\sigma}$ & \cellcolor{purple!5} 1.96$\mathrm{\sigma}$ & \cellcolor{purple!5} 1.16$\mathrm{\sigma}$ & \cellcolor{purple!20} 9.52$\mathrm{\sigma}$ & \cellcolor{purple!20} 7.59$\mathrm{\sigma}$ & \cellcolor{purple!5} 1.02$\mathrm{\sigma}$ \\
         & Model 2 &\cellcolor{purple!20} 9.54$\mathrm{\sigma}$ & N/A & N/A & \cellcolor{purple!20} 11.25$\mathrm{\sigma}$& N/A & N/A & \cellcolor{purple!20} 11.60$\mathrm{\sigma}$& \cellcolor{purple!20} 6.44$\mathrm{\sigma}$ &  N/A \\
        & Model 3 & \cellcolor{purple!20} 8.24$\mathrm{\sigma}$ & N/A & N/A & \cellcolor{purple!20} 7.07$\mathrm{\sigma}$ & N/A & N/A & \cellcolor{purple!20} 11.60$\mathrm{\sigma}$ & \cellcolor{purple!20} 5.87$\mathrm{\sigma}$ &  N/A \\
        \hline
    \end{tabular}

\label{tab:sigma-detection}
\end{table*}
\begin{table*}
    \centering
    \caption{Derived posterior values for the logarithmic concentration of PAHs (log~X$_{\text{PAH}}$) in the atmospheric models of planets under various conditions. The analysis is segmented by three distinct C/O ratios (0.3, 0.55, and 1.0), across three temperature regimes (800~K, 1200~K, and 1600~K), and for three logX$_{\text{PAH}}$ values (-5, -6, -7). The posterior values are provided for Models 1, 2, and 3. Cells marked 'N/A' signify scenarios where retrievals are not performed, adhering to the precedent that retrievals are only conducted for set-ups with a prior detection of PAHs, as  highlighted by the green and yellow shaded regions.}
    \renewcommand{\arraystretch}{1.5}

    \resizebox{\textwidth}{!}{
    \begin{tabular}{c|cccc|ccc|ccc}
        & & \multicolumn{3}{c}{T=800K} & \multicolumn{3}{c}{T=1200K} & \multicolumn{3}{c}{T=1600K} \\
        \hline 
        \diagbox{C/O}{$\mathrm{\log X_{PAH}}$} & & -5 & -6 & -7 & -5 & -6 & -7 & -5 & -6 & -7 \\ \hline

        0.3 & Model 1 & -5.33 $^{+0.36}_{-0.35}$ & -6.29 $^{+0.31}_{-0.29}$ & -12.99 $^{+4.88}_{-4.33}$ & -4.89 $^{+0.37}_{-0.29}$ & -6.31 $^{+0.13}_{-0.11}$ & -6.84 $^{+0.13}_{-0.13}$  & -9.22 $^{+3.63}_{-7.18}$ & -13.01 $^{+4.57}_{-4.38}$ & -5.65 $^{+0.33}_{-0.41}$ \\
         & Model 2 & -5.34 $^{+0.39}_{-0.33}$ & -6.26 $^{+0.34}_{-0.31}$ & N/A & -4.92 $^{+0.36}_{-0.27}$ & -6.30 $^{+0.15}_{-0.13}$ & -6.83 $^{+0.14}_{-0.15}$ & N/A & N/A & N/A \\
         & Model 3 &  -5.12 $^{+0.42}_{-0.43}$ & -6.24 $^{+0.32}_{-0.31}$ & N/A & -4.74 $^{+0.45}_{-0.36}$ & -6.29 $^{+0.15}_{-0.14}$ & N/A & N/A & N/A & N/A \\
         \hline
        0.55 & Model 1 & -5.85 $^{+0.24}_{-0.19}$ & -6.00 $^{+0.26}_{-0.33}$ & -12.43 $^{+4.75}_{-4.89}$ &   -5.05 $^{+0.39}_{-0.29}$  & -6.28 $^{+0.34}_{-0.25}$ & -6.95 $^{+0.27}_{-0.22}$ & -5.16 $^{+0.23}_{-0.20}$ &   -5.26 $^{+0.27}_{-0.35}$ & -5.75 $^{+0.33}_{-0.41}$ \\
         & Model 2 & -5.92 $^{+0.26}_{-0.18}$ &   -5.92 $^{+0.26}_{-0.33}$ & N/A & -4.96 $^{+0.48}_{-0.35}$  & -6.32 $^{+0.28}_{-0.22}$ & -6.98 $^{+0.21}_{-0.19}$  & -5.14 $^{+0.24}_{-0.22}$ &   -5.37$^{+0.28}_{-0.30}$ & -5.69$^{+0.33}_{-0.39}$ \\
         & Model 3 & -5.78 $^{+0.36}_{-0.23}$ & -5.88 $^{+0.27}_{-0.33}$ & N/A & -5.07 $^{+0.23}_{-0.26}$  &  -6.32 $^{+0.36}_{-0.31}$ & -6.95 $^{+0.27}_{-0.22}$ & -5.31 $^{+0.28}_{-0.17}$ & -11.81$^{+4.40}_{-6.14}$ & -13.42$^{+4.94}_{-4.40}$ \\
         \hline
        1.0 & Model 1 &  -5.02 $^{+0.39}_{-0.38}$ & -13.75 $^{+4.07}_{-3.86}$ & -12.55 $^{+4.56}_{-5.12}$ & -4.92 $^{+0.40}_{-0.37}$ & -6.86 $^{+0.27}_{-6.32}$ & -13.84 $^{+4.26}_{-4.00}$ & -4.90 $^{+0.29}_{-0.22}$ &   -5.90 $^{+0.13}_{-0.11}$ & -13.24 $^{+4.34}_{-4.14}$ \\
        & Model 2 &   -5.11 $^{+0.39}_{-0.35}$ & N/A & N/A & -4.97 $^{+0.46}_{-0.33}$ & N/A &N/A & -4.84 $^{+0.31}_{-0.24}$ & -5.89 $^{+0.16}_{-0.12}$ & N/A \\
         & Model 3 & -5.02 $^{+0.38}_{-0.40}$ & N/A & N/A & -4.84 $^{+0.47}_{-0.38}$ & N/A &N/A & -4.84 $^{+0.31}_{-0.24}$ & -5.89 $^{+0.16}_{-0.12}$ & N/A
        \\
        \hline
    \end{tabular}%
     }

    \label{tab:pahs_jwst}
\end{table*}

\begin{table*}
\centering

\def\arraystretch{1.5}
\tiny\centering
\setlength{\tabcolsep}{1.4mm}
\scriptsize
\centering
\caption{Comparative Analysis of reduced $\chi^2$ across different atmospheric models for various planets. This figure presents a systematic comparison of $\chi^2$ for different planetary atmospheric set-ups, delineated into three models: (1) an atmosphere with exclusively PAHs contributing to haze; (2) an atmosphere featuring both PAHs and a power law cloud distribution; (3) an expanded version of Model 2 including a cloud deck. The columns ’incl’ and ’not’ represent the retrievals made including PAHs and removing them, respectively. Cells marked ’N/A’ signify scenarios where retrievals were not performed,
adhering to the precedent that retrievals were only conducted for set-ups with a prior detection of PAHs.}
{\fontsize{7.6}{10}\selectfont
\resizebox{\textwidth}{!}{%
\begin{tabular}{c|c*{17}{p{0.6cm}|}c}
       & & \multicolumn{6}{c}{T=800K} & \multicolumn{6}{c}{T=1200K} & \multicolumn{6}{c}{T=1600K} \\
        \hline 
        \diagbox{C/O}{$\mathrm{\log X_{PAH}}$} & & \multicolumn{2}{c|}{-5}  & \multicolumn{2}{c|}{-6}  & \multicolumn{2}{c|}{-7}  & \multicolumn{2}{c|}{-5}  & \multicolumn{2}{c|}{-6}  & \multicolumn{2}{c|}{-7} & \multicolumn{2}{c|}{-5}  & \multicolumn{2}{c|}{-6}  & \multicolumn{2}{c}{-7}  \\ \hline
         & & \cellcolor{lightgray!20} incl &  not & \cellcolor{lightgray!20}incl & not & \cellcolor{lightgray!20}incl & not & \cellcolor{lightgray!20}incl & not & \cellcolor{lightgray!20}incl & not & \cellcolor{lightgray!20}incl & not & \cellcolor{lightgray!20}incl & not & \cellcolor{lightgray!20}incl & not & \cellcolor{lightgray!20}incl & not \\ \hline

        0.3 & Model 1 & \cellcolor{lightgray!20} 0.993 &  1.29 & \cellcolor{lightgray!20}1.03 & 1.11 & \cellcolor{lightgray!20}1.05 & 1.03 & \cellcolor{lightgray!20}0.98 & 1.92 & \cellcolor{lightgray!20}1.29 & 2.68 & \cellcolor{lightgray!20}1.38 & 1.55 & \cellcolor{lightgray!20}1.30 & 1.32 & \cellcolor{lightgray!20}1.24 & 1.24 & \cellcolor{lightgray!20}1.60 & 1.68 \\
         & Model 2 & \cellcolor{lightgray!20}1.00  & 1.30 & \cellcolor{lightgray!20}1.03 & 1.12 & \cellcolor{lightgray!20}N/A & N/A & \cellcolor{lightgray!20}0.99 & 1.93 & \cellcolor{lightgray!20}1.30 & 1.69 & \cellcolor{lightgray!20}1.40 & 1.55 & \cellcolor{lightgray!20}N/A & N/A & \cellcolor{lightgray!20}N/A & N/A & \cellcolor{lightgray!20}N/A & N/A \\
         & Model 3 & \cellcolor{lightgray!20} 1.00 & 1.20 & \cellcolor{lightgray!20}1.04 & 1.13 & \cellcolor{lightgray!20}N/A & N/A & \cellcolor{lightgray!20}0.99 & 1.72 & \cellcolor{lightgray!20}1.30 & 1.65& \cellcolor{lightgray!20}1.40 & 1.55 & \cellcolor{lightgray!20}N/A & N/A & \cellcolor{lightgray!20}N/A & N/A & \cellcolor{lightgray!20}N/A & N/A \\
        \hline
        0.55 & Model 1 & \cellcolor{lightgray!20} 1.06 & 1.35 & \cellcolor{lightgray!20}1.05 & 1.15 & \cellcolor{lightgray!20}1.02 & 1.04 & \cellcolor{lightgray!20}1.05 & 1.45 & \cellcolor{lightgray!20}1.12 & 1.30 & \cellcolor{lightgray!20}1.45 & 1.52 & \cellcolor{lightgray!20}1.14 & 1.72 & \cellcolor{lightgray!20}1.43 & 1.69 & \cellcolor{lightgray!20}1.83 & 2.00 \\
          & Model 2 & \cellcolor{lightgray!20} N/A &  N/A & \cellcolor{lightgray!20}1.06 & 1.16 & \cellcolor{lightgray!20}N/A & N/A & \cellcolor{lightgray!20}1.06 & 1.46 & \cellcolor{lightgray!20}1.13 & 1.31 & \cellcolor{lightgray!20}1.42 & 1.53 & \cellcolor{lightgray!20}1.15 & 1.35 & \cellcolor{lightgray!20} 1.44& 1.73 & \cellcolor{lightgray!20}1.84 & 2.02  \\
        & Model 3 & \cellcolor{lightgray!20} N/A & N/A & \cellcolor{lightgray!20}1.06 & 1.14 & \cellcolor{lightgray!20}N/A & N/A & \cellcolor{lightgray!20}1.03 & 1.47 & \cellcolor{lightgray!20}1.13 & 1.32 & \cellcolor{lightgray!20}1.42 & 1.53 & \cellcolor{lightgray!20}1.16 & 2.03 & \cellcolor{lightgray!20} 1.44 & 1.42 &\cellcolor{lightgray!20} 1.82 & 1.81 \\  
          \hline
        1.0 & Model 1 & \cellcolor{lightgray!20} 1.07 & 1.51 & \cellcolor{lightgray!20}0.88 & 0.88 & \cellcolor{lightgray!20}1.10 & 1.09 & \cellcolor{lightgray!20}\cellcolor{lightgray!20}0.94 & 1.39 & \cellcolor{lightgray!20}1.03 & 1.08 & \cellcolor{lightgray!20}1.26 & 1.25 & \cellcolor{lightgray!20}1.05 & 1.37 & \cellcolor{lightgray!20}1.21 & 1.45 & \cellcolor{lightgray!20}1.37 & 1.37 \\
         & Model 2 & \cellcolor{lightgray!20}1.08 & 1.42 & \cellcolor{lightgray!20}N/A & N/A & \cellcolor{lightgray!20}N/A & N/A & \cellcolor{lightgray!20}0.95 & 1.40 & \cellcolor{lightgray!20}N/A & N/A & \cellcolor{lightgray!20}N/A & N/A & \cellcolor{lightgray!20}1.06 & 1.57 & \cellcolor{lightgray!20}1.30 & 1.46 & \cellcolor{lightgray!20}N/A & N/A \\
        & Model 3 & \cellcolor{lightgray!20}1.09 & 1.29 & \cellcolor{lightgray!20}N/A & N/A & \cellcolor{lightgray!20}N/A & N/A & \cellcolor{lightgray!20}0.95 & 1.11 & \cellcolor{lightgray!20}N/A & N/A & \cellcolor{lightgray!20}N/A & N/A & \cellcolor{lightgray!20}1.06 & 1.52 & \cellcolor{lightgray!20}1.24 & 1.35 & \cellcolor{lightgray!20}N/A & N/A \\
        \hline
\end{tabular}%
}}
\label{tab:reduced_chi2}
\end{table*}


As illustrated in Figure \ref{fig:forward_model} for 800~K, it is observed that the contribution of PAHs is significantly more pronounced at low C/O ratios between 0.3 and 0.55, especially in the optical slope, and becomes negligible at high C/O ratios. This pattern is confirmed as the retrieved PAH abundance decreases (Figure \ref{fig:forward_model_6} and \ref{fig:forward_model_7}), with their contribution tending to vanish, aligning with previously mentioned findings. Upon increasing the temperature to 1200~K, this trend becomes more pronounced for C/O ratios between 0.3 and 0.55 and all input PAH abundances, consistent with the studies by \citealt{Dubey_2023}, suggesting that PAH formation reaches its peak. However, at 1600K and under low C/O conditions, the formation of TiO becomes dominant, overshadowing the PAH features in the spectrum (0.6-1 $\mu$m), particularly in the optical slope. In contrast, under high C/O conditions, TiO features contribute less, allowing PAHs to be more influential.

\subsubsection{Initial Model Assessment}

Initially, we assess the fit of our atmospheric models by comparing the reduced $\chi^2$ values of the baseline scenarios (without PAHs) against those augmented with PAHs, as detailed in Table \ref{tab:reduced_chi2}. Generally, models that include PAHs show reduced $\chi^2$ values nearer to unity, suggesting better fits to the observed data. 

Furthermore, while interpreting these findings, it should be considered that in our cloud-free model (see Model 1 in Table \ref{tab:baselines}), the apparent detection of PAHs might be misleading, as the optical slope typically associated with PAHs can also be attributed to other photochemically-produced hazes, such as those involved in Rayleigh scattering \citep{Ohno_2020, Gao_2021}.
Other types of hazes, composed of tiny aerosol particles, can produce similar optical characteristics, thereby complicating the identification of PAHs from the optical slope alone. 
Nonetheless, the PAH signature also includes the 3.3 $\mu$m feature in the NIR. This feature, particularly at high abundances, is likely the primary reason for detecting PAHs, as demonstrated in \citealt{grubel2024detectability}.

Nevertheless, similar reduced $\chi^2$ values are observed under specific conditions: 800~K with C/O ratios of 0.3 and 0.55, and a PAH abundance of $10^{-7}$, as well as at a C/O=1.0 with X$\mathrm{_{PAH}=10^{-6}}$; 1200~K with C/O=1.0 and X$\mathrm{_{PAH}=10^{-7}}$; and 1600~K with C/O=0.3 and PAH abundances of $10^{-5}$ and $10^{-6}$, and also at a C/O=1.0 with X$\mathrm{_{PAH}=10^{-7}}$. These similarities suggest that both the baseline and PAH-augmented models are comparably effective in representing the observed data.

To further validate our findings, we incorporate Bayes factor analysis as depicted in Table \ref{tab:sigma-detection}. In our investigation, we have transformed the Bayes factors into sigma-significance for clearance. Here, we consider all cases with high 'sigma'-significance (>5$\sigma$) to be significant, and therefore PAHs could be detectable by transmission spectroscopy under the specific conditions presented.

\subsubsection{Model 1: Cloud-Free atmosphere}

\textbf{Constant [Fe/H]=0, varying C/O ratio:} 
\\

In the cloud-free atmospheric model at 800~K and a C/O ratio of 0.3, NIRSpec PRISM can detect PAHs within a 1-$\sigma$ confidence level for abundances of $10^{-5}$ and $10^{-6}$, with very strong  (>5$\sigma$) and strong (>3.6$\sigma$) detection probabilities. Nevertheless, it encounters difficulties in detecting PAHs at around $10^{-7}$, where the detection probability is inconclusive (<2$\sigma$), as shown in Table \ref{tab:sigma-detection}. The cause of this is likely the overlap between PAHs and $\mathrm{CH_4}$ around the 3.3$\mu m$ feature, becoming more pronounced for lower PAH abundances (Fig. \ref{fig:forward_model_7}). 

At C/O ratios of 0.55 and 1.0, similar challenges are observed, leading to inconclusive detection probabilities (<2$\sigma$) at PAH abundances of $10^{-7}$ for 0.55 and $10^{-6}$-$10^{-7}$ for 1.0. We notice that in the presence of Na, the posterior of PAHs for the lowest input is underestimated $\mathrm{-13.84^{+4.26}_{-4.00}}$, and on the contrary, $\mathrm{\ln X_{Na}}$ is overestimated ($\mathrm{-2.60^{+0.88}_{-0.86}}$). It can be seen at Fig. \ref{fig:forward_model_7}, that Na might mimic the optical slope of PAHs, obscuring its features. The non-detection of C/O 0.3 is probably of the same nature as with C/O of 0.55. 
However, PAHs are detectable at abundance levels of $10^{-5}$ and $10^{-6}$ under these ratios, albeit with varying confidence levels. For instance, at C/O 0.55, the instrument provides very strong and strong detection probabilities for abundances of $10^{-5}$ and $10^{-6}$, respectively. Conversely, at a C/O ratio of 1.0, PAHs are only detectable at an abundance level of $10^{-5}$, with a very strong detection probability (>5$\sigma$).

As the temperature rises to 1200~K, PAH detection remains consistent across all C/O ratios at an abundance of $10^{-5}$. Notably, at C/O ratios of 0.3 and 0.55, the detectability of PAHs expands to include abundance levels of $10^{-6}$ and $10^{-7}$ within a 1-2$\sigma$ confidence level, accompanied by strong detection probabilities exceeding 5$\sigma$. However, with an increase in the C/O ratio to 1.0, although PAHs exhibit their characteristic optical slope signature, it becomes softer compared to lower ratios. Resulting in the non-detection of PAHs by NIRSpec PRISM at a C/O ratio of 1.0. These findings are consistent with the results of equilibrium chemistry models presented by \citealt{Dubey_2023}.

At 1600~K, PAHs are predominantly detected at higher C/O ratios (0.55 and 1.0): specifically, at $10^{-5}$ and  $10^{-6}$ (C/O=1.0), and down to $10^{-7}$ (C/O=0.55). In these cases, PAHs are ascertained with high detection probability (>5$\sigma$). However, unlike at other temperatures where PAHs were detectable even at the highest abundances, for T=1600~K and C/O=0.3, detection of PAHs would not be feasible, even at the highest abundance. This is possibly due to the masking of PAHs by other molecules with similar characteristics, such as TiO (retrieved value in Table \ref{tab:pahs_jwst}, -5.52$^{+0.35}_{-0.33}$ compared to the input value of -7.06 ) in the optical range.

The observed deviations in PAH values, though constrained, are particularly notable at a C/O ratio of 0.55. This significance is likely due to the transitional region where planetary transmission spectra shift from being methane- or carbon-bearing species-dominated to water-dominated as the C/O ratio decreases (\citealp{Molaverdikhani2019a}).
\\
\\

\textbf{Constant C/O=0.55, varying [Fe/H]:} 
\\
At 800~K and [Fe/H]=-1, NIRSpec PRISM can detect PAHs  within a 1-$\sigma$ confidence level for abundances of $10^{-5}$ and $10^{-6}$, but both with very strong (>5$\sigma$) detection probability, having difficulties detecting them at $10^{-7}$, with an inconclusive (<2$\sigma$) detection probability. Again, the most probable reason for this non-detection is caused likely by the overlap between PAHs and $\mathrm{CH_4}$ around the 3.3$\mu m$ feature, becoming more pronounced for lower PAH abundances (Fig. \ref{fig:3metal_7}). 
On the other hand, at [Fe/H] of 1, we could only detect PAHs at an abundance of $10^{-5}$ with a very strong detection probability and inconclusive for $10^{-6}$ and $10^{-7}$.

At 1200~K and [Fe/H] of -1, PAHs could be detected for an abundance level of $10^{-5}$, $10^{-6}$ and $10^{-7}$ with a very strong confidence level for the first two and strong for the latter. All abundances of PAHs could be detected for any of the three level cases for a metallicity of 1.

As we increase to 1600 ~K, similar behaviour is found as in the previous temperature.
\subsubsection{PAH and Power law haze}
\textbf{Constant [Fe/H]=0, varying C/O ratio:} 
\\
Incorporating power law clouds into the atmospheric model introduces additional complexity, aiding in differentiating cases where the detection of PAHs could be mistaken for hazes. In our baseline Model 1, which mimics the presence of hazes, the inclusion of power-law cloud parameters allows for a detailed examination of the robustness of PAH detectability. Specifically, our analysis indicates that the inclusion of these parameters does not significantly alter the detection significance of PAHs across most scenarios (as shown in Table \ref{tab:sigma-detection}), except in the case of 10$^{-7}$ where it shifts from not significantly detected to N/A. This suggests that the detectability of PAHs remains largely independent of the power law cloud model parameters \\

\textbf{Constant C/O=0.55, varying [Fe/H]:} 
\\
Incorporating variations in metallicity reveals similar detection behavior for PAHs as observed in the [Fe/H]=0 case, particularly at [Fe/H]=-1 for temperatures of 800~K and 1600~K. In these scenarios, PAHs are detectable at abundances of $10^{-5}$ and $10^{-6}$, both yielding very strong detection probabilities (>5$\sigma$). At 1200~K, PAHs are also robustly detected at these same abundances, while the detection confidence at an abundance of $10^{-7}$ diminishes slightly, remaining at a strong level (>2$\sigma$). As metallicity increases to [Fe/H]=1, there is a marked reduction in detection confidence for lower PAH abundances, indicating increased difficulty in constraining PAHs under higher metallicity conditions.

\subsubsection{PAH, Power law haze, and Grey clouds}
\textbf{Constant [Fe/H]=0, varying C/O ratio:}
\\
Extending the analysis to include gray clouds along with power law clouds should provide a deeper insight into the detectability of PAHs. However, it does not significantly change the detectability outcomes in most cases compared to Model 2. The detectability constraints for PAHs remain largely consistent, with notable deviations only in specific scenarios: at a temperature of T=1600~K and a carbon-to-oxygen ratio (C/O) of 0.55, particularly for PAH abundances of $10^{-6}$ and $10^{-7}$ (<2$\sigma$). These exceptions could be attributed to the complexities in the transition of the C/O ratio, as previously noted. \\


\textbf{Constant C/O=0.55, varying [Fe/H]:} 
\\
At the lowest temperature (800 K), the detection probability of PAHs at an abundance of $10^{-6}$ increases from strong to very strong. At 1200 K, the detection confidence for a PAH abundance of $10^{-7}$ drops slightly, moving from very strong to strong. Meanwhile, at 1600 K, the detection remains inconclusive for PAH abundances of both $10^{-6}$ and $10^{-7}$. For the highest metallicity value ([Fe/H] = 1), PAHs are not detectable at the lowest (800 K) and highest (1600 K) temperatures, indicating that higher metallicities make it more difficult to constrain PAHs. This reduced detectability at high metallicities and temperature extremes is likely due to the increased abundance of non-H molecules such as CO, CO$_2$, CH$_4$, and TiO \citep{Molaverdikhani2019a, Line2021}, which obscure PAH signatures. Specifically, at C/O = 0.55 and [Fe/H] = 1, molecules like CH$_4$ (at 800 K) and TiO (at 1600 K) dominate, further limiting the detection of PAHs (see Fig. 5). However, at 1200 K, PAHs are still detectable with very strong confidence, regardless of the PAH abundance. However, at 1200 K, PAHs are still detectable with very strong confidence, regardless of the PAH abundance.

Overall, the addition of gray clouds alongside power law haze demonstrates that while PAH detectability at a C/O ratio of 0.55 remains generally robust, increasing metallicity introduces additional challenges, particularly at temperature extremes. At both 800 K and 1600 K, PAHs become progressively harder to detect as metallicity increases, suggesting that a combination of high metallicity and either very low or very high temperatures complicate the identification of PAH features. However, at intermediate temperatures (1200 K), PAH detection remains viable across different abundances, even at high metallicity levels, indicating that this temperature range presents a favorable window for identifying PAH features.

\section{Conclusions}\label{sec:conclusions}

In our work, we have investigated the detectability of PAHs on a variety of exoplanet atmospheres at three C/O ratios of 0.3, 0.55 and 1.0, three metallicities of -1, 0 and 1, three temperatures of 800, 1200 and 1600 K, and three PAH abundances of 10$^{-5}$, 10$^{-6}$ and 10$^{-7}$. We simulated the expected observed transit spectra of this planets with the instrument NIRSpec on board the JWST, using PandExo package \citep{batalha2017pasp}. We used the retrieval package petitRADTRANS \citep{molliere2019, molliere2020} to run retrievals and compared the results to the model input parameters. The analysis across various planetary conditions and atmospheric models has allowed us to constrain several key parameters that have the largest influence on the detectability of PAHs, which will help design future observational campaigns with JWST.


Our results indicate that the detectability of PAHs is significantly influenced by the atmospheric temperature, C/O ratio, and the presence of clouds within the atmosphere. 
At a temperature of 1200~K, and particularly for planets with a C/O ratio of 0.3 and 0.55, PAHs are consistently detectable across a range of PAH abundances, down to  10$^{-7}$ (>5$\sigma$), highlighting this specific atmospheric condition as a prime candidate for PAH detectability. The same scenario repeats for 1600~K with a C/O ratio of 0.55 and 1.0, and only at metallicities of [Fe/H]= -1 and 0. At higher metallicities ([Fe/H]=1), the detection of PAHs becomes more challenging for both 800 and 1600~K, whereas at 1200~K, PAHs remain detectable across all three metallicities.

Our analysis suggests that PAHs can be detected even in the presence of clouds. If PAHs are identified through all three models (1, 2, and 3) with consistent detection significance across the models, we could report the feasibility of PAH detection under those conditions. 

For colder planets $\sim$800 K with carbon-rich atmospheres, where CH$_4$ dominates, the PAH signature at 3.3$\mu m$ is masked out, rendering its detection more challenging. On the other hand, warmer planets $\sim$1600 K and low C/O, PAHs are not detected (<2$\sigma$) even for the highest PAH abundance.

In conclusion, our investigation into the detectability of PAHs with one transit observation on JWST with NIRSpec instrument, across a spectrum of exoplanetary atmospheres reveals that these molecules are most likely to be detected, if present, in environments with temperatures around ~1200 K, and with C/O ratios of 0.3 and 0.55, as well as in low C/O ratios in atmospheres of colder planets and carbon-rich atmospheres of warmer planets. This methodology could be adapted for other dedicated facilities with similar wavelength coverage, such as Twinkle and Ariel, by increasing the number of transits to achieve a comparable signal-to-noise ratio.

\section*{Acknowledgements}
This research was supported by the Excellence Cluster ORIGINS, which is funded by the Deutsche Forschungsgemeinschaft (DFG, German Research Foundation) under Germany’s Excellence Strategy – EXC- 2094 – 390783311 (http://www.universe-cluster.de/). The simulations have been carried out on the computing facilities of the Computational Center for Particle and Astrophysics (C2PAP). We would like to express our gratitude to the anonymous reviewer for their insightful feedback, which helped us refine our analysis and enhance the overall depth of the paper. We also thank the editor for their efficient management of the review process and support throughout.

\section*{Data Availability}

The data supporting the findings of this study are available upon reasonable request from the corresponding author, including the 45 forward models calculated with pRT and the synthetic observations with PandExo. 


\bibliographystyle{mnras}
\bibliography{ms} 



\newpage
\appendix

\section{Methodology: figures and tables}
\begin{table}
    \caption{Medium of mass fraction inputs used to create the synthetic spectra for with the configuration T=800, 1200 and 1600K. The values presented here are the median of the mass fraction, for pressures between 0.01mbar and 1 bar.} 
    \centering
    \renewcommand{\arraystretch}{1.5}
    \begin{tabular}{lcccc}
    \hline \hline
    \textbf{Parameter} & \textbf{C/O=0.3} & \textbf{C/O=0.55} & \textbf{C/O=1.0} \\
    T=800K & & & \\
    \hline
    log$\mathrm{_{10}(X_{H_2O}}$) & -1.99 & -2.30 & 2.68 \\
    log$\mathrm{_{10}(X_{CH_4}}$) & -2.51 & 2.50 & -2.50 \\
    log$\mathrm{_{10}(X_{H_2S}}$) & -3.51 & -3.51 & -3.51 \\
    log$\mathrm{_{10}(X_{K}}$) & -12.8 & -13.24 & ... \\
    \hline \hline
    T=1200K & & & \\
    \hline
    log$\mathrm{_{10}(X_{H_2O}}$) & -2.17 & -2.82 & -4.60 \\
    log$\mathrm{_{10}(X_{CH_4}}$) & ... & -4.68 & -2.89 \\
    log$\mathrm{_{10}(X_{K}}$) & ... & -5.91 & ... \\
    log$\mathrm{_{10}(X_{CO}}$) & -2.26 & ... & -2.49 \\
    log$\mathrm{_{10}(X_{CO_2}}$) & -4.86 & -5.46 & ... \\
    log$\mathrm{_{10}(X_{Na}}$) & ... & ... & -4.60 \\
    \hline \hline
    T=1600K & & & \\
    \hline
    log$\mathrm{_{10}(X_{H_2O}}$) & -2.11 & -2.82 & ... \\
    log$\mathrm{_{10}(X_{CH_4}}$) & ... & ... & -4.34 \\
    log$\mathrm{_{10}(X_{CO}}$) & ... & -2.26 & -2.28 \\
    log$\mathrm{_{10}(X_{CO_2}}$) & -5.48 & -6.05 & ... \\
    log$\mathrm{_{10}(X_{TiO}}$) & -7.06 & -6.96 & -9.17 \\
    log$\mathrm{_{10}(X_{VO}}$) & -7.06 & ... & ... \\
    log$\mathrm{_{10}(X_{C_2H_2}}$) & ... & ... & -6.19 \\
    \hline
    \label{tab:T800K}
    \end{tabular}
\end{table}

\begin{figure*}
\centering
\includegraphics[width=\textwidth]{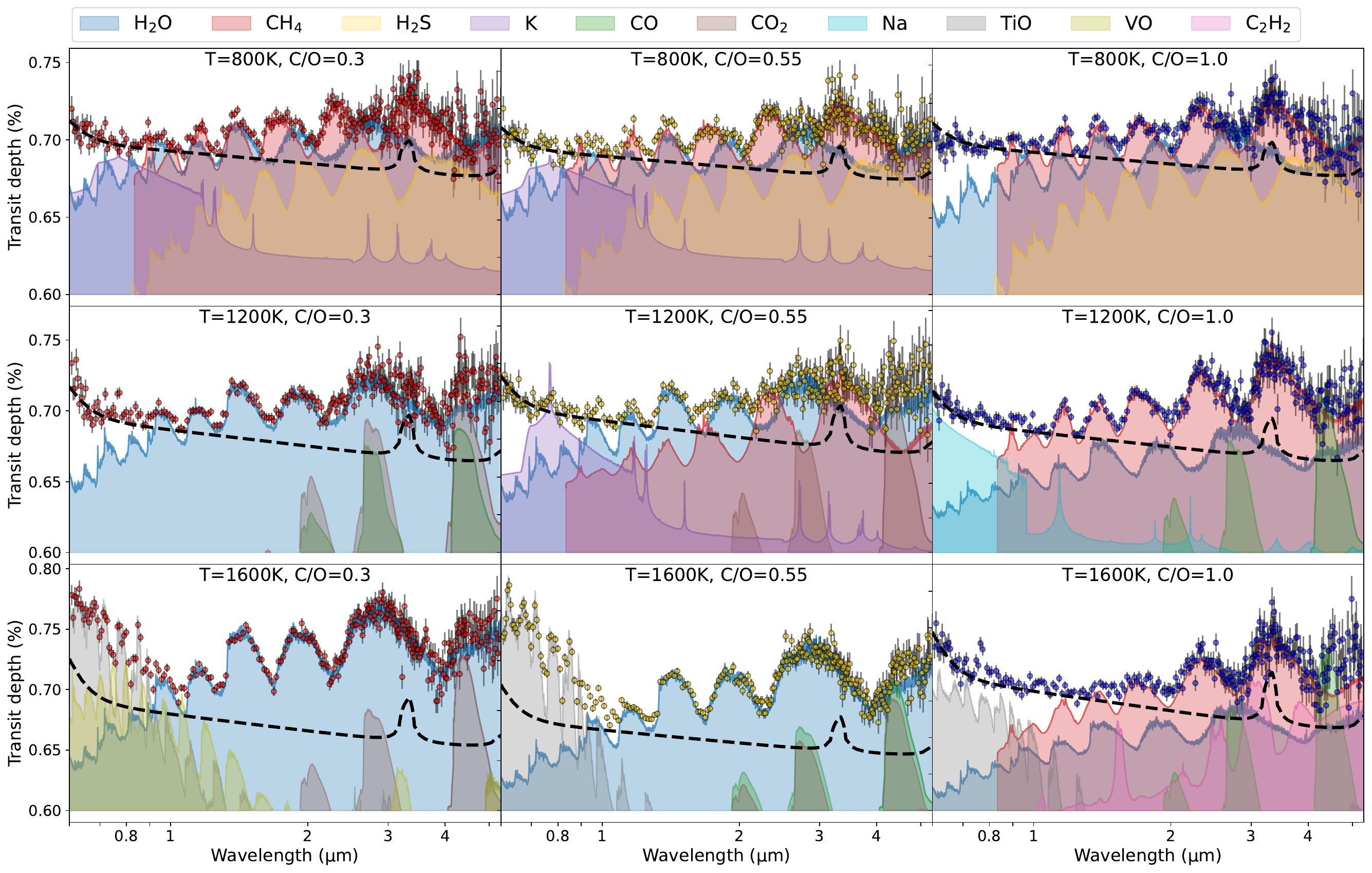}
\caption{Synthetic spectra for $\mathrm{X_{PAH}}$=10$^{-6}$ at temperatures of 800~K (top), 1200~K (middle), and 1600~K (bottom). The forward-modeled spectra (depicted as straight lines) and PandExo-simulated observations (represented by dotted points) for each planet are illustrated. The C/O ratios of 0.3 (left) are displayed in red, 0.55 (center) in yellow, and 1.0 (right) in blue. Various colors are assigned to indicate the contribution of each molecule to the model spectrum. While the primary features are attributed to H$_2$O, CH$_4$, CO, and CO$_2$, the spectrum also showcases a diverse range of other molecules.}
\label{fig:forward_model_6}
\end{figure*}

\begin{figure*}
\centering
\includegraphics[width=\textwidth]{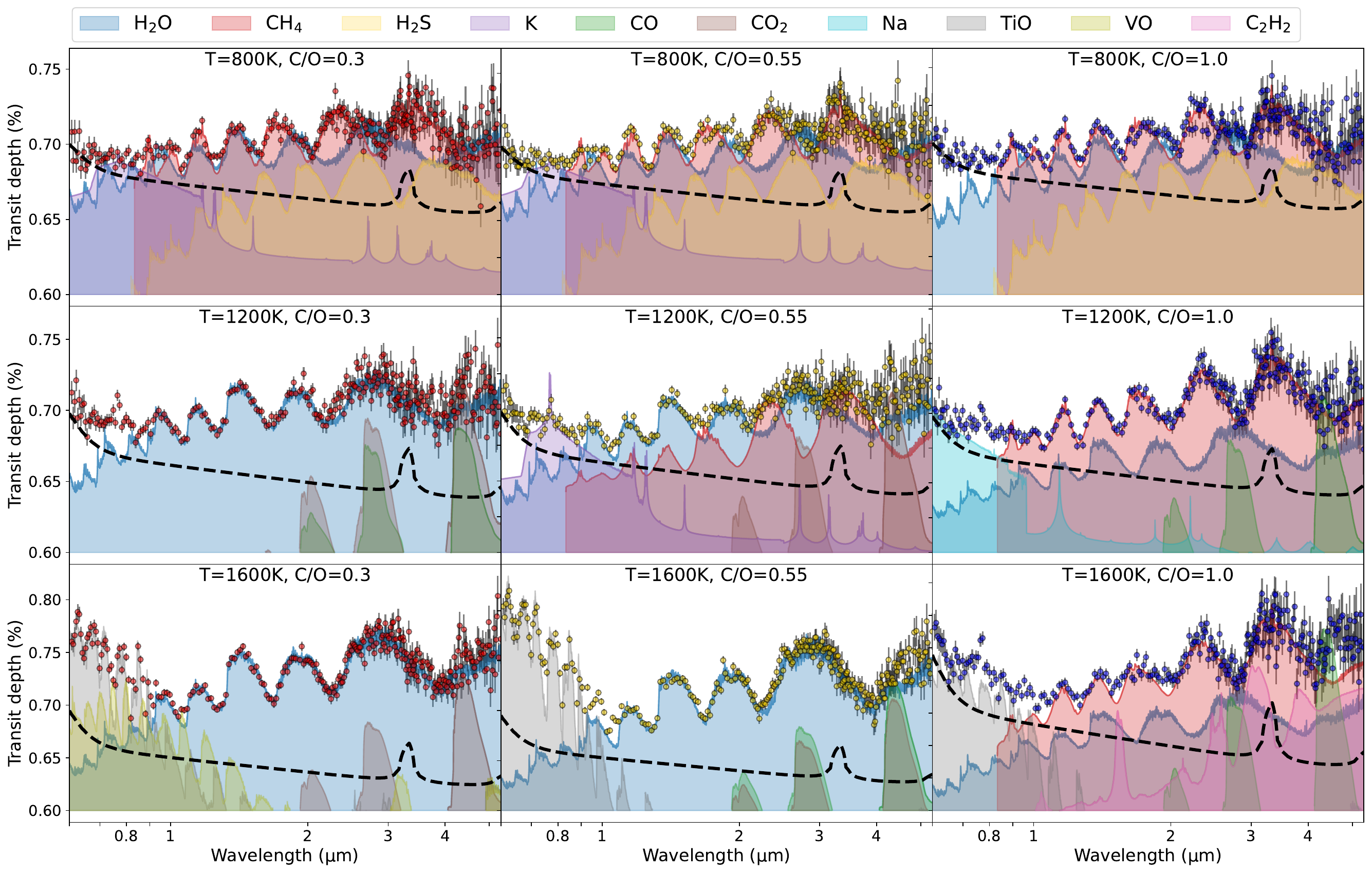}
\caption{Synthetic spectra for $\mathrm{X_{PAH}}$=10$^{-7}$ at temperatures of 800~K (top), 1200~K (middle), and 1600~K (bottom). The forward-modeled spectra (depicted as straight lines) and PandExo-simulated observations (represented by dotted points) for each planet are illustrated. The C/O ratios of 0.3 (left) are displayed in red, 0.55 (center) in yellow, and 1.0 (right) in blue. Various colors are assigned to indicate the contribution of each molecule to the model spectrum. While the primary features are attributed to H$_2$O, CH$_4$, CO, and CO$_2$, the spectrum also showcases a diverse range of other molecules.}
\label{fig:forward_model_7}
\end{figure*}

\begin{figure*}
\centering
\includegraphics[width=\textwidth]{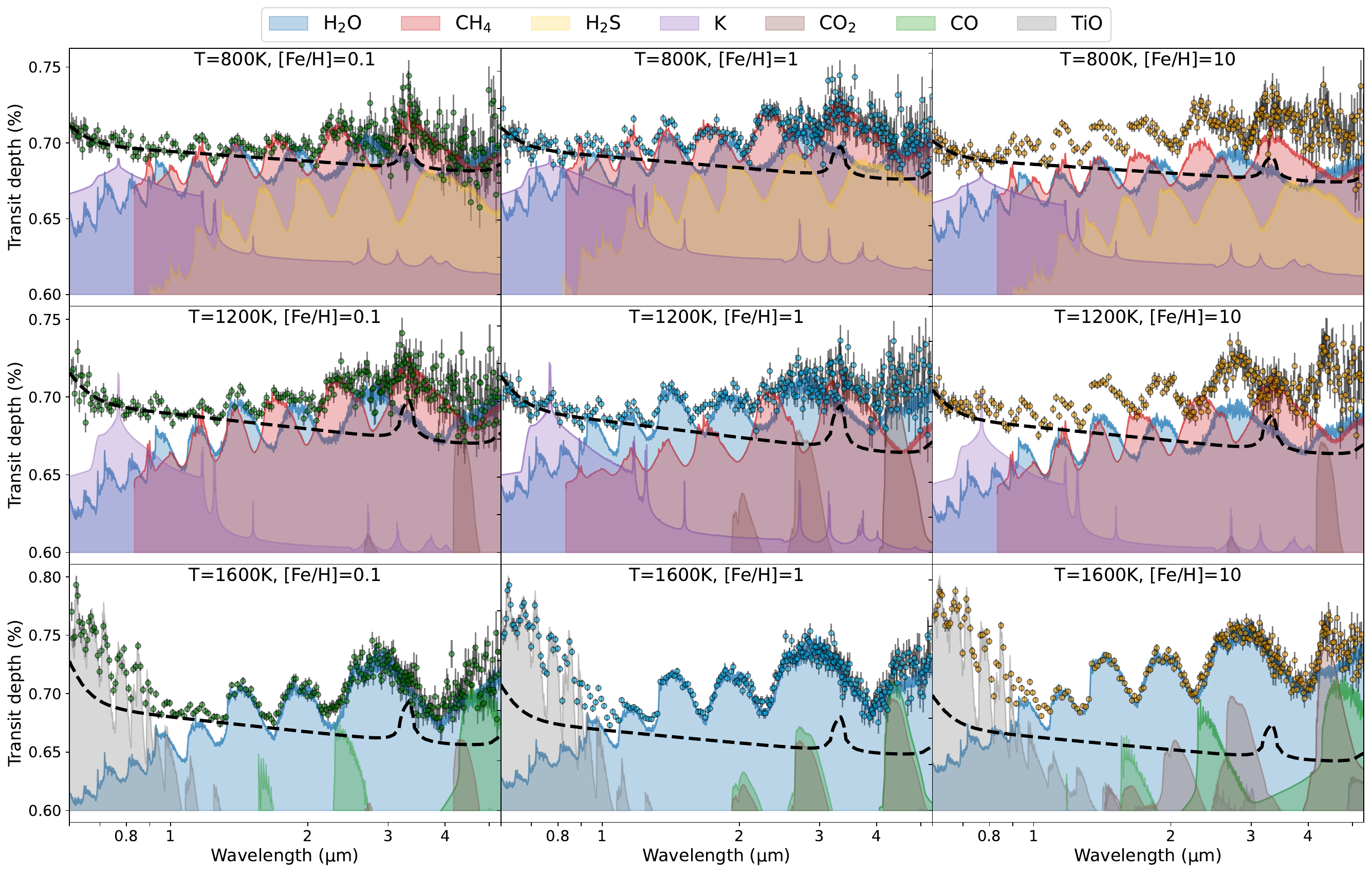}
\caption{Synthetic spectra for a $\mathrm{X_{PAH}=10^{-6}}$ at temperatures of 800~K (top), 1200~K (middle), and 1600~K (bottom). Similar to Fig. \ref{fig:forward_model_6}, with constant C/O (0.55) and varying metallicities ([Fe/H]) are displayed for 0.1 (left) in green, 1 (middle) in cyan and 10 (right) in yellow.}

\label{fig:3metal_6}
\end{figure*}

\begin{figure*}
\centering
\includegraphics[width=\textwidth]{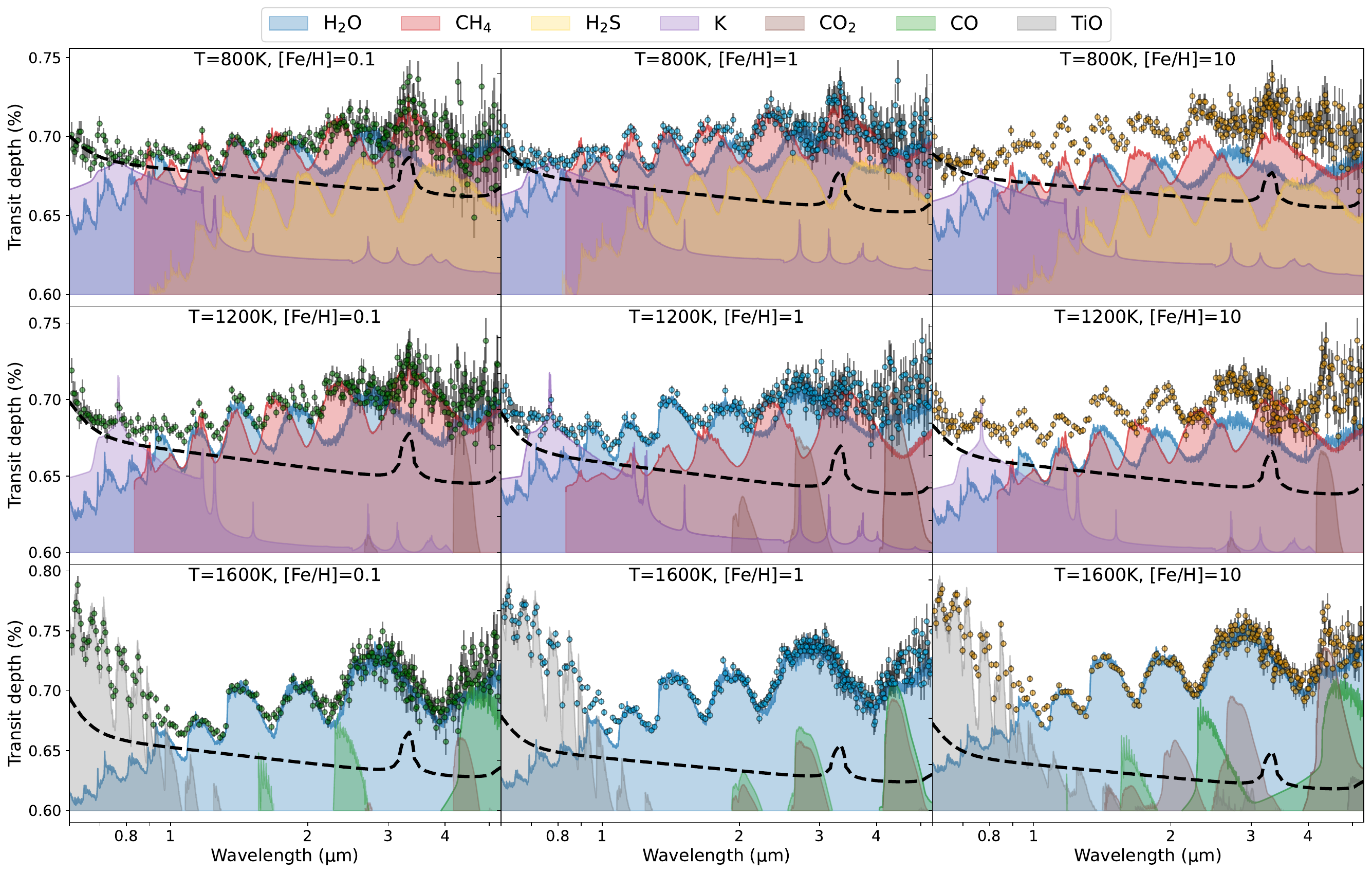}
\caption{$\mathrm{X_{PAH}=10^{-7}}$ at temperatures of 800~K (top), 1200~K (middle) and 1600~K (bottom). Similar to Fig. \ref{fig:forward_model_7}, with constant C/O (0.55) and varying metallicities ([Fe/H]) are displayed for 0.1 (left) in green, 1 (middle) in cyan and 10 (right) in yellow.}

\label{fig:3metal_7}
\end{figure*}

\section{Results: figures and tables}
\begin{table*}
\centering
\caption{Derived abundance Posterior Probabilities for a PAH input abundance of X$_{\text{PAH}}$ = 10$^{-5}$. The table enumerates the posterior distributions of abundance parameters for various atmospheric constituents across different C/O ratios and temperature conditions. Each entry details the median logarithmic abundance with corresponding 1-sigma uncertainties for temperatures T = 800K, 1200K, and 1600K, and for C/O ratios of 0.3, 0.55, and 1.0. Ellipses (...) indicate the absence of these molecules in the retrievals.  The results include posteriors for major volatiles, such as H$_2$O, CH$_4$, and CO, alongside trace species like K, Na, TiO, and VO. Additional parameters, such as surface gravity (log(g)), planetary radius (R$\mathrm{_p}$), and effective temperature (T$_{\mathrm{eff}}$), are also reported, offering a comprehensive overview of the planetary characteristics inferred from the data.}
\renewcommand{\arraystretch}{1.5}
\resizebox{\textwidth}{!}{%
\begin{tabular}{lcccc ccccccc cccc}
\hline \hline
& \multicolumn{3}{c}{T=800K} & & \multicolumn{3}{c}{T=1200K}& & \multicolumn{3}{c}{T=1600K} \\
\cline{2-4} \cline{6-8} \cline{10-12}  
\textbf{Parameter} &  \textbf{C/O=0.3} & \textbf{C/O=0.55} & \textbf{C/O=1.0} & & \textbf{C/O=0.3} & \textbf{C/O=0.55} & \textbf{C/O=1.0} & & \textbf{C/O=0.3} & \textbf{C/O=0.55} & \textbf{C/O=1.0} \\
\hline
Input $\mathrm{H_2O}$ & -1.99 & -2.30 & -2.68 & & -2.17 & -2.82 & -4.60 & & -2.11 & -2.61 & ... \\
log$\mathrm{_{10}(X_{H_2O}}$)  & -2.63 $^{+0.50}_{-0.54}$ & -3.38 $^{+0.38}_{-0.36}$ & -2.84 $^{+0.59}_{-0.68}$ & & -2.26 $^{+0.37}_{-0.34}$ & -3.64 $^{+0.42}_{-0.37}$ & -4.24 $^{+0.52}_{-0.45}$ & & -1.88 $^{+0.35}_{-0.33}$ & -3.00 $^{+0.23}_{-0.21}$ & ... \\ \hdashline

Input $\mathrm{CH_4}$ & -2.51 & -2.50 & -2.50 & & ... & -4.68 & -2.89 & & ... & ... & -4.34 \\
log$\mathrm{_{10}(X_{CH_4}}$) & -2.59 $^{+0.35}_{-0.35}$ & -3.30 $^{+0.24}_{-0.20}$ & -2.23 $^{+0.40}_{-0.38}$ & & ... & -4.77 $^{+0.42}_{-0.36}$ & -2.71 $^{+0.40}_{-0.37}$ & & ... & ... & -4.71 $^{+0.30}_{-0.25}$ \\
\hdashline
Input $\mathrm{H_2S}$ & -3.51 & -3.51 & -3.51 & & ... & ... & ... & & ... & ... & ... \\
log$\mathrm{_{10}(X_{H_2S}}$) & -11.45$^{+5.07}_{-5.01}$ & -10.61$^{+6.02}_{-5.66}$ & -11.43 $^{+5.43}_{-5.33}$ & & ... & ... & ... & & ... & ... & ... \\ \hdashline

Input K & -12.8 & -13.24 & ... & & ... & -5.91. & ... & & ... & ... & ... \\
log$\mathrm{_{10}(X_{K}}$)  & -13.28 $^{+3.91}_{-3.86}$ & -13.47 $^{+3.92}_{-3.90}$ & ... & & ... & -13.07 $^{+4.17}_{-3.94}$ & ... & & ... & ... & ...\\ \hdashline

Input CO & ... & ... & ... & & -2.26 & ... & -2.49 & & ... & -2.26 & -2.28 \\
log$\mathrm{_{10}(X_{CO}}$) & ... & ... & ... & & -11.24 $^{+5.27}_{-5.32}$ & ... &  -3.21 $^{+1.19}_{-8.75}$ & & ... & -1.94 $^{+0.37}_{-0.37}$ & -2.25 $^{+0.50}_{-0.45}$ \\ \hdashline

Input CO$_2$& ... & ... & ... & & -4.86 & -5.46 & ... & & -5.48 & -6.05 & ... \\
log$\mathrm{_{10}(X_{CO_2}}$)  & ... & ... & ... & & -4.52 $^{0.43}_{-0.41}$ & -5.14 $^{+0.48}_{-0.44}$ & ... & & -4.82 $^{+0.39}_{-0.40}$ & -12.83 $^{+4.44}_{-4.36}$ & ... \\ \hdashline

Input Na & ... & ... & ... & & ... & ... & -4.60 & & ... & ... & ... \\
log$\mathrm{_{10}(X_{Na}}$) &... & ... & ... & & ... & ... & -11.22 $^{+5.91}_{-0.54}$ & & ... & ... & ...\\ \hdashline

Input TiO & ... & ... & ... & & ... & ... & ... & & -7.06 & -6.96. & -9.17\\
log$\mathrm{_{10}(X_{TiO}}$)  & ... & ... & ... & & ... & ... & ... & & -5.52 $^{+0.35}_{-0.33}$ & -5.86 $^{+0.23}_{-0.22}$ & -8.76 $^{+1.37}_{-7.54}$ \\ \hdashline

Input VO & ... & ... & ... & & ... & ... & ... & & -7.06 & ... & ... \\
log$_{10}$(X$\mathrm{_{VO}}$) &... & ... & ... & & ... & ... & ... & & -6.77 $^{+0.36}_{-0.35}$ & ... & ... \\ \hdashline

Input $\mathrm{{C_2H_2}}$& ... & ... & ... & & ... & ... & ... & & ... & ... & -6.19 \\
log$\mathrm{_{10}(X_{C_2H_2}}$)  &... & ... & ... & & ... & ... & ... & & ... & ... &  -13.03 $^{+4.64}_{-4.24}$ \\ 

\hline
\textit{Other parameters} &  &  &  & &  & &  & &  &  & \\

log(g) (cgs) &  3.03 $^{+0.05}_{-0.05}$ & 2.94 $^{+0.05}_{-0.05}$ & 3.01 $^{+0.05}_{-0.05}$ & & 3.08 $^{+0.04}_{-0.03}$ & 3.31 $^{+0.07}_{-0.07}$ & 3.04 $^{+0.04}_{-0.04}$ & & 3.08 $^{+0.02}_{-0.02}$ & 2.96 $^{+0.01}_{-0.01}$ & 3.19 $^{+0.07}_{-0.01}$ \\

R$\mathrm{_{p}(R_J)}$ &  0.84 $^{+0.02}_{-0.02}$ & 0.84 $^{+0.01}_{-0.01}$ & 0.83 $^{+0.02}_{-0.02}$ & & 0.832 $^{+0.003}_{-0.004}$ & 0.836 $^{+0.003}_{-0.007}$ & 0.832 $^{+0.004}_{-0.004}$ & & 0.836 $^{+0.005}_{-0.006}$ & 0.836 $^{+0.005}_{-0.006}$ & 0.834 $^{+0.003}_{-0.003}$\\

T$_{\mathrm{eff}}$(K) &  557 $^{+42}_{-041}$ & 521 $^{+45}_{-45}$ & 510 $^{+42}_{-43}$ & & 1076 $^{+84}_{-82}$ & 1415 $^{+211}_{-205}$ & 856 $^{+54}_{-55}$ & & 1435 $^{+57}_{-49}$ & 1718 $^{+33}_{-32}$ & 1726 $^{+272}_{-298}$\\

\hline
\end{tabular}%
}

\label{tab:all_results_jwst_5}
\end{table*}
\begin{table*}
    \centering
    \caption{Comparative Analysis of "sigma" significance across atmospheric models for various planets with three different metallicities: 0.1, 1 and 10$\times$ solar. This figure presents a systematic comparison of Bayes factors for different planetary atmospheric set-ups, delineated into three models: (1) an atmosphere with exclusively PAHs contributing to haze; (2) an atmosphere featuring both PAHs and a power law cloud distribution; (3) an expanded version of Model 2 including a cloud deck. Calculations of B$_{01}$ for Models 2 and 3 were limited to planets with significant detection evidence in Model 1. The strength of detection is visually differentiated by shades of pink, categorizing the detection probability as very strong ("sigma" > 5$\mathrm{\sigma}$), strong ("sigma" > 3.6$\mathrm{\sigma}$), moderate ("sigma" > 2.7$\mathrm{\sigma}$), weak ("sigma" > 2.0$\mathrm{\sigma}$), or inconclusive ("sigma" < 2$\mathrm{\sigma}$).}
    \renewcommand{\arraystretch}{1.5}
    \begin{tabular}{c|cccc|ccc|ccc}
        
        & & \multicolumn{3}{c}{T=800K} & \multicolumn{3}{c}{T=1200K} & \multicolumn{3}{c}{T=1600K} \\
        \hline 
        \diagbox{[Fe/H]}{$\mathrm{\log X_{PAH}}$} & & -5 & -6 & -7 & -5 & -6 & -7 & -5 & -6 & -7 \\ \hline

        -1 & Model 1 & \cellcolor{purple!20}9.25$\mathrm{\sigma}$ & \cellcolor{purple!20}7.14$\mathrm{\sigma}$ & \cellcolor{purple!5} 1.23$\mathrm{\sigma}$ & \cellcolor{purple!20}10.67$\mathrm{\sigma}$ &\cellcolor{purple!20}8.54$\mathrm{\sigma}$ & \cellcolor{purple!10}3.92$\mathrm{\sigma}$ & \cellcolor{purple!20}18.00$\mathrm{\sigma}$ & \cellcolor{purple!20}10.58$\mathrm{\sigma}$  & \cellcolor{purple!10}4.10$\mathrm{\sigma}$\\
         & Model 2 & \cellcolor{purple!20}9.15$\mathrm{\sigma}$ & \cellcolor{purple!20}7.13$\mathrm{\sigma}$ & N/A &\cellcolor{purple!20} 10.69$\mathrm{\sigma}$ & \cellcolor{purple!20}8.53$\mathrm{\sigma}$ & \cellcolor{purple!10}3.87$\mathrm{\sigma}$ & \cellcolor{purple!20}18.05$\mathrm{\sigma}$ & \cellcolor{purple!20}10.50$\mathrm{\sigma}$ & \cellcolor{purple!10}4.13$\mathrm{\sigma}$ \\
         & Model 3 & \cellcolor{purple!20}9.14$\mathrm{\sigma}$  & \cellcolor{purple!20}7.02$\mathrm{\sigma}$ & N/A & \cellcolor{purple!20}10.67$\mathrm{\sigma}$ & \cellcolor{purple!20}8.53$\mathrm{\sigma}$ & \cellcolor{purple!10}3.92$\mathrm{\sigma}$ & \cellcolor{purple!20}12.69$\mathrm{\sigma}$ & \cellcolor{purple!10}1.37$\mathrm{\sigma}$ & \cellcolor{purple!5}0.95$\mathrm{\sigma}$ \\
        \hline

        0 & Model 1 & \cellcolor{purple!20} 8.92$\mathrm{\sigma}$ & \cellcolor{purple!10} 4.65$\mathrm{\sigma}$ & \cellcolor{purple!5}0.93$\mathrm{\sigma}$ & \cellcolor{purple!20}10.30$\mathrm{\sigma}$ & \cellcolor{purple!20} 6.85$\mathrm{\sigma}$ & \cellcolor{purple!20} 5.12$\mathrm{\sigma}$ & \cellcolor{purple!20} 12.61$\mathrm{\sigma}$ & \cellcolor{purple!20} 8.01$\mathrm{\sigma}$ & \cellcolor{purple!20} 6.47$\mathrm{\sigma}$ \\
          & Model 2 & \cellcolor{purple!20} 9.39$\mathrm{\sigma}$ & \cellcolor{purple!10} 4.70$\mathrm{\sigma}$ & N/A & \cellcolor{purple!20}10.34$\mathrm{\sigma}$ & \cellcolor{purple!20}6.72$\mathrm{\sigma}$ & \cellcolor{purple!20} 5.06$\mathrm{\sigma}$ & \cellcolor{purple!20} 15.27$\mathrm{\sigma}$ &  \cellcolor{purple!20} 7.98$\sigma$ & \cellcolor{purple!20} 6.66$\sigma$  \\
        & Model 3 & \cellcolor{purple!20} 8.83$\sigma$ & \cellcolor{purple!10} 4.49$\mathrm{\sigma}$ & N/A  &\cellcolor{purple!20}10.36$\mathrm{\sigma}$ & \cellcolor{purple!20}6.93$\mathrm{\sigma}$ & \cellcolor{purple!20} 5.02$\mathrm{\sigma}$ & \cellcolor{purple!20} 6.61$\mathrm{\sigma}$ & \cellcolor{purple!5} 1.67$\sigma$ & \cellcolor{purple!5} 0.96$\sigma$ \\ 
          \hline
        1 & Model 1 & \cellcolor{purple!20}5.57$\mathrm{\sigma}$ & \cellcolor{purple!5}1.67$\mathrm{\sigma}$ & \cellcolor{purple!5}1.02$\mathrm{\sigma}$ & \cellcolor{purple!20}13.04$\mathrm{\sigma}$ & \cellcolor{purple!20}6.82$\mathrm{\sigma}$ & \cellcolor{purple!20}5.45$\mathrm{\sigma}$ & \cellcolor{purple!20}10.70$\mathrm{\sigma}$ & \cellcolor{purple!20}6.95$\mathrm{\sigma}$ & \cellcolor{purple!20}5.02$\mathrm{\sigma}$ \\
          & Model 2 & \cellcolor{purple!20}5.51$\mathrm{\sigma}$ & N/A & N/A & \cellcolor{purple!20}12.98$\mathrm{\sigma}$ & \cellcolor{purple!20}6.75$\mathrm{\sigma}$ & \cellcolor{purple!20}5.55$\mathrm{\sigma}$ & \cellcolor{purple!20} 10.64$\mathrm{\sigma}$ &  \cellcolor{purple!20} 6.88$\sigma$ & \cellcolor{purple!20} 6.11$\sigma$  \\
        & Model 3 & \cellcolor{purple!5}2.25$\sigma$ & N/A & N/A  &\cellcolor{purple!20}12.66$\mathrm{\sigma}$ & \cellcolor{purple!20}6.76$\mathrm{\sigma}$ & \cellcolor{purple!20}5.34$\mathrm{\sigma}$ & \cellcolor{purple!5} 0.93$\mathrm{\sigma}$ & \cellcolor{purple!5}0.93$\sigma$ & \cellcolor{purple!5} 0.93$\sigma$ \\ 
          \hline

        \hline
    \end{tabular}
\label{tab:sigma-detection-metal}
\end{table*}
\begin{table*}
    \centering
    \caption{Derived posterior values for the logarithmic concentration of PAHs (log~X$_{\text{PAH}}$) in the atmospheric models of planets under various conditions. The analysis is segmented by three distinct metallicities (0.1, 1, and 10$\times$ solar) and C/O=0.55, across three temperature regimes (800~K, 1200~K, and 1600~K), and for three logX$_{\text{PAH}}$ values (-5, -6, -7). The posterior values are provided for Models 1, 2, and 3. Cells marked 'N/A' signify scenarios where retrievals are not performed, adhering to the precedent that retrievals are only conducted for set-ups with a prior detection of PAHs, as  highlighted by the green and yellow shaded regions.}
    \renewcommand{\arraystretch}{1.5}

    \resizebox{\textwidth}{!}{
    \begin{tabular}{c|cccc|ccc|ccc}
        & & \multicolumn{3}{c}{T=800K} & \multicolumn{3}{c}{T=1200K} & \multicolumn{3}{c}{T=1600K} \\
        \hline 
        \diagbox{[Fe/H]}{$\mathrm{\log X_{PAH}}$} & & -5 & -6 & -7 & -5 & -6 & -7 & -5 & -6 & -7 \\ \hline

        -1 & Model 1 & -4.49 $^{+0.85}_{-0.78}$ & -5.65 $^{+0.47}_{-0.33}$ & -11.64 $^{+4.14}_{-5.83}$ & -4.73 $^{+0.63}_{-0.47}$ & -6.04 $^{+0.48}_{-0.27}$ & -7.09 $^{+0.24}_{-0.25}$  & -5.00 $^{+0.33}_{-0.22}$ & -6.01 $^{+0.11}_{-0.10}$ & -6.68 $^{+0.13}_{-0.16}$ \\
         & Model 2 & -4.90 $^{+0.58}_{-0.45}$ & -5.69 $^{+0.42}_{-0.31}$ & N/A & -4.79 $^{+0.60}_{-0.46}$ & -6.06 $^{+0.39}_{-0.25}$ & -7.11 $^{+0.25}_{-0.23}$ & -4.81 $^{+0.50}_{-0.34}$ & -6.02 $^{+0.11}_{-0.09}$ & -6.68 $^{+0.14}_{-0.15}$ \\
         & Model 3 &  -4.82 $^{+0.61}_{-0.48}$ & -5.61 $^{+0.51}_{-0.35}$ & N/A & -4.57 $^{+0.66}_{-0.59}$ & -6.01 $^{+0.56}_{-0.27}$ & -7.09 $^{+0.25}_{-0.24}$ & -4.78 $^{+0.50}_{-0.35}$ & -10.78 $^{+4.61}_{-6.24}$ & -13.80 $^{+4.44}_{-4.26}$ \\
         \hline
        0 & Model 1 & -5.85 $^{+0.24}_{-0.19}$ & -6.00 $^{+0.26}_{-0.33}$ & -12.43 $^{+4.75}_{-4.89}$ &   -5.05 $^{+0.39}_{-0.29}$  & -6.28 $^{+0.34}_{-0.25}$ & -6.95 $^{+0.27}_{-0.22}$ & -5.16 $^{+0.23}_{-0.20}$ &   -5.26 $^{+0.27}_{-0.35}$ & -5.75 $^{+0.33}_{-0.41}$ \\
         & Model 2 & -5.92 $^{+0.26}_{-0.18}$ &   -5.92 $^{+0.26}_{-0.33}$ & N/A & -4.96 $^{+0.48}_{-0.35}$  & -6.32 $^{+0.28}_{-0.22}$ & -6.98 $^{+0.21}_{-0.19}$  & -5.14 $^{+0.24}_{-0.22}$ &   -5.37$^{+0.28}_{-0.30}$ & -5.69$^{+0.33}_{-0.39}$ \\
         & Model 3 & -5.78 $^{+0.36}_{-0.23}$ & -5.88 $^{+0.27}_{-0.33}$ & N/A & -5.07 $^{+0.23}_{-0.26}$  &  -6.32 $^{+0.36}_{-0.31}$ & -6.95 $^{+0.27}_{-0.22}$ & -5.31 $^{+0.28}_{-0.17}$ & -11.81$^{+4.40}_{-6.14}$ & -13.42$^{+4.94}_{-4.40}$ \\
         \hline
        1 & Model 1 & -5.65 $^{+0.27}_{-0.38}$ & -7.38 $^{+0.73}_{-7.68}$ & -13.41 $^{+4.49}_{-4.42}$ &   -4.89 $^{+0.46}_{-0.37}$ & -6.30 $^{+0.40}_{-0.28}$ & -6.76 $^{+0.24}_{-0.22}$ & -4.77 $^{+0.30}_{-0.42}$ & -5.08 $^{+0.24}_{-0.31}$ & -5.74 $^{+0.26}_{-0.34}$ \\
         & Model 2 & -5.68 $^{+0.28}_{-0.37}$ & N/A & N/A & -4.97 $^{+0.38}_{-0.33}$ & -6.06 $^{+0.39}_{-0.25}$ & -6.76 $^{+0.24}_{-0.22}$ & -4.82 $^{+0.30}_{-0.40}$ & -5.09 $^{+0.24}_{-0.29}$ & -5.71$^{+0.24}_{-0.31}$ \\
         & Model 3 & -5.66 $^{+0.27}_{-0.35}$ & N/A  & N/A & -4.87 $^{+0.46}_{-0.37}$ & -6.30 $^{+0.40}_{-0.27}$ & -6.75 $^{+0.24}_{-0.27}$ & -12.35 $^{+5.79}_{-5.16}$ & -13.34$^{+4.83}_{-4.38}$ & -13.41$^{+4.53}_{-4.46}$ \\
         \hline
    \end{tabular}%
     }

    \label{tab:pahs_jwst_feh}
\end{table*}

\begin{table*}
\caption{Mass fraction posteriors from the set-ups with $X_{\text{PAH}}=10^{-6}$}
\centering
\renewcommand{\arraystretch}{1.5}
\resizebox{\textwidth}{!}{%
\begin{tabular}{lcccc ccccccc cccc}
\hline \hline
& \multicolumn{3}{c}{T=800~K} & & \multicolumn{3}{c}{T=1200~K}& & \multicolumn{3}{c}{T=1600~K} \\
\cline{2-4} \cline{6-8} \cline{10-12}  
\textbf{Parameter} &  \textbf{C/O=0.3} & \textbf{C/O=0.55} & \textbf{C/O=1.0} & & \textbf{C/O=0.3} & \textbf{C/O=0.55} & \textbf{C/O=1.0} & & \textbf{C/O=0.3} & \textbf{C/O=0.55} & \textbf{C/O=1.0} \\
\hline
Input $\mathrm{H_2O}$ & -1.99 & -2.30 & -2.68 & & -2.17 & -2.82 & -4.60 & & -2.11 & -2.61 & ... \\
log$\mathrm{_{10}(X_{\text{H}_2\text{O}})}$  & -1.94 $^{+0.37}_{-0.35}$ & -1.68 $^{+0.30}_{-0.40}$ & -3.74 $^{+0.279}_{-0.31}$ & & -2.92 $^{+0.16}_{-0.14}$ & -2.91 $^{+0.29}_{-0.25}$ & -4.26 $^{+0.32}_{-0.31}$ & & -1.29 $^{+0.20}_{-0.40}$ & -2.37 $^{+0.23}_{-0.31}$ & ... \\ \hdashline

Input $\mathrm{CH_4}$ & -2.51 & -2.50 & -2.50 & & ... & -4.68 & -2.89 & & ... & ... & -4.34 \\
log$\mathrm{_{10}(X_{\text{CH}_4})}$ & -2.58 $^{+0.27}_{-0.25}$ & -2.30 $^{+0.23}_{-0.28}$ & -3.11 $^{+0.14}_{-0.13}$ & & ... & -4.41 $^{+0.27}_{-0.24}$ &  -3.13 $^{+0.28}_{-0.17}$ & & ... & ... & -4.26 $^{+0.20}_{-0.18}$ \\ \hdashline

Input $\mathrm{H_2S}$ & -3.51 & -3.51 & -3.51 & & ... & ... & ... & & ... & ... & ... \\
log$\mathrm{_{10}(X_{\text{H}_2\text{S}})}$ & -11.67$^{+5.08}_{-5.00}$ & -10.72$^{+5.81}_{-5.51}$ & -11.46 $^{+5.55}_{-5.23}$ & & ... & ... & ... & & ... & ... & ... \\ \hdashline

Input K & -12.8 & -13.24 & ... & & ... & -5.91. & ... & & ... & ... & ... \\
log$\mathrm{_{10}(X_{\text{K}})}$  & -14.45 $^{+3.38}_{-0.36}$ & -11.07 $^{+3.70}_{-5.25}$ & ... & & ... & -5.99 $^{+0.56}_{-0.47}$ & ... & & ... & ... & ...\\ \hdashline

Input CO & ... & ... & ... & & -2.26 & ... & -2.49 & & ... & -2.26 & -2.28 \\
log$\mathrm{_{10}(X_{\text{CO}})}$ & ... & ... & ... & & -10.99 $^{+6.36}_{-5.74}$ & ... &  -3.38 $^{+0.73}_{-6.38}$ & & ... & -1.65 $^{+0.33}_{-0.46}$ & -2.14 $^{+0.34}_{-0.30}$ \\ \hdashline

Input CO$_2$& ... & ... & ... & & -4.86 & -5.46 & ... & & -5.48 & -6.05 & ... \\
log$\mathrm{_{10}(X_{\text{CO}_2})}$  & ... & ... & ... & & -5.34 $^{+0.27}_{-0.28}$ & -5.05 $^{+0.41}_{-0.42}$ & ... & & -4.53 $^{+0.30}_{-0.46}$ & -11.78 $^{+5.24}_{-5.06}$ & ... \\  \hdashline

Input Na & ... & ... & ... & & ... & ... & -4.60 & & ... & ... & ... \\
log$\mathrm{_{10}(X_{\text{Na}})}$ &... & ... &  ... & & ... & ... & -3.73 $^{+0.73}_{-0.47}$ & & ... & ... & ...\\ \hdashline

Input TiO & ... & ... & ... & & ... & ... & ... & & -7.06 & -6.96. & -9.17\\
log$\mathrm{_{10}(X_{\text{TiO}})}$  & ... & ... & ... & & ... & ... & ... & & -5.04 $^{+0.19}_{-0.39}$ & -5.27 $^{+0.23}_{-0.31}$ & -7.89 $^{+0.29}_{-0.27}$ \\ \hdashline

Input VO & ... & ... & ... & & ... & ... & ... & & -7.06 & ... & ... \\
log$\mathrm{_{10}(X_{\text{VO}})}$ &... & ... & ... & & ... & ... & ... & & -6.29 $^{+0.27}_{-0.40}$ & ... & ... \\ \hdashline

Input $\mathrm{C_2H_2}$& ... & ... & ... & & ... & ... & ... & & ... & ... & -6.19 \\
log$\mathrm{_{10}(X_{\text{C}_2\text{H}_2})}$  &... & ... & ... & & ... & ... & ... & & ... & ... & -13.16 $^{+3.97}_{-3.94}$ \\ \hdashline

\hline
\textit{Other parameters} &  &  &  & &  & &  & &  &  & \\

log(g) (cgs) &  3.01 $^{+0.04}_{-0.04}$ & 3.06 $^{+0.04}_{-0.04}$ & 3.01 $^{+0.05}_{-0.05}$ & & 3.12 $^{+0.03}_{-0.03}$ & 3.16 $^{+0.04}_{-0.04}$ & 3.04 $^{+0.04}_{-0.04}$ & & 3.03 $^{+0.02}_{-0.02}$ & 2.96 $^{+0.01}_{-0.01}$ & 3.32 $^{+0.04}_{-0.04}$ \\

R$\mathrm{_{p}(R_J)}$ &  0.83 $^{+0.00}_{-0.00}$ & 0.83 $^{+0.00}_{-0.00}$ & 0.84 $^{+0.00}_{-0.00}$ & & 0.84 $^{+0.00}_{-0.00}$ & 0.83 $^{+0.00}_{-0.00}$ & 0.832 $^{+0.004}_{-0.004}$ & & 0.82 $^{+0.01}_{-0.00}$ & 0.82 $^{+0.01}_{-0.01}$ & 0.83 $^{+0.00}_{-0.00}$\\

T$_{\mathrm{eff}}$(K) &  547 $^{+37}_{-34}$ & 601 $^{+39}_{-40}$ & 504 $^{+39}_{-36}$ & & 1165 $^{+68}_{-80}$ & 1042 $^{+85}_{-68}$ & 856 $^{+54}_{-55}$ & & 1605 $^{+36}_{-45}$ & 1749 $^{+27}_{-25}$ & 1933 $^{+155}_{-175}$\\

\hline
\label{tab:all_results_jwst_6}
\end{tabular}%
}
\end{table*}

\begin{table*}
\caption{Mass fraction posteriors from the set-ups with $X_{\text{PAH}}=10^{-7}$.}
\centering
\renewcommand{\arraystretch}{1.5}
\resizebox{\textwidth}{!}{%
\begin{tabular}{lcccc ccccccc cccc}
\hline \hline
& \multicolumn{3}{c}{T=800~K} & & \multicolumn{3}{c}{T=1200~K}& & \multicolumn{3}{c}{T=1600~K} \\
\cline{2-4} \cline{6-8} \cline{10-12}  
\textbf{Parameter} &  \textbf{C/O=0.3} & \textbf{C/O=0.55} & \textbf{C/O=1.0} & & \textbf{C/O=0.3} & \textbf{C/O=0.55} & \textbf{C/O=1.0} & & \textbf{C/O=0.3} & \textbf{C/O=0.55} & \textbf{C/O=1.0} \\
\hline
Input $\mathrm{H_2O}$ & -1.99 & -2.30 & -2.68 & & -2.17 & -2.82 & -4.60 & & -2.11 & -2.61 & ... \\
log$\mathrm{_{10}(X_{\text{H}_2\text{O}})}$  & -2.43 $^{+0.20}_{-0.19}$ & -2.47 $^{+0.28}_{-0.23}$ & -2.54 $^{+0.21}_{-0.34}$ & & -2.98 $^{+0.12}_{-0.12}$ & -3.30 $^{+0.27}_{-0.24}$ & -3.66 $^{+0.46}_{-0.50}$ & & -1.10 $^{+0.07}_{-0.13}$ & -2.66 $^{+0.28}_{-0.41}$ & ... \\ \hdashline

Input $\mathrm{CH_4}$ & -2.51 & -2.50 & -2.50 & & ... & -4.68 & -2.89 & & ... & ... & -4.34 \\
log$\mathrm{_{10}(X_{\text{CH}_4})}$ & -2.79 $^{+0.15}_{-0.13}$ & -2.61 $^{+0.18}_{-0.14}$ & -2.68 $^{+0.15}_{-0.20}$ & & ... & -4.83 $^{+0.30}_{-0.30}$ & -2.51 $^{+0.42}_{-0.41}$ & & ... & ... & -3.95 $^{+0.12}_{-0.11}$ \\ \hdashline

Input $\mathrm{H_2S}$ & -3.51 & -3.51 & -3.51 & & ... & ... & ... & & ... & ... & ... \\
log$\mathrm{_{10}(X_{\text{H}_2\text{S}})}$ & -3.15$^{+0.54}_{-9.13}$ & -2.30$^{+0.32}_{-0.33}$ & -11.92 $^{+5.31}_{-4.52}$ & & ... & ... & ... & & ... & ... & ... \\ \hdashline

Input K & -12.8 & -13.24 & ... & & ... & -5.91. & ... & & ... & ... & ... \\
log$\mathrm{_{10}(X_{\text{K}})}$  & -14.39 $^{+3.42}_{-3.45}$ & -11.93 $^{+2.81}_{-5.17}$ & ... & & ... & -6.55 $^{+0.43}_{-0.37}$ & ... & & ... & ... & ...\\ \hdashline

Input CO & ... & ... & ... & & -2.26 & ... & -2.49 & & ... & -2.26 & -2.28 \\
log$\mathrm{_{10}(X_{\text{CO}})}$ & ... & ... & ... & & -3.35 $^{+0.73}_{-8.37}$ & ... &  -1.92 $^{+0.56}_{-0.95}$ & & ... & -1.53 $^{+0.36}_{-0.59}$ & -1.37 $^{+0.21}_{-0.26}$ \\ \hdashline

Input CO$_2$& ... & ... & ... & & -4.86 & -5.46 & ... & & -5.48 & -6.05 & ... \\
log$\mathrm{_{10}(X_{\text{CO}_2})}$  & ... & ... & ... & & -6.49$^{+0.64}_{-7.76}$ & -5.85 $^{+0.42}_{-0.42}$ & ... & & -12.63$^{+5.47}_{-4.89}$ & -12.63 $^{+5.24}_{-5.06}$ & ... \\  \hdashline
Input Na & ... & ... & ... & & ... & ... & -4.60 & & ... & ... & ... \\
log$\mathrm{_{10}(X_{\text{Na}})}$ &... & ... &  ... & & ... & ... & -2.68 $^{+0.88}_{-0.86}$ & & ... & ... & ...\\ \hdashline

Input TiO & ... & ... & ... & & ... & ... & ... & & -7.06 & -6.96. & -9.17\\
log$\mathrm{_{10}(X_{\text{TiO}})}$  & ... & ... & ... & & ... & ... & ... & & -4.97 $^{+0.07}_{-0.13}$ & -5.57 $^{+0.28}_{-0.40}$ & -7.52 $^{+0.12}_{-0.12}$ \\ \hdashline

Input VO & ... & ... & ... & & ... & ... & ... & & -7.06 & ... & ... \\
log$\mathrm{_{10}(X_{\text{VO}})}$ &... & ... & ... & & ... & ... & ... & & -6.22 $^{+0.13}_{-0.16}$ & ... & ... \\ \hdashline
Input $\mathrm{{C_2H_2}}$& ... & ... & ... & & ... & ... & ... & & ... & ... & -6.19 \\
log$\mathrm{_{10}(X_{\text{C}_2\text{H}_2})}$  &... & ... & ... & & ... & ... & ... & & ... & ... & -12.76 $^{+4.81}_{-4.56}$ \\

\hline
\textit{Other parameters} &  &  &  & &  & &  & &  &  & \\

log(g) (cgs) & 2.98 $^{+0.04}_{-0.04}$ & 3.00 $^{+0.04}_{-0.04}$ & 3.04 $^{+0.04}_{-0.05}$ & & 3.08 $^{+0.03}_{-0.03}$ & 3.11 $^{+0.03}_{-0.03}$ & 3.08 $^{+0.02}_{-0.02}$ & & 2.97 $^{+0.02}_{-0.01}$ & 2.94 $^{+0.01}_{-0.01}$ & 3.31 $^{+0.03}_{-0.03}$ \\

R$\mathrm{_{p}(R_J}$) &  0.84 $^{+0.00}_{-0.00}$ & 0.83 $^{+0.00}_{-0.00}$ & 0.83 $^{+0.00}_{-0.00}$ & & 0.84 $^{+0.00}_{-0.00}$ & 0.84 $^{+0.00}_{-0.00}$ & 0.83 $^{+0.00}_{-0.00}$ & & 0.81 $^{+0.00}_{-0.00}$ & 0.83 $^{+0.01}_{-0.01}$ & 0.83 $^{+0.00}_{-0.00}$\\

T$_{\mathrm{eff}}$(K) &  552 $^{+40}_{-35}$ & 533 $^{+38}_{-34}$ & 568 $^{+27}_{-47}$ & & 1137 $^{+59}_{-63}$ & 1172 $^{+70}_{-82}$ & 981 $^{+33}_{-32}$ & & 1512 $^{+48}_{-41}$ & 1800 $^{+36}_{-32}$ & 1593 $^{+107}_{-103}$\\

\hline
\label{tab:all_results_jwst_7}
\end{tabular}%
}
\end{table*}


\begin{figure*}  
  \centering  
  \begin{minipage}{0.9\linewidth}
    \includegraphics[width=\textwidth]{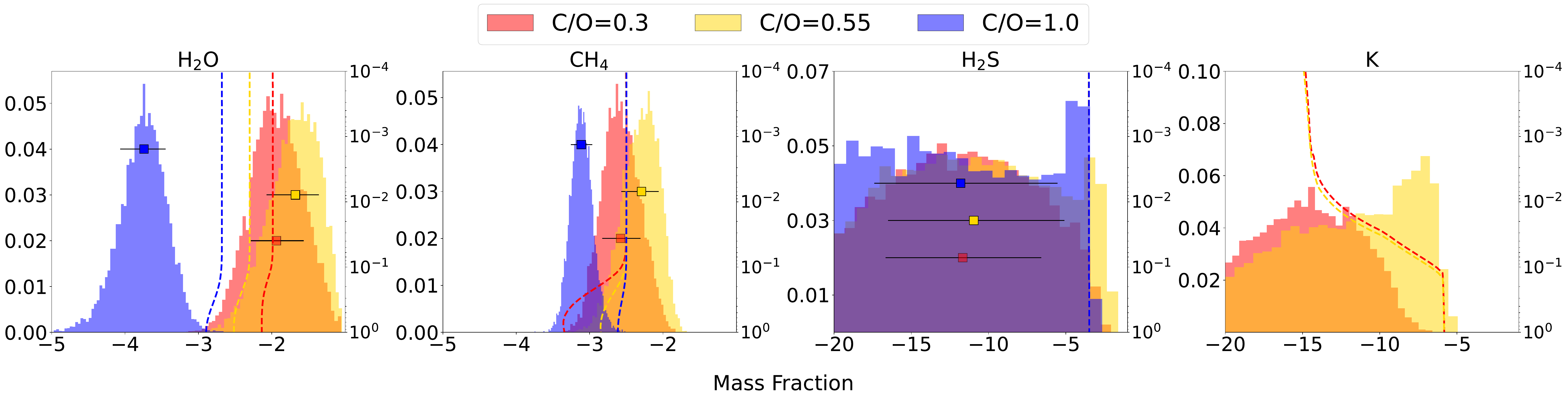}
    \subcaption{T=800~K.}
    \label{fig:800-6-results_dist}
  \end{minipage}
  \begin{minipage}{0.9\linewidth}
    \includegraphics[width=\textwidth]{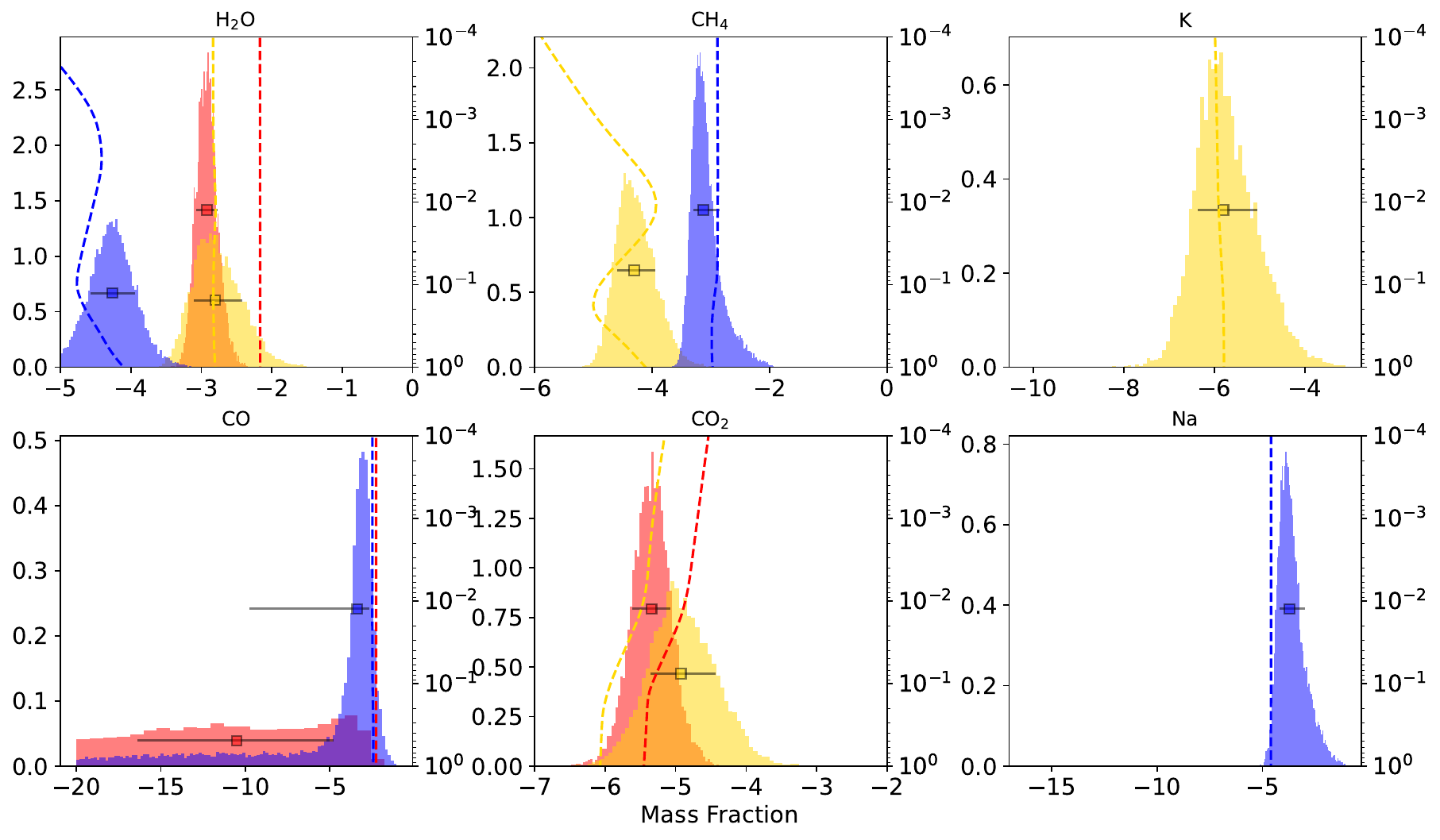}
    \subcaption{T=1200~K}
    \label{fig:1200-6-results_dist}
  \end{minipage}
  \begin{minipage}{0.9\linewidth}
    \includegraphics[width=\textwidth]{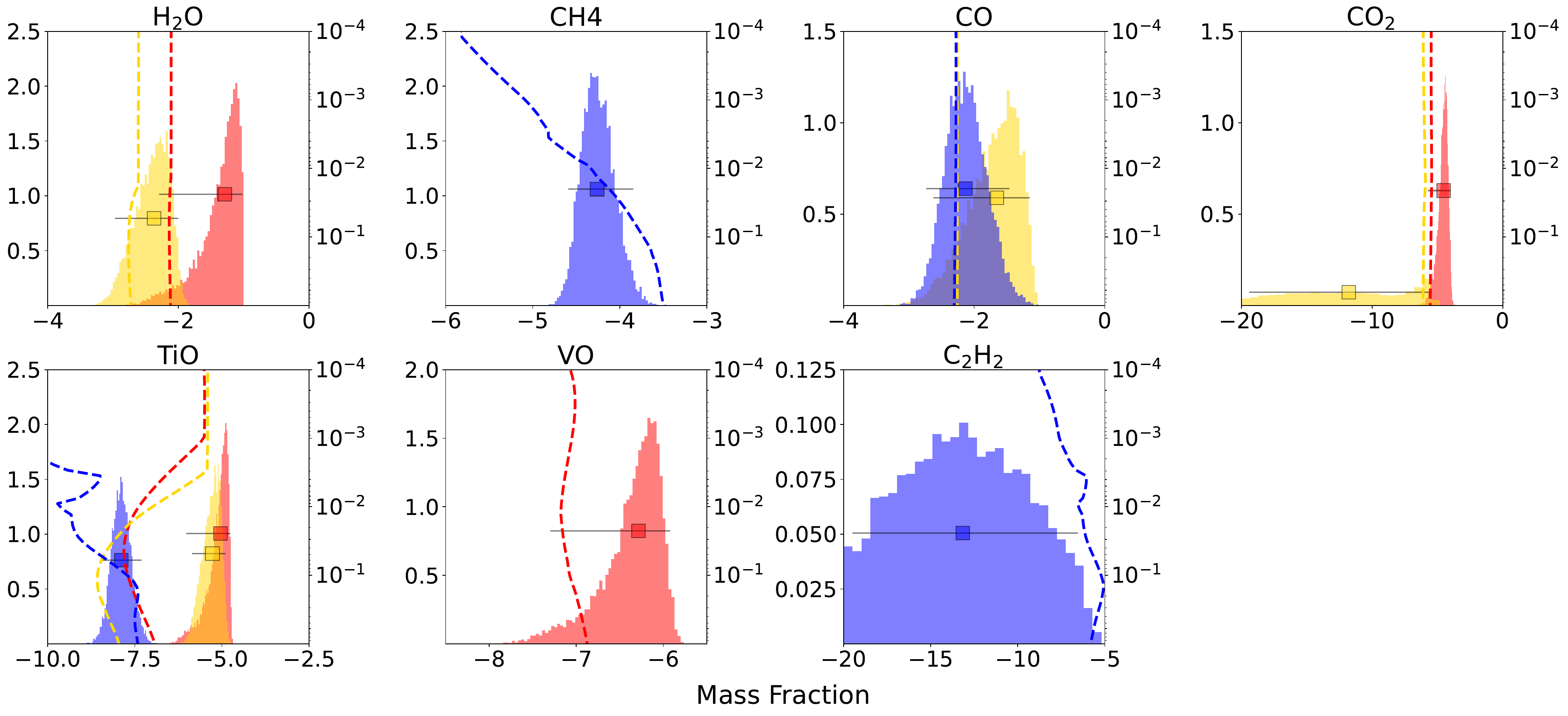}
    \subcaption{T=1600~K}
    \label{fig:1600-6-results_dist}
  \end{minipage}
  \caption{Retrieved posterior probability distributions (left y-axis) and pressure profiles (right y-axis) for mass fractions of different molecule's planet set-up described in Section \ref{sec:methodology} for $\mathrm{X_{PAH}=10^{-6}}$. C/O ratios 0.3, 0.55 and 1.0 are depicted in red, yellow and 1.0, respectively. Mass fraction inputs at different pressures are shown in dashed lines. The error bar denotes each distribution's median and corresponding 1$\sigma$ interval. The abundance estimates are shown in Table \ref{tab:all_results_jwst_6}}
  \label{fig:posteriors6}
\end{figure*}
\begin{figure*}  
  \centering  
  \begin{minipage}{0.9\linewidth}
    \includegraphics[width=\textwidth]{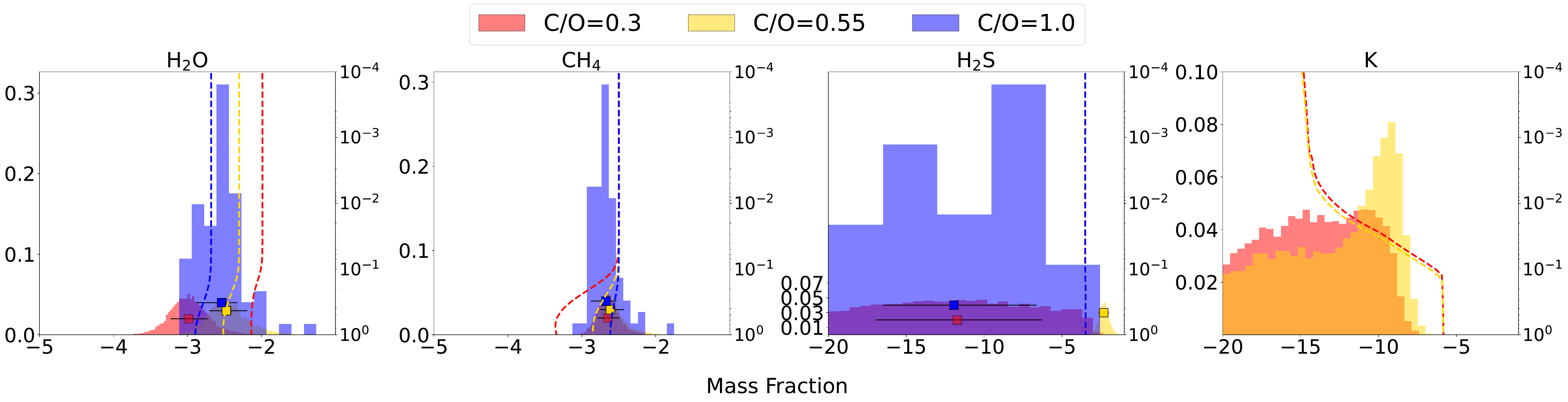}
    \subcaption{T=800~K.}
    \label{fig:800-7-results_dist}
  \end{minipage}
  \begin{minipage}{0.9\linewidth}
    \includegraphics[width=\textwidth]{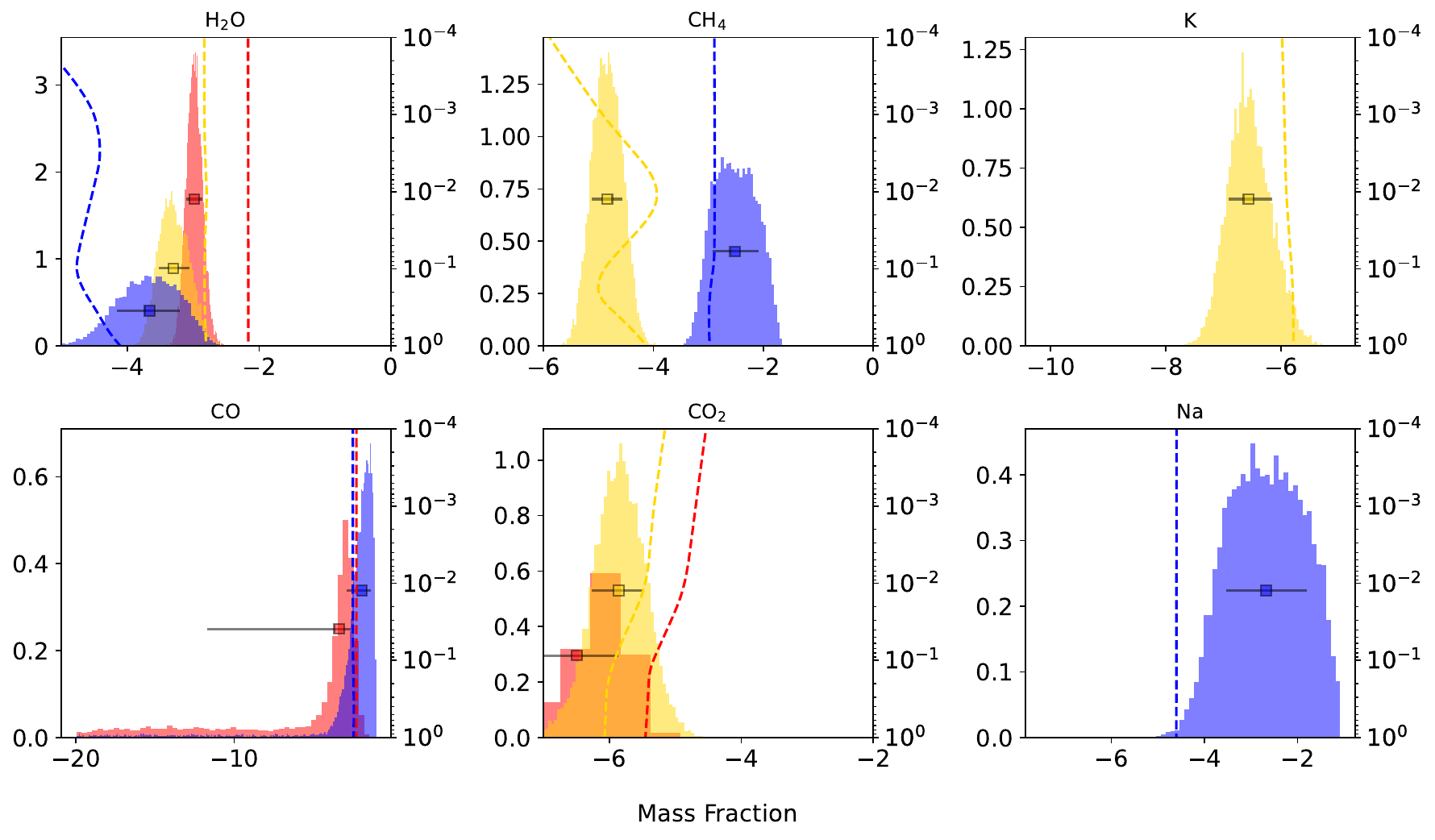}
    \subcaption{T=1200~K}
    \label{fig:1200-7-results_dist}
  \end{minipage}
  \begin{minipage}{0.9\linewidth}
    \includegraphics[width=\textwidth]{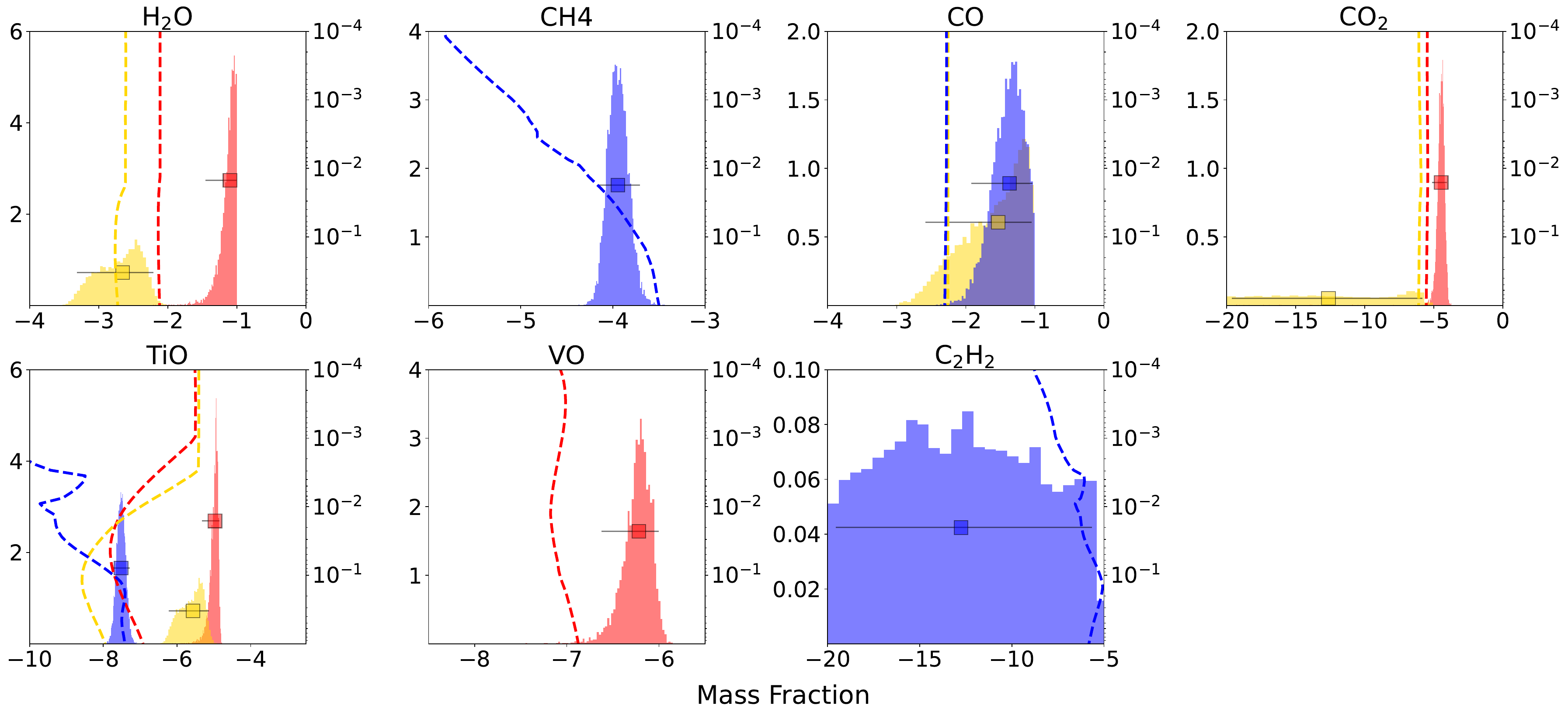}
    \subcaption{T=1600~K}
    \label{fig:1600-7-results_dist}
  \end{minipage}
  \caption{Retrieved posterior probability distributions (left y-axis) and pressure profiles (right y-axis) for mass fractions of different molecule's planet set-up described in Section \ref{sec:methodology} for $\mathrm{X_{PAH}=10^{-7}}$. C/O ratios 0.3, 0.55 and 1.0 are depicted in red, yellow and 1.0, respectively. Mass fraction inputs at different pressures are shown in dashed lines. The error bar denotes each distribution's median and corresponding 1$\sigma$ interval. The abundance estimates are shown in Table \ref{tab:all_results_jwst_7}}
  \label{fig:posteriors7}
\end{figure*}


\begin{figure*}

  \begin{minipage}[b]{\linewidth}
    \centering
    \includegraphics[width=\textwidth]{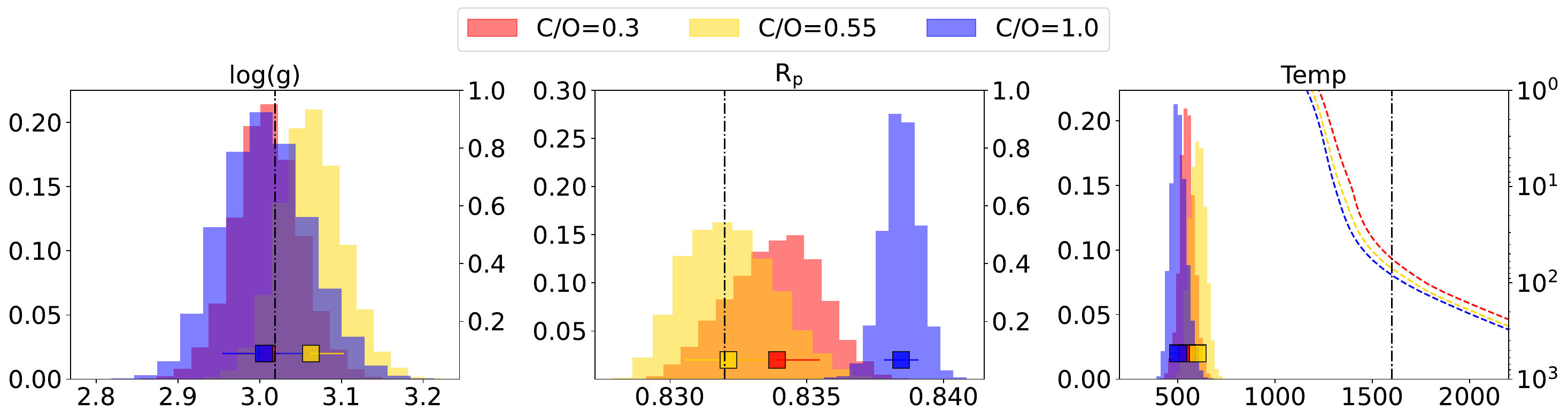}
    \subcaption{X$\mathrm{_{PAH}}$ of 10$^{-6}$}
    \label{fig:800-6-temps}
  \end{minipage}
  
  \begin{minipage}[b]{\linewidth}
    \centering
    \includegraphics[width=\textwidth]{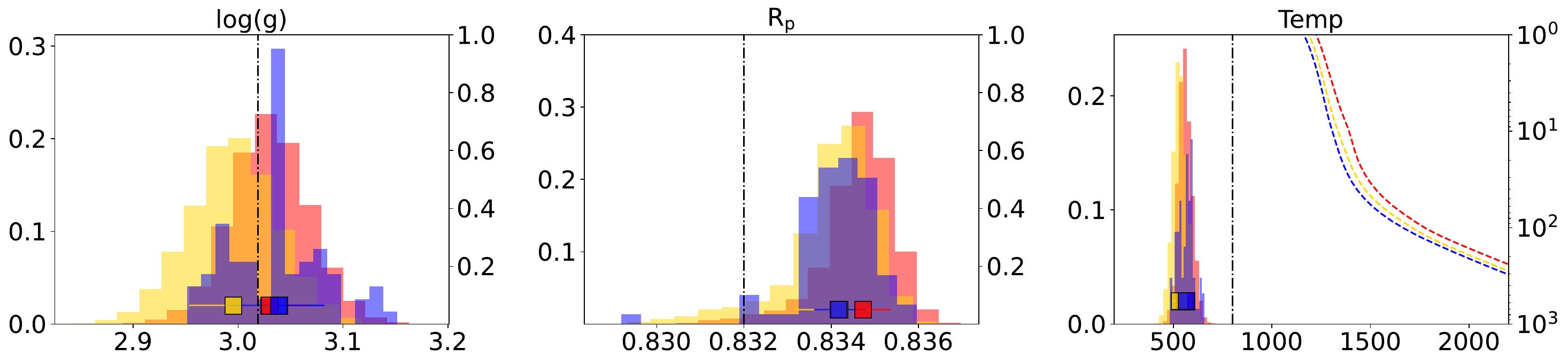}
    \subcaption{X$\mathrm{_{PAH}}$ of 10$^{-7}$}
    \label{fig:800-7-temps}
  \end{minipage}
  \caption{Distribution probabilities of log(g), R$_{p}$ and Temperature for planets with temperature of 800~K, C/O from 0.3 to 1.0, and X$\mathrm{_{PAH}}$ of 10$^{-6}$ and 10$^{-7}$. The dashed black lines correspond to the input parameters.}
\end{figure*}
\begin{figure*}
  \begin{minipage}[b]{\linewidth}
    \centering
    \includegraphics[width=\textwidth]{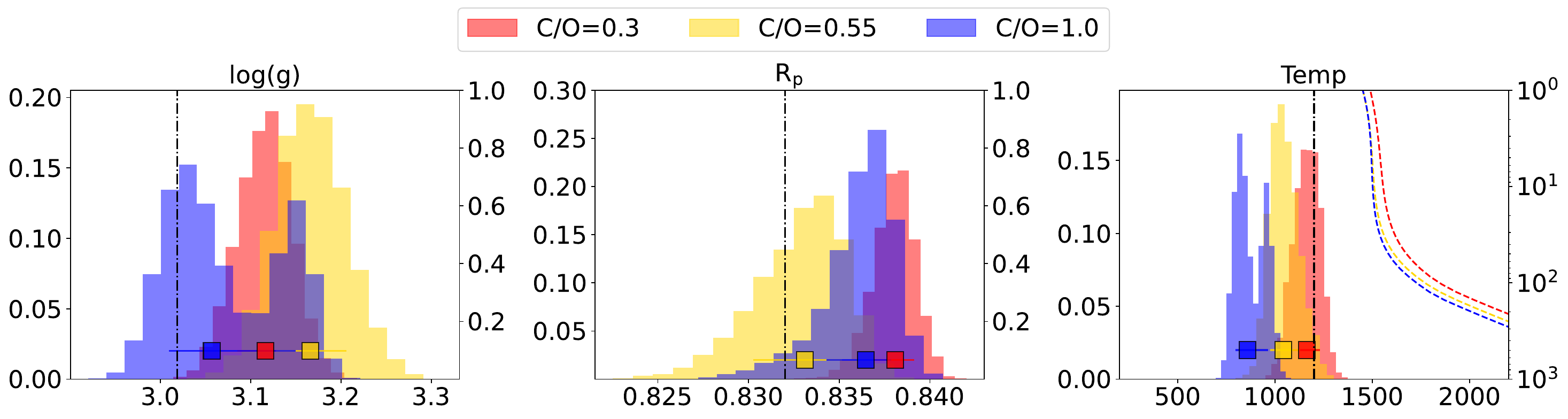}
    \subcaption{X$\mathrm{_{PAH}}$ of 10$^{-7}$}
    \label{fig:1200-6-temps}
  \end{minipage}

  \begin{minipage}[b]{\linewidth}
    \centering
    \includegraphics[width=\textwidth]{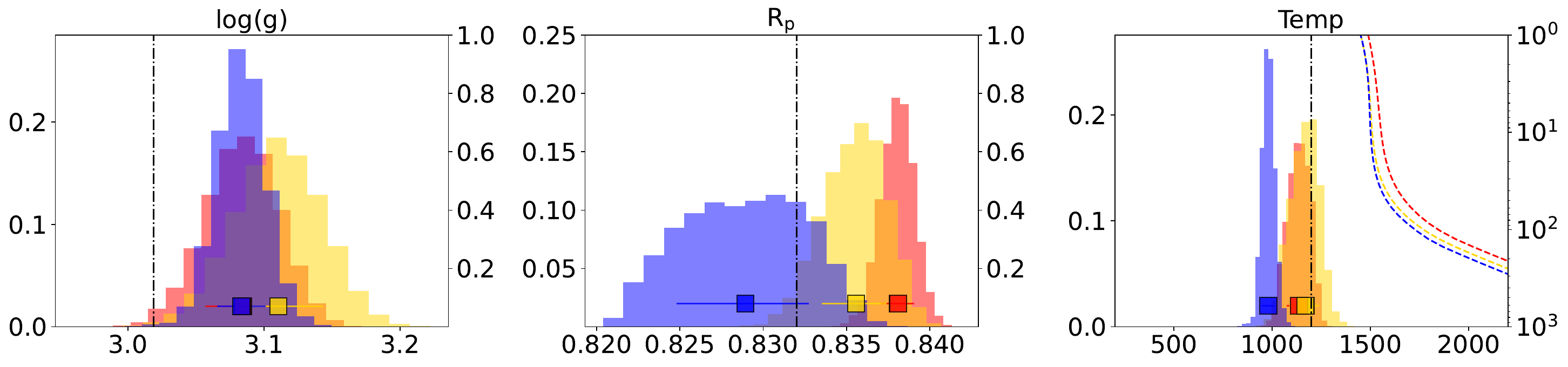}
    \subcaption{X$\mathrm{_{PAH}}$ of 10$^{-6}$}
    \label{fig:1200-7-temps}
  \end{minipage}
\caption{Distribution probabilities of log(g), R$_{p}$ and Temperature for planets with temperature of 1200~K, C/O from 0.3 to 1.0, and X$\mathrm{_{PAH}}$ of 10$^{-6}$ and 10$^{-7}$. The dashed black lines correspond to the input parameters.}

\end{figure*}
\begin{figure*}

  \begin{minipage}[b]{\linewidth}
    \centering
    \includegraphics[width=\textwidth]{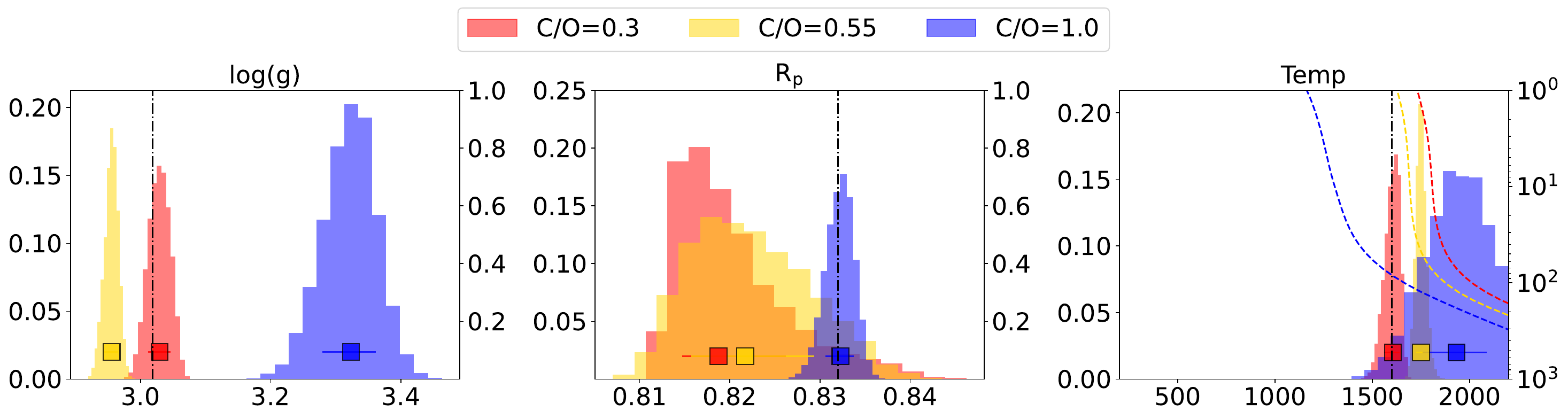}
    \subcaption{}{}
    \label{fig:1600-6-temps}
  \end{minipage}
  
  \begin{minipage}[b]{\linewidth}
    \centering
    \includegraphics[width=\textwidth]{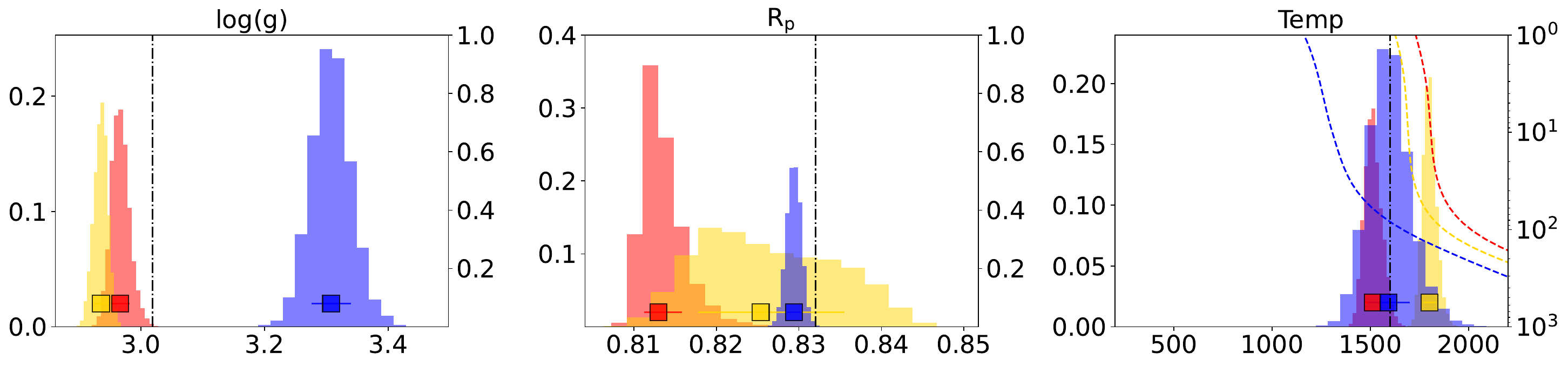}
   \subcaption{X$\mathrm{_{PAH}}$ of 10$^{-7}$}
    \label{fig:1600-7-temps}
  \end{minipage}
  \caption{Distribution probabilities of log(g), R$_{p}$ and Temperature for planets with temperature of 1600~K, C/O from 0.3 to 1.0, and X$\mathrm{_{PAH}}$ of 10$^{-6}$ and 10$^{-7}$. The dashed black lines correspond to the input parameters.}
\end{figure*}

\bsp	
\label{lastpage}
\end{document}